\documentclass[12pt]{article}
\pdfoutput=1
\usepackage{amsfonts}
\usepackage{amsmath, amssymb}
\usepackage{youngtab}
\usepackage{hyperref}
\usepackage{psfrag}
\usepackage{feynmp}
\usepackage[pdftex]{graphicx}
\usepackage{braket}
\usepackage{bm}
\usepackage{subfigure}

\allowdisplaybreaks[1]

\Yboxdim{5pt}

\newcommand{\nn}{\nonumber}

\makeatletter
    
    \@addtoreset{equation}{section}
  \makeatother

\renewcommand{\title}[1]{\vbox{\center\LARGE{#1}}\vspace{5mm}}
\renewcommand{\author}[1]{\vbox{\center\large#1}\vspace{5mm}}

\begin{document}
\bibliographystyle{utphys}


\begin{titlepage}
\begin{center}
\vspace{5mm}
\hfill {\tt 
UT-Komaba-19-2 
}\\
\hfill {\tt 
IPMU19-0058
}\\
\vspace{20mm}

\title{
Wall-crossing and operator ordering
\\
\vspace{2mm}
for 't Hooft operators in $\mathcal{N}=2$ gauge theories
}
\vspace{7mm}

Hirotaka Hayashi$^a$, Takuya Okuda$^b$, and Yutaka Yoshida$^c$

\vspace{6mm}

$^a$Department of Physics, School of Science, 
Tokai University\\
Hiratsuka-shi, 
Kanagawa 259-1292, Japan\\
\vskip 1mm
\href{mailto:h.hayashi@tokai.ac.jp}{\tt h.hayashi@tokai.ac.jp}
\vspace{3mm}

$^b$Graduate School of Arts and Sciences, University of Tokyo\\
Komaba, Meguro-ku, Tokyo 153-8902, Japan \\
\vskip 1mm
\href{mailto:takuya@hep1.c.u-tokyo.ac.jp}{\tt takuya@hep1.c.u-tokyo.ac.jp}

\vspace{3mm}
$^c$Kavli IPMU (WPI), UTIAS, University of Tokyo \\
 Kashiwa, Chiba 277-8583, Japan \\
\href{mailto:yutaka.yoshida@ipmu.jp}{\tt yutaka.yoshida@ipmu.jp}

\end{center}

\vspace{10mm}
\abstract{
\noindent
We study half-BPS 't Hooft line operators in 4d $\mathcal{N}=2$ $U(N)$ gauge theories on $S^1\times\mathbb{R}^3$ with an $\Omega$-deformation.
The recently proposed brane construction of 't Hooft operators shows that non-perturbative contributions to their correlator are identified with the Witten indices of quiver supersymmetric quantum mechanics.
For the products of minimal 't Hooft operators, a chamber in the space of Fayet-Iliopoulos parameters in the quantum mechanics corresponds to an ordering of the operators inserted along a line.
These considerations lead us to conjecture that the Witten indices can be read off from the Moyal products of the expectation values of the minimal 't Hooft operators, and also that wall-crossing occurs in the quantum mechanics only when the ordering of the operators changes.
We confirm the conjectures by explicitly computing the Witten indices for the products of two and three minimal 't Hooft operators in all possible chambers.
}
\vfill

\end{titlepage}

\tableofcontents

\section{Introduction}


In quantum field theories, not only local operators but also non-local operators are important for understanding physics. The 't~Hooft line operator, which was first introduced in \cite{tHooft:1977nqb}, is a prime example of such an extended operator in a gauge theory. The 't Hooft line operator with a magnetic charge ${\bm B}$ on a curve is defined by requiring that the gauge field configuration near the curve is given by a Dirac monopole singularity 
\begin{equation}
\frac12 F_{\mu\nu} dx^\mu \wedge dx^\nu \sim \frac{\bm B}{2} {\rm vol}_{S^2} 
\end{equation}
and the path integral is performed in the presence of the singularity. Here the magnetic charge $\bm B$ is an element of the cocharacter lattice of a gauge group and ${\rm vol}_{S^2}$ represents the (radius-independent) volume form on the two-sphere that surrounds the codimension-3 operator. 
The expectation value of a circular 't~Hooft operator serves as an order parameter for gauge symmetry breaking.
The supersymmetric version of the 't~Hooft operator, as defined in~\cite{Kapustin:2005py}, has also played significant roles in supersymmetric gauge theories~\cite{Kapustin:2006pk,Alday:2009aq,Alday:2009fs,Drukker:2009id,Gaiotto:2010be}.

In this paper we study half-BPS 't~Hooft operators in four-dimensional (4d) $\mathcal{N}=2$ gauge theories on $S^1\times \mathbb{R}^3$. The BPS condition requires that a scalar field also obeys a singular boundary condition. Furthermore, the magnetic charge must satisfy a Dirac quantization condition for the matter fields to be single-valued and the magnetic charge lattice is in general a sublattice of the cocharacter lattice. 
The expectation values of the half-BPS 't~Hooft operators wrapped along the circle in $S^1\times \mathbb{R}^3$, with an $\Omega$-deformation in a two-dimensional plane $\mathbb{R}^2 \subset \mathbb{R}^3$, were computed in~\cite{Ito:2011ea} by supersymmetric localization.
The expectation value of the 't~Hooft operator $T_{\bm B}$ with a magnetic charge ${\bm B}$ of a gauge group $G$ 
takes the form
\begin{equation}\label{eq:TB-vev}
\Braket{T_{\bm B} } = \mathop{\sum_{\bm{v} \in \bm{B}+ \Lambda_{\text{cort}}}}_{|\bm v|\leq |\bm B|} e^{\bm{v}\cdot \bm{b}} Z_\text{1-loop}(\bm{v}) Z_\text{mono}(\bm B,\bm v) \,,
\end{equation}
where $\Lambda_{\text{cort}}$ is the coroot lattice and $|\cdot|$ denotes the norm given by the Killing form. 
The expectation value is given by a sum over the sectors parameterized by the magnetic charge~${\bm v}$, for which ${\bm b}$ is a complexified fugacity.
The various sectors occur due to monopole screening where 't~Hooft-Polyakov monopoles screen the charge of the singular monopole. The monopole screening may yield non-perturbative contributions $Z_{\text{mono}}$ in \eqref{eq:TB-vev}, which was introduced in \cite{Gomis:2011pf,Ito:2011ea}.

For the quantitative understanding of 't~Hooft operators,
 it is important to explicitly compute $Z_{\text{mono}}$.
In \cite{Ito:2011ea} the factors~$Z_\text{mono}$ were computed for $G=U(N)$ using Kronheimer's correspondence~\cite{Kronheimer:MTh} between singular monopoles and instantons on a Taub-NUT space. From this perspective, $Z_\text{mono}$ is thought of as the monopole analog of the five-dimensional instanton partition function~\cite{Nekrasov:2002qd}.
Recently, it has been found in \cite{Brennan:2018yuj, Brennan:2018moe, Brennan:2018rcn,Brennan:2019hzm} that the monopole screening contribution $Z_\text{mono}$ can be identified with the Witten index
of 
a suitable supersymmetric quantum mechanics (SQM). 
The SQM may be read off from a brane realization of the 't~Hooft operators. 
The relevant SQMs come with real Fayet-Iliopoulos (FI) parameters~$\zeta_a$.

The reference \cite{Brennan:2018rcn} focused on the gauge group $SU(N)$ and evaluated the monopole screening contributions at the origin of the space of the FI parameters where non-compact Coulomb branches develop.
They conjecture that for $SU(N)$ one needs to include the contributions from the BPS ground states on the non-compact Coulomb and mixed branches in addition to those of the BPS ground states on Higgs branches. 
It was found in examples that the inclusion of such contributions computed in the Born-Oppenheimer approximation
 reproduces the predictions from the AGT correspondence~\cite{Alday:2009aq,Alday:2009fs,Drukker:2009id,Passerini:2010pr,Gomis:2010kv,Ito:2011ea}.
Later the authors of~\cite{Assel:2019iae} proposed that the contributions from the non-compact branches are automatically included if one uses a modified SQM which arises by completing the brane configuration using a 5-brane web; they confirmed their proposal by working out more examples.

In this paper, we are interested in theories with gauge group $U(N)$ rather than $SU(N)$, and in the SQMs with FI-parameters away from the origin.
These gauge groups have the same monopole moduli space, and their monopole screening contributions are described by essentially the same SQMs.
A $U(N)$ gauge theory possesses two minimal 't~Hooft operators $T_\Box$ and $T_{\overline \Box}$%
\footnote{%
There is a subtlety in this statement when the number $N_F$ of hypermultiplets in the fundamental representation is odd as we will discuss in Section~\ref{sec:NFsmaller}.
We mostly focus on the case $N_F=2N$ in the paper.
}, which correspond respectively to the fundamental and anti-fundamental representations of the Langlands dual group, which is again $U(N)$.
We take the gauge group of the 4d theory engineered by the brane configuration to be $U(N)$ rather than $SU(N)$; the minimal operators can be realized by branes, and they do not have counterparts in the $SU(N)$ theory because its Langlands dual group $SU(N)/\mathbb{Z}_N$ does not admit a representation that corresponds to the fundamental or anti-fundamental representation of $U(N)$.

We wish to consider an 't~Hooft operator $T_{\bm B}$ given as the product
\begin{equation}\label{prod.ops}
 T_{\bm B} =  T_1(s_1) \cdot T_2(s_2) \cdot \ldots \cdot T_{\ell}(s_\ell) 
 \quad \text{ in path integral},
\end{equation}
where each $T_a$ is either $T_\Box$ or $T_{\overline{\Box}}$.
We let $(x^1, x^2, x^3)$ be the orthonormal coordinates of the~$\mathbb{R}^3$ and turn on an $\Omega$-deformation in the $(x^1, x^2)$-space.
The $\Omega$-deformation requires $T_a$'s ($a=1,\ldots,\ell$) to be inserted on the 3-axis ($x^1=x^2=0$).
The parameter~$s_a$ in~(\ref{prod.ops}) is the $x^3$-value ($x^3=s_a$) of the $a$-th operator, and we assume that $s_a$'s are distinct.
The expectation value of the operator \eqref{prod.ops.ordered}, or equivalently the correlator of $T_a$'s, depends only on the ordering of $s_a$'s~\cite{Gaiotto:2010be,Ito:2011ea}; a small change in $s_a$ does not change the correlator. If we take $x^3$ as the Euclidean time, the ordering of $s_a$'s specified by a permutation $\sigma\in S_\ell$ as
\begin{equation} \label{ordering-position}
 s_{\sigma(1)} > s_{\sigma(2)} >\ldots > s_{\sigma(\ell)} 
\end{equation}
 translates to the time ordering of the corresponding operators $\widehat T_a$'s in the canonical quantization
 \begin{equation}\label{prod.ops.ordered}
 \widehat T_{\sigma(1)} \cdot \widehat T_{\sigma(2)} \cdot \ldots  \cdot \widehat T_{\sigma(\ell)}.
\end{equation}

A useful fact shown in~\cite{Ito:2011ea} is that the expectation value of a product of operators is given by the so-called Moyal product~\cite{Ito:2011ea}, 
denoted by $*$,
of the expectation values of the individual operators.
We thus have the equality
\begin{align}
\Braket{T_1 \cdot T_2 \cdot \ldots  \cdot T_\ell  } 
=
\Braket{T_{\sigma(1)}} * \Braket{T_{\sigma(2)}} * \ldots * \Braket{T_{\sigma(\ell)}} .
\end{align}
The possible dependence on the ordering of $s_a$'s is realized by the non-commutativity of the Moyal product.

Dependence on the ordering also appears in the Witten index of SQM. 
The brane construction of the product of minimal 't~Hooft operators implies that the difference $s_{a + 1} - s_a$ (in appropriate units) is nothing but an FI parameter $\zeta_a$ for the SQM \cite{Brennan:2018yuj, Brennan:2018moe, Brennan:2018rcn}:
  $\zeta_a=s_{a + 1} - s_a$ ($a=1,\ldots,\ell-1$).
Let us set $\bm\zeta:=(\zeta_a)_{a=1}^{\ell-1}$.
The permutation $\sigma$ above determines a chamber in the space of FI parameters
\begin{equation} \label{def-FI-chamber}
\big\{\bm\zeta \in \mathbb{R}^{\ell-1} | s_{\sigma(1)} > s_{\sigma(2)} >\ldots > s_{\sigma(\ell)} \big\}
\quad
=:
\quad
\text{{\it FI-chamber} specified by }\sigma.
\end{equation}
The chambers are separated by walls, which we call FI-walls; each wall is part of a hyperplane ($0=s_b - s_a = \zeta_a+\ldots +\zeta_{b-1}$) specified by an ordered pair $(a,b)$ satisfying $a<b$.
Across this wall the ordering of the position parameters $s_a$ and $s_b$ changes, and a Coulomb or mixed branch develops.
The Witten index may change discretely, {\it i.e.}, wall-crossing may occur, as we vary $\bm\zeta$ from one FI-chamber to another across the wall, only if $T_a$ and $T_b$ are operators of different types ({\it i.e.}, $T_\Box$ and $T_{\overline \Box}$), not operators of the same type ({\it i.e.}, two $T_\Box$'s or two $T_{\overline\Box}$'s).

We will see that $\Braket{ T_\Box} \ast \Braket{ T_{\overline \Box}}$ is indeed different from $\Braket{ T_{\overline \Box}} \ast \Braket{ T_\Box} $ in the theory with $2N$ hypermultiplets in the fundamental representation (flavors).
This implies that the relevant SQM should exhibit wall-crossing; we will confirm this explicitly%
\footnote{%
In~\cite{Assel:2019iae} the authors also observed the correspondence between the ordering and wall-crossing for the product of $T_\Box$ and $T_{\overline \Box}$.
We complete their analysis by including the product of two minimal operators of the same type.  We further investigate the products of three minimal operators of all types, which exhibit a richer structure.
}.
The expectation value of the product operator~(\ref{prod.ops}) with $\ell\geq 3$ is similarly given by the Moyal product of $\Braket{ T_\Box}$'s and~$\Braket{ T_{\overline\Box}}$'s.
These considerations lead us to the following conjectures%
\footnote{%
A primitive version of the conjectures and some evidence were presented by T.O. in the workshop ``Representation Theory, Gauge Theory, and Integrable Systems'' held at the Kavli IPMU in  February, 2019.
}.
\begin{center}
\begin{tabular}{c p{14cm}}
(i) &The Witten indices of the SQMs coincide with the $Z_\text{mono}$'s read off from the Moyal products of $\Braket{ T_\Box}$'s and $\Braket{ T_{\overline\Box}}$'s.
\\
\vspace{-3mm}
\\
(ii) & Wall-crossing can occur in the SQMs only across the FI-walls where the ordering of  $T_\Box$ and $T_{\overline\Box}$ changes.
\end{tabular}
\end{center}
We emphasize that in (ii), even if a discrete change occurs as the ordering of $T_\Box$ and $T_{\overline\Box}$ changes, in general only the SQMs for some $\bm{v}$'s exhibit wall-crossing while those for the other $\bm{v}$'s do not.
The main result of this paper is the demonstration of the conjectures by the explicit computation of the Witten indices relevant for two and three minimal 't~Hooft operators, in all possible FI-chambers.
While the Moyal product is easy to calculate,  the computation of the Witten indices requires the evaluation of the Jeffrey-Kirwan (JK) residues~\cite{MR1318878} of the poles in the one-loop determinants~\cite{Hwang:2014uwa, Cordova:2014oxa, Hori:2014tda}; in some cases we encounter the so-called degenerate poles that are subtle to deal with.
Nonetheless the results of the JK residue calculation exactly match those of the Moyal products.

Compared with instanton counting which also exhibits wall-crossing when the stability parameters are varied~\cite{MR2517812,MR2793270,Ito:2013kpa,Hwang:2014uwa,Ohkawa:2015dam,Ohkawa:2018vow}, for 't Hooft operators the physical meaning of the stability parameters (FI parameters) is simpler; they are the positions of the operators.

The paper is organized as follows. 
In Section~\ref{sec:SQM} we review the basics of 't~Hooft operators and the Moyal product.
We explain how to read off the SQMs for 't~Hooft operators in a $U(N)$ theory from appropriate brane configurations.
We also describe the correspondence between the FI parameters of the SQMs and the position parameters of the operators.
In Sections~\ref{sec:twoops} and~\ref{sec:threeops}, we demonstrate our conjectures by comparing the Moyal products and the Witten indices of the SQMs for various examples in the theory with $N_F=2N$ flavors.
Section~\ref{sec:NFsmaller} studies the case with $N_F<2N$. 
We conclude with discussions in Section~\ref{sec:discussion}.
Appendix~\ref{sec:formulas} collects useful facts and formulae for computing the monopole screening contributions.
In Appendix~\ref{sec:branes}, we review and slightly generalize the brane construction of 't~Hooft operators with monopole screening found in~\cite{Brennan:2018yuj,Brennan:2018moe,Brennan:2018rcn}.

\section{Wall-crossing and operator ordering}
 \label{sec:SQM}


In this section, we point out a relation between the non-commutativity in the Moyal product of the expectation values of 't~Hooft operators and the wall-crossing phenomena in the SQMs for monopole screening contributions.  We begin by recalling the basics of 't~Hooft operators, the Moyal product, and the brane realization of the SQMs.

\subsection{Basics of 't~Hooft operators}\label{sec:preliminaries}
We consider the expectation values of half-BPS 't~Hooft line operators in a 4d $\mathcal{N}=2$ supersymmetric gauge theory with a compact gauge group $G$ on $S^1 \times \mathbb{R}^3$. The line operators are put along the $S^1$ direction. 
An 't~Hooft operator $T_{\bm B}$ is labelled by its magnetic charge~${\bm B}$.
The charge ${\bm B}$ is an element of the cocharacter lattice $\Lambda_{\text{cochar}}$ and is required to obey the Dirac quantization condition, {\it i.e.},  the pairings of ${\bm B}$ and the highest weights of the matter representations must be integers%
\footnote{%
The cocharacter lattice is the dual of the character lattice, which is the lattice of all weights that appear in all representations of $G$.
}.
Two magnetic charges related by a Weyl group action give rise to the same 't~Hooft operator.
The vev (vacuum expectation value) of the 't~Hooft operator, as given in~(\ref{eq:TB-vev}), contains various monopole screening sectors specified by ${\bm v} \in {\bm B} + \Lambda_{\text{cort}}$ with $|{\bm v}| \leq |{\bm B}|$, where $\Lambda_{\text{cort}}$ is the coroot lattice and the inner product that defines the norm is given by the Killing form of the Lie algebra $\mathfrak{g}$ of $G$.

In this paper, we focus on the 4d $\mathcal{N}=2$ $U(N)$ gauge theory with $N_F \leq 2N$ hypermultiplets in the fundamental representation.
We will refer to such a theory as a $U(N)$ SQCD. For the $U(N)$ group, its Langlands or GNO dual group is again $U(N)$. We may identify the Cartan subalgebra
 of $U(N)$ with the $N$-dimensional Euclidean space $\mathbb{R}^N$ by the map 
\begin{align}
\text{diag}(z_1,\ldots,z_N)\mapsto \sum_{i=1}^Nz_i{\bm e}_i,
\end{align}
where ${\bm e}_i$  ($i = 1, \ldots, N$)%
\footnote{%
The same symbol ${\bm e}_i$ (or ${\bm e}_a$) will denote an orthonormal basis of a Euclidean space other than $\mathbb{R}^N$.
} form an orthonormal basis of $\mathbb{R}^N$. 
The Weyl group acts by permuting the $z_i$'s. The cocharacter lattice is then given by 
\begin{align}
\Lambda_{\text{cochar}} = \bigoplus_{i=1}^N\mathbb{Z}{\bm e}_i \label{cocharacter}
\end{align}
and the coroot lattice by
\begin{align}
\Lambda_{\text{cort}} = \bigoplus_{i=1}^{N-1}\mathbb{Z}({\bm e}_i - {\bm e}_{i+1}).\label{Lambda-cort}
\end{align}
The magnetic charges ${\bm B}$ and ${\bm v}$ take values in \eqref{cocharacter} and the inner product of the vectors is given by the Euclidean metric%
\footnote{%
To compare wtih~\cite{Brennan:2018yuj,Brennan:2018rcn} note the following.
$\Lambda_{\text{cort}}$ in~(\ref{Lambda-cort}) is also the coroot lattice of $SU(N)$.
The cocharacter lattice of $SU(N)$ is the sublattice $\{\sum_i z_i=0\}$ of $\Lambda_{\text{cochar}}$ in~(\ref{cocharacter}).
 Defining
${\bm z}_{U(1)} := (1/N) \sum_i z_i \sum_j {\bm e}_{j}$ and ${\bm z}_{SU(N)}:= {\bm z} - {\bm z}_{U(1)}$ for ${\bm z}=\sum_i z_i {\bm e}_i$,
we have $|{\bm v}| \leq |{\bm B}| \Leftrightarrow |{\bm v}_{SU(N)}| \leq |{\bm B}_{SU(N)}|$ for ${\bm v}\in {\bm B} + \Lambda_{\text{cort}}$.
}.

We are interested in discrete changes of the vev $\langle T_{\bm B}\rangle$, or more precisely of the monopole screening contributions $Z_\text{mono}({\bm B},{\bm v})$, in the $U(N)$ SQCD with $N_F=2N$ flavors.
In Sections~\ref{sec:twoops} and~\ref{sec:threeops} we will employ two methods for computing $Z_\text{mono}$. 
The first method is the Moyal product of the expectation values of minimal 't~Hooft operators.
The second is the localization computation of the Witten index
of an SQM, which may be read off from a brane configuration corresponding to the monopole screening sector. 
The discrete changes are visible in both methods.
In the rest of this section we will describe relevant materials for the two methods and explain a relation between them.

\subsection{Moyal product}
\label{sec:Moyal}

A $U(N)$ gauge theory has two minimal 't~Hooft operators. 
One is specified by ${\bm B} = {\bm e}_i$ and the other by ${\bm B} = -{\bm e}_i$, where the value of $i$ is irrelevant because different values are related by the Weyl group action.
We denote the operators by $T_{\Box}$ and $T_{\overline{\Box}}$ respectively%
\footnote{%
Since the Langlands dual group of $SU(N)$ is $SU(N)/\mathbb{Z}_N$, 't~Hooft operators corresponding to $B=\pm {\bm e}_i$ do not exist in an $SU(N)$ gauge theory. 
}.

We focus on the $U(N)$ SQCD with $N_F = 2N$ flavors on $S^1 \times \mathbb{R}^3$ with an $\Omega$-deformation. 
We denote the Cartesian coordinates of $S^1 \times \mathbb{R}^3$ by $(\tau, x^1, x^2, x^3)$, where $\tau$ is the coordinate along the $S^1$.   In this theory the operators $T_{\Box}$ and $T_{\overline{\Box}}$ can coexist%
\footnote{%
Section~\ref{sec:NFsmaller} discusses subtleties for $N_F$ odd.
}.
We introduce an $\Omega$-deformation with parameter $\epsilon_+$.
We refer to the reference \cite{Ito:2011ea}, where $\epsilon_+$ was denoted by~$\lambda$, for the details of  the $\Omega$-deformation, the definition of the line operator vev, and  supersymmetric localization.
The vevs of $T_{\Box}$ and $T_{\overline{\Box}}$ do not receive monopole screening contributions since the fundamental and the anti-fundamental representations are minuscule, {\it i.e.}, their weights form a single Weyl orbit~\cite{Kapustin:2006pk}.
 Hence their vevs can be computed from the knowledge of the one-loop determinants, and for $T_{\Box}$ we have explicitly
\begin{align}
\Braket{T_{\Box}} = \sum_{i=1}^{N}e^{b_i}Z_i({\bm a}), \label{TF}
\end{align}
where 
\begin{align}
Z_i({\bm a}) = \left(\frac{\prod_{f=1}^{2N}2\sinh\frac{a_i -m_f}{2}}{\prod_{1\leq j (\neq \text{ fixed }i) \leq N}2\sinh\frac{a_i - a_j + \epsilon_+}{2}2\sinh\frac{-a_i + a_j + \epsilon_+}{2}}\right)^{\frac{1}{2}}. \label{Zi}
\end{align}
We defined complex variables%
\footnote{%
The tuple $(\theta_m^i)$ here equals $\Theta'$ (rather than $\Theta$) defined in footnote~10 of~\cite{Ito:2011ea} (arXiv version 3).
}
\begin{align}
a_i = i (2\pi R)A_{\tau, \infty}^i - (2\pi R)\text{Re}(\Phi_{\infty}^i)
 , \qquad b_i = i\theta_m^i + \frac{8\pi^2 R}{g^2}\text{Im}(\Phi_{\infty}^i) +\frac{\theta}{2\pi}a_i 
 \label{aandb}
\end{align}
for $i=1, \ldots, N$.
Here $R$ is the radius of the circle $S^1$, $g$ the gauge coupling constant,
 $A^i_{\tau, \infty}$'s  the asymptotic values of the gauge holonomy along the $S^1$,  $\Phi^i_{\infty}$'s the asymptotic values of the Coulomb branch moduli,  $\theta_m^i$'s the chemical potentials for the magnetic charges, and~$\theta$ the theta angle%
\footnote{%
Here $\theta$ is the coefficient of $\frac{i}{8\pi^2} \int {\rm Tr}(F\wedge F)$ in the Euclidean action.
Gauge group~$U(N)$ admits another theta angle, the coefficient of $\frac{i}{8\pi^2} \int ({\rm Tr}\,F)\wedge ({\rm Tr}\,F)$, which we ignore because it plays no role in this paper.
}. 
In addition $m_f$'s are the complex parameters defined by expressions similar to those for $a_i$, with $A_{\tau, \infty}^i$ replaced by flavor holonomies and with $\Phi^i_{\infty}$ replaced by mass parameters. 
Similarly the vev of $T_{\overline{\Box}}$ is given by
\begin{align}
\Braket{T_{\overline{\Box}}} = \sum_{i=1}^{N}e^{-b_i}Z_i({\bm a}). \label{TAF}
\end{align}

We define the Moyal product of one function $f({\bm a}, {\bm b})$ and another $g({\bm a}, {\bm b})$ that depend on parameters ${\bm a}=(a_1,\ldots,a_N)$ and ${\bm b}=(b_1,\ldots,b_N)$ in \eqref{aandb} 
to be
\begin{align}
(f\ast g)({\bm a}, {\bm b}) =\exp\left[-\epsilon_+ \sum_{k=1}^N\left(\partial_{b_k}\partial_{a'_k} - \partial_{a_k}\partial_{b'_k}\right)\right]f({\bm a}, {\bm b})g({\bm a}', {\bm b}')\Big|_{{\bm a}' = {\bm a}, {\bm b}' = {\bm b}}. \label{moyal}
\end{align}
It is straightforward to show that the Moyal product is associative.

A product of minimal operators can be written by
\begin{align}
T_{\bm B} = T_1(s_1) \cdot T_2(s_2) \cdot \ldots \cdot T_{\ell}(s_\ell) , \label{op.product}
\end{align}
where each $T_{a}$ ($a=1, \ldots, \ell$) is either $T_{\Box}$ or $T_{\overline{\Box}}$. 
The operators are inserted at separated points $s_a$ on the $x^3$-axis.
The magnetic charge ${\bm B}$ of the product operator is ${\bm B} = n_\Box {\bm e}_N - n_{\overline\Box} {\bm e}_1$ up to the Weyl group action, where $n_\Box$ ($n_{\overline\Box}$) is the number of $T_\Box$'s ($T_{\overline\Box}$'s) in the product.
The expectation value of $T_{\bm B}$ may be calculated as the Moyal product of the vevs of $T_a$ \cite{Ito:2011ea}  as
\begin{align}
\Braket{T_{\bm B}} = \Braket{T_{\sigma(1)}} \ast \Braket{T_{\sigma(2)}} \ast \ldots \ast \Braket{T_{\sigma(\ell)}}, \label{TB1toell}
\end{align}
where $\sigma$ is the permutation such that $s_{\sigma(1)} > s_{\sigma(2)} > \ldots > s_{\sigma(\ell)}$.

The magnetic charge ${\bm B}$ of the product operator $T_{\bm B }$ corresponds to a weight in a non-minuscule representation of the dual group. 
This implies that its vev $\Braket{T_{\bm B}}$ contains contributions  $Z_\text{mono}({\bm B},{\bm v})$'s from monopole screening. 
A useful technique we will use in~Sections~\ref{sec:twoops} and~\ref{sec:threeops} is to read off $Z_\text{mono}$ in~$\Braket{T_{\bm B}}$ from the right hand side of~(\ref{TB1toell}), which is easy to compute.
The Moyal product of two different functions is in general non-commutative. 
While the one-loop determinants, summarized in Appendix~\ref{sec:locfor4d}, that appear in~$\Braket{T_{\bm B}}$ are fixed solely by the matter content of the theory and are in particular independent of the ordering of $T_a$'s, the $Z_\text{mono}$'s in $\Braket{T_{\bm B}}$ on the other hand may or may not depend on the ordering.

\subsection{SQMs from branes}
\label{sec:SQMbrane}

We now turn to 
the SQMs obtained by the dimensional reduction of 2d $\mathcal{N}=(0, 4)$ supersymmetric field theories.
The moduli space of a suitable SQM coincides with the component~$\mathcal{M}({\bm B},{\bm v})$ of the monopole moduli space corresponding to a screening sector ${\bm v}$ in the expansion~(\ref{eq:TB-vev}) of $\Braket{ T_{\bm B}}$. 
In particular the Witten index of the SQM equals the monopole screening contribution~$Z_\text{mono}({\bm B}, {\bm v})$.
In order to read off the SQM, we use the brane realization of monopole screening~\cite{Cherkis:1997aa, Brennan:2018yuj, Brennan:2018moe,Brennan:2018rcn}.

Let us begin with a brane realization of the gauge theory itself. Let $x^{\mu}$ ($\mu = 0, 1, 2, \cdots, 9$) be the coordinates on $\mathbb{R}^{1,9}$.
A 4d $\mathcal{N}=2$ $U(N)$ gauge theory may be realized on $N$ D3-branes\footnote{The worldvolume theory on the D3-branes contains an $\mathcal{N}=2$ adjoint hypermultiplet which originates from an $\mathcal{N}=4$ vector multiplet. As in~\cite{Brennan:2018rcn} we assume that an appropriate supergravity background gives an infinitely large mass to the adjoint hypermultiplet and  that it is integrated out.} in type IIB string theory. 
The D3-branes are localized in the $(x^4, x^5, x^6, x^7, x^8, x^9)$-space. The location of the D3-branes in the $(x^4, x^5)$-space gives complex Coulomb branch moduli~$\Phi^i_\infty$ in~(\ref{aandb}) and  we choose the D3-branes to be located at $x^6=x^7=x^8=x^9=0$.
$N_F = 2N$ hypermultiplets in the fundamental representation can be added by introducing $2N$ D7-branes localized in the $(x^4, x^5)$-space. In this case the location of a D7-brane in the $(x^4, x^5)$-space gives two real mass parameters. 
We choose the location in the $x^5$-direction to be~${\rm Re}(\Phi^i_\infty)$ for a D3-brane, and to be the real mass parameter in ${\rm Re}(m_f)$ for a D7-brane.
The combined system yields the 4d $\mathcal{N}=2$ $U(N)$ gauge theory with $2N$ flavors on the worldvolume of the D3-branes.

\begin{figure}[t]
\centering
\subfigure[]{\label{fig:sampletHooft1}
\includegraphics[width=8cm]{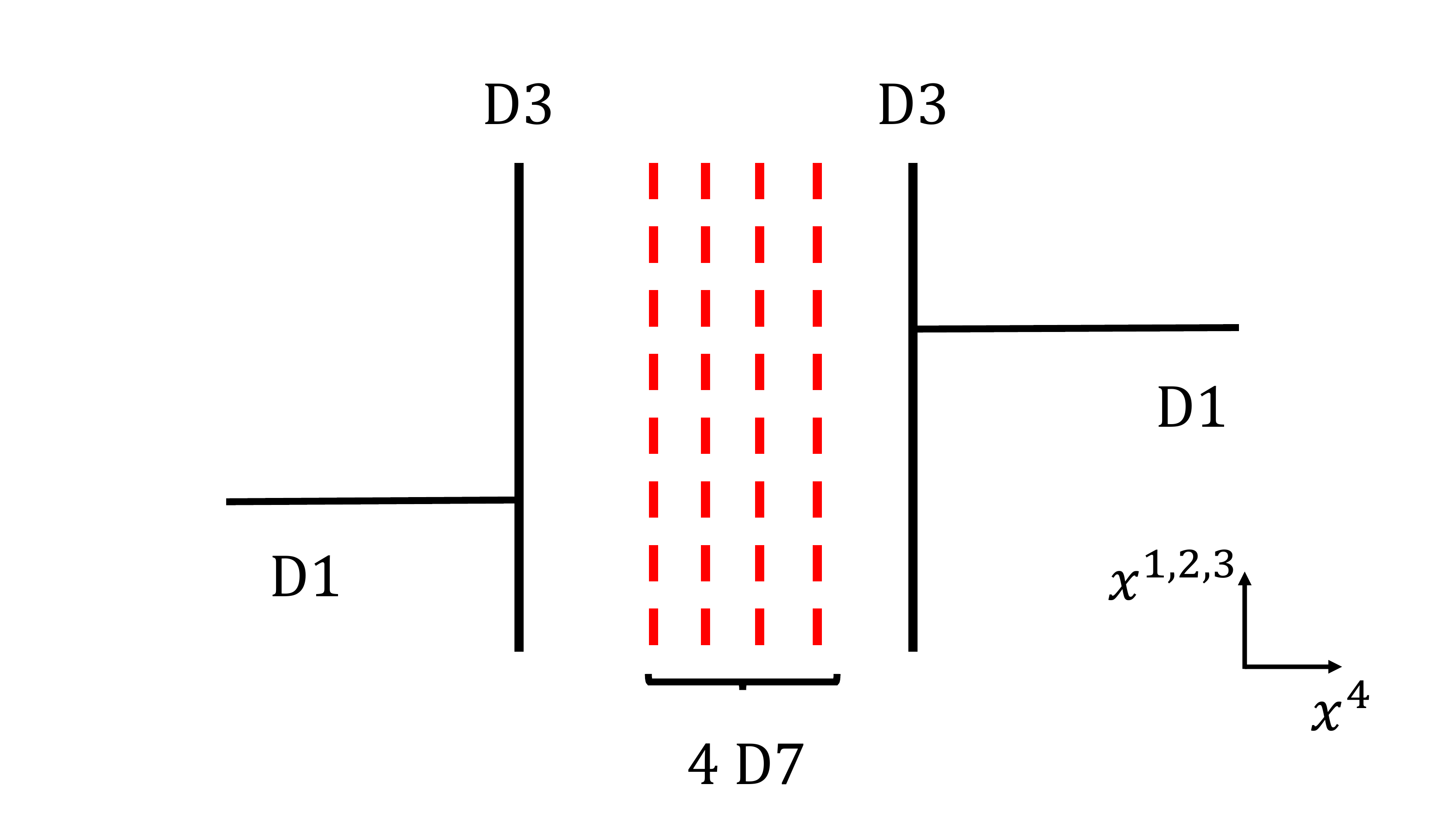}}
\subfigure[]{\label{fig:sampletHooft2}
\includegraphics[width=8cm]{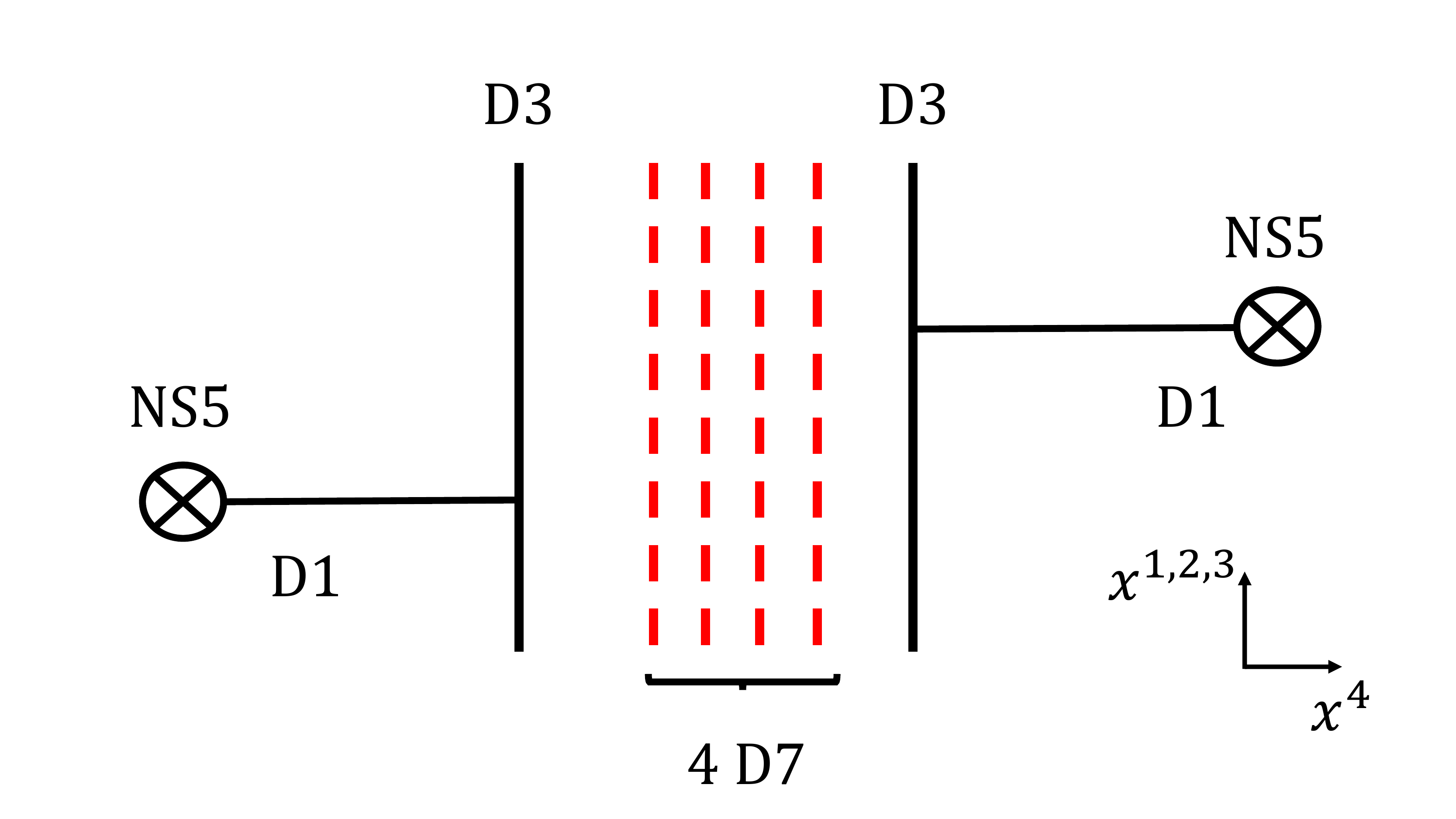}}
\caption{(a): A brane configuration for the 't~Hooft operator with magnetic charge ${\bm B} = -{\bm e}_1 + {\bm e}_2$ in the $U(2)$ theory with four flavors. 
Vertical (horizontal) solid lines represent D3-branes (D1-branes).
 Dotted red lines represent D7-branes. (b): Another configuration for the same operator. Each semi-infinite D1-brane in Figure~\ref{fig:sampletHooft1} is now terminated by an NS5-brane ($\otimes$). }
\label{fig:sampletHooft}
\end{figure}

We can introduce 't~Hooft line operators by adding semi-infinite D1-branes ending on the D3-branes. 
The D1-branes are localized at $x^{5, 6, 7, 8, 9} = 0$ and at a point in the $(x^1, x^2, x^3)$-space. 
A collection of semi-infinite D1-branes ending on the D3-branes introduces
an 't~Hooft operator.
We can read off the magnetic charge 
from the configuration of D1- and D3-branes.
We label the D3-branes so that their locations $x^4_i$ ($i=1,\ldots,N$) in the $x^4$-direction are in the non-decreasing order\footnote{The D3-branes are assumed to be at generic positions in the $(x^4, x^5)$-space.
}: 
$x^4_1< x^4_2<\ldots< x^4_N$.
Suppose that $k^{\rm L}_i$ ($k^{\rm R}_i$) D1-branes end on the $i$-th D3-brane from the left (right).
We choose the convention such that the magnetic charge
on the D3 worldvolume is
\begin{align}
\text{magnetic charge }=
 \sum_{i=1}^N( k^{\rm R}_i -k^{\rm L}_i ){\bm e}_i
 \in \Lambda_{\text{cochar}}
 . \label{mcharge}
\end{align}
As an example, a configuration for ${\bm B} = -{\bm e}_1 + {\bm e}_2$ in the $U(2)$ theory with four flavors is depicted in Figure~\ref{fig:sampletHooft1}.
Note that D7-branes can cross D3-branes in the figure, with no actual collision in the 10d spacetime, because the two types of branes are both point-like in the $(x^4, x^5)$-space. Therefore, we can put D7-branes on any side of the D3-branes. 
In Figure \ref{fig:sampletHooft1} we choose to put the four D7-branes between the two D3-branes.

A crucial step~\cite{Brennan:2018yuj} in specifying the precise 't~Hooft operator and toward deriving the SQM is to terminate the semi-infinite D1-branes by NS5-branes. 
In this picture a product of 't~Hooft operators is realized by a collection of D1-branes each of which ends on an NS5-brane and a D3-brane. 
 Supersymmetry imposes the s-rule \cite{Hanany:1996ie}: for each pair of a D3-brane and an NS5-brane, there can be at most one D1-brane ending on them%
\footnote{%
Due to the s-rule a single NS5-brane, with $K$ D1-branes ($1\leq K\leq N$) attached and placed on the right of the stack of all the D3- and D7-branes, gives an 't~Hooft operator with magnetic charge $\sum_{i=1}^K \bm{e}_i$, which seems to correspond to the $K$-th exterior power of the fundamental representation.
Similarly an NS5-brane placed on the left of the stack seems to give an operator with charge $-\sum_{i=1}^K \bm{e}_i$ corresponding to the $K$-th exterior power of the anti-fundamental representation.
}.
The minimal operator~$T_{\Box}$ ($T_{\overline\Box}$) corresponds to an NS5-brane on the right (left) side of the stack of all the D3- and D7-branes, with a single D1-brane stretching between the NS5-brane and a D3-brane.
 The NS5-branes are localized in the $(x^1, x^2, x^3, x^4)$-space. 
The brane configuration with the NS5-branes for the same example, $T_{\bm{B}=-{\bm e}_1+{\bm e}_2}=T_{\Box}\cdot T_{\overline\Box}$, is depicted in Figure~\ref{fig:sampletHooft2}.

\begin{figure}[t]
\centering
\subfigure[]{\label{fig:sampletHooft3}
\includegraphics[width=8cm]{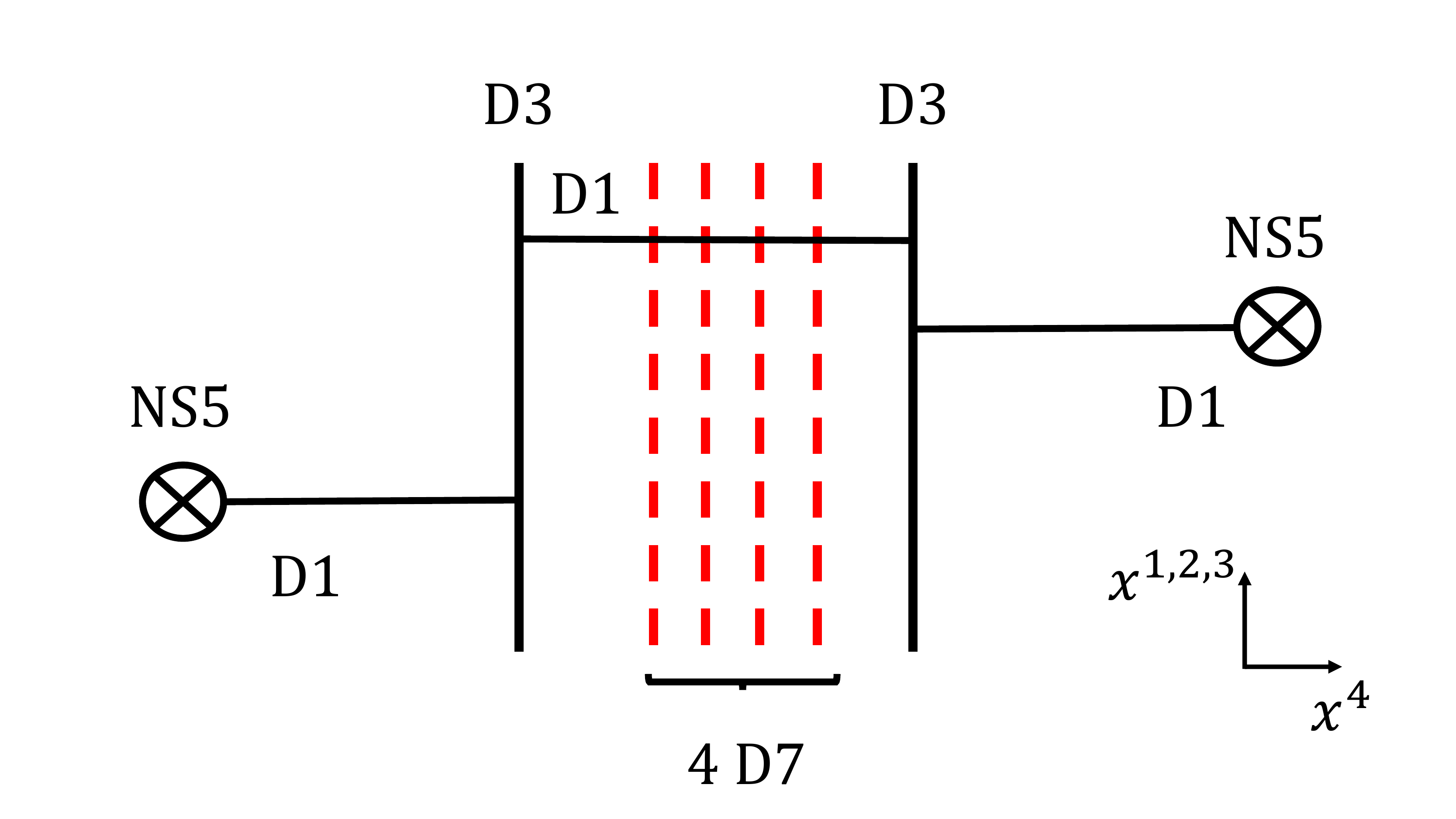}}
\subfigure[]{\label{fig:sampletHooft4}
\includegraphics[width=8cm]{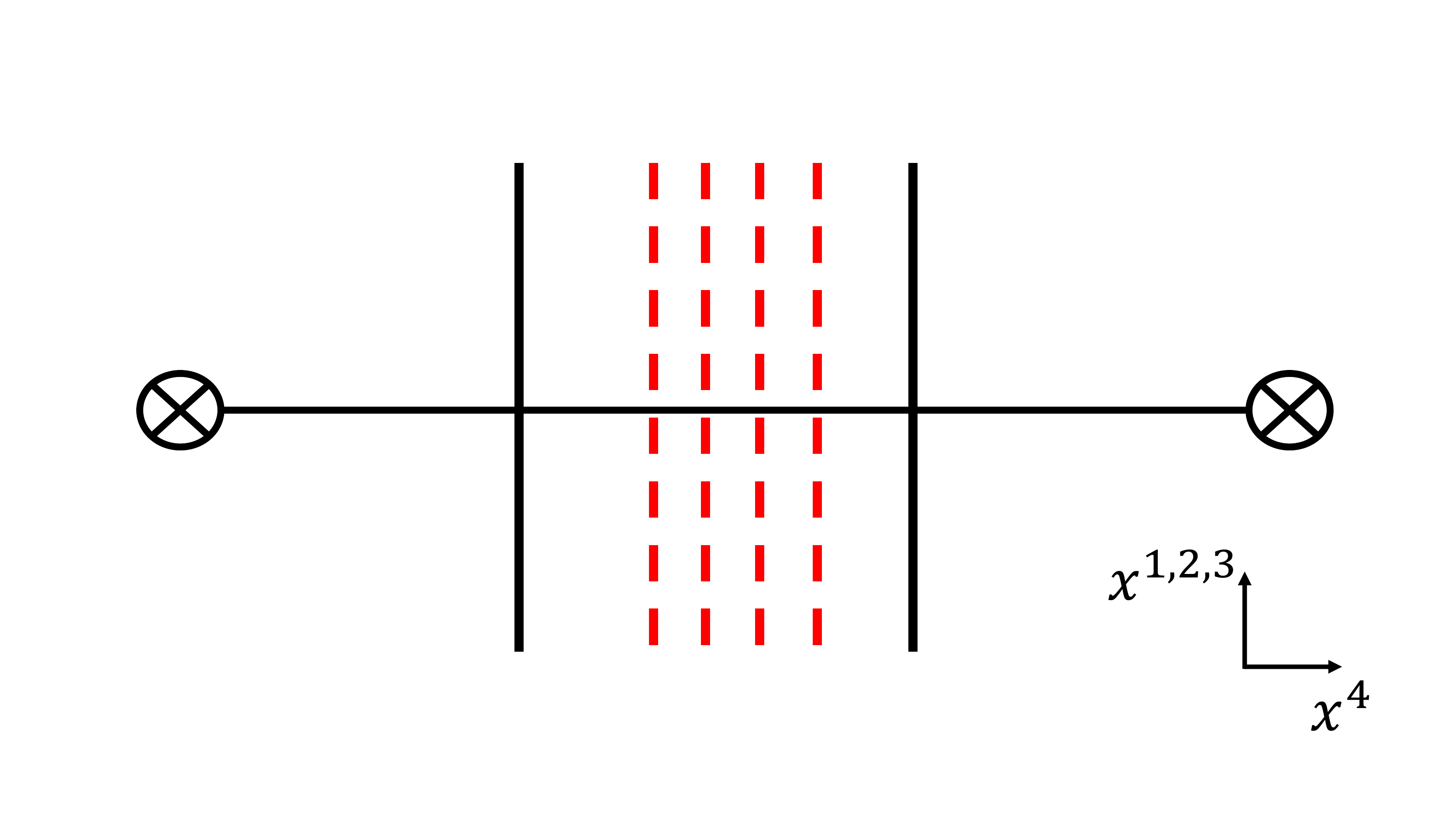}}
\caption{(a): The brane configuration where a finite D1-brane is added to the configuration in Figure \ref{fig:sampletHooft2}. (b): Tuning the position of the D1-brane and two NS5-branes in Figure \ref{fig:sampletHooft3} so that they are at the same point in the $(x^1, x^2, x^3)$-space. The magnetic charge of the line operator is completely screened and the configuration describes the ${\bm v} = {\bm 0}$ sector. }
\label{fig:sampletHooftp}
\end{figure}

To reach the sector specified by the coweight~${\bm v}$ in the expansion~(\ref{eq:TB-vev}), we let smooth monopoles screen the singular monopoles.
Smooth monopoles correspond to finite D1-branes stretched between D3-branes~\cite{Diaconescu:1996rk}. 
These D1-branes are localized at $x^{5, 6, 7, 8, 9} = 0$ because they are attached to D3-branes, but can move freely as point-like objects in the $(x^1, x^2, x^3)$-space.
A smooth monopole screens the  singular monopole precisely when the corresponding finite D1-brane between D3-branes reconnects with a D1-brane between an NS5-brane and a D3-brane.
This is only possible when the $x^{1,2,3}$-position of the finite D3-brane coincides with that of an NS5-brane.
For example Figure~\ref{fig:sampletHooft3} depicts a finite D1-brane between the two D3-branes of Figure~\ref{fig:sampletHooft2}. 
By tuning the positions of the finite D1-brane and the NS5-branes we reach the configuration in Figure~\ref{fig:sampletHooft4}. 
In this example the magnetic charge of the operator is completely screened, {\it i.e.}, ${\bm v}=0$, since no D1-brane ends on D3-branes. 
The s-rule restricts the possible brane configurations that realize monopole screening.

\begin{table}[t]
\begin{center}
\begin{tabular}{c|c|ccc|c|c|cccc}
&0&1&2&3&4&5&6&7&8&9
\\
\hline
D3& $\times$ & $\times$ & $\times$&$\times$&&&&&
\\
D7 & $\times$&$\times$&$\times$&$\times$&&&$\times$&$\times$&$\times$&$\times$
\\
\hline
D1&$\times$ &&&&$\times$&&
\\
NS5 &$\times$&&&&&$\times$&$\times$&$\times$&$\times$&$\times$
\end{tabular}
\end{center}
\caption{The directions in which branes extend are indicated by $\times$. }
\label{table:brane-directions}
\end{table}
Table~\ref{table:brane-directions} summarizes the embedding of the branes in $\mathbb{R}^{1,9}$.
As stated above, a D7-brane can cross a D3-brane in the figure, without an actual collision in the 10d spacetime, as they are both point-like in the $(x^4, x^5)$-space. 
On the other hand, an NS5-brane cannot cross a D3-brane or a D7-brane without a collision in 10d.
In particular when an NS5-brane crosses a D3-brane, a D1-brane between the 5- and 3-branes is created or destroyed, {\it i.e.}, the Hanany-Witten (HW) transition \cite{Hanany:1996ie} occurs. 
The infrared physics on the D1-branes is, however, not altered.

\begin{figure}[tb]
\centering
\subfigure[]{\label{fig:permute1}
\includegraphics[width=5cm]{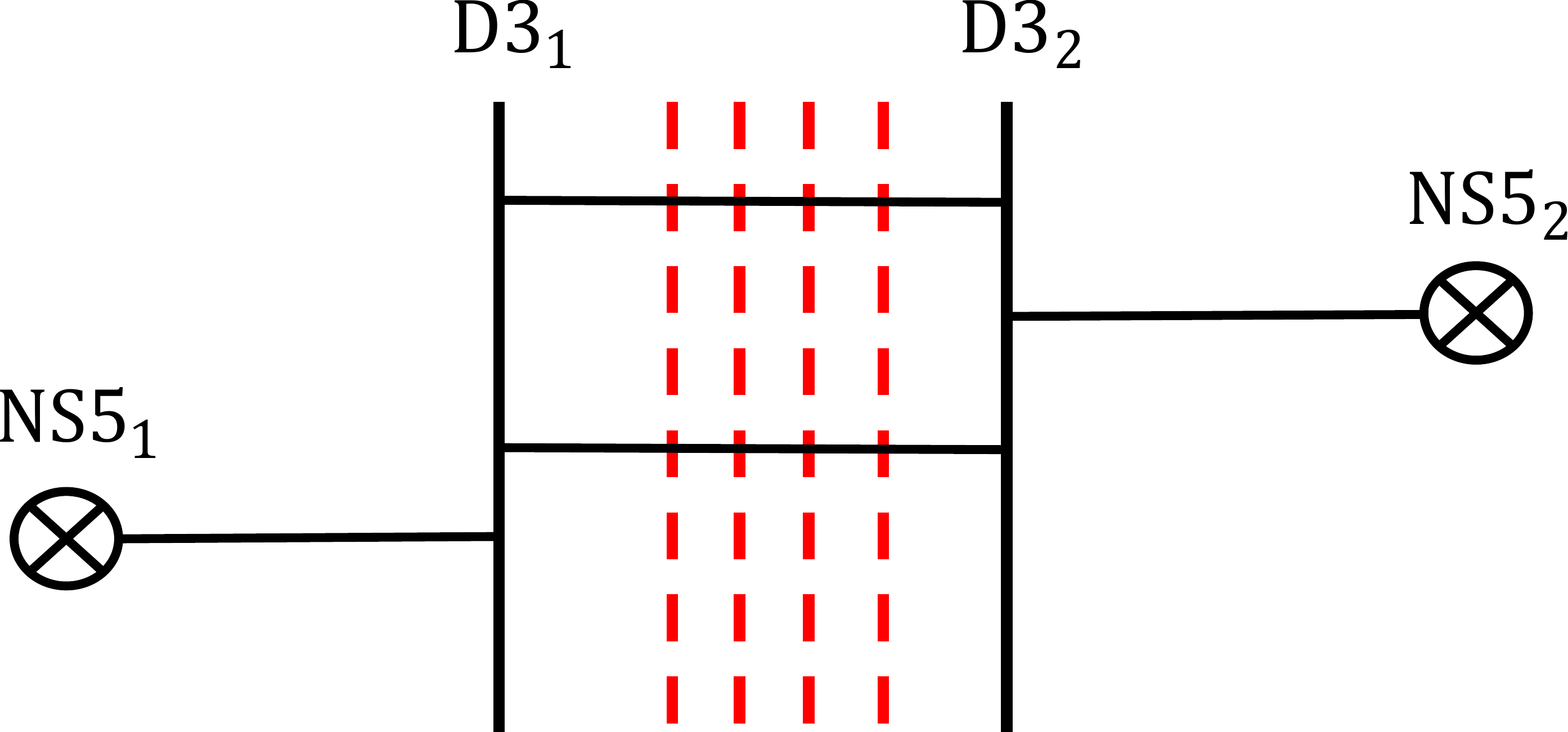}}
\subfigure[]{\label{fig:permute2}
\includegraphics[width=5cm]{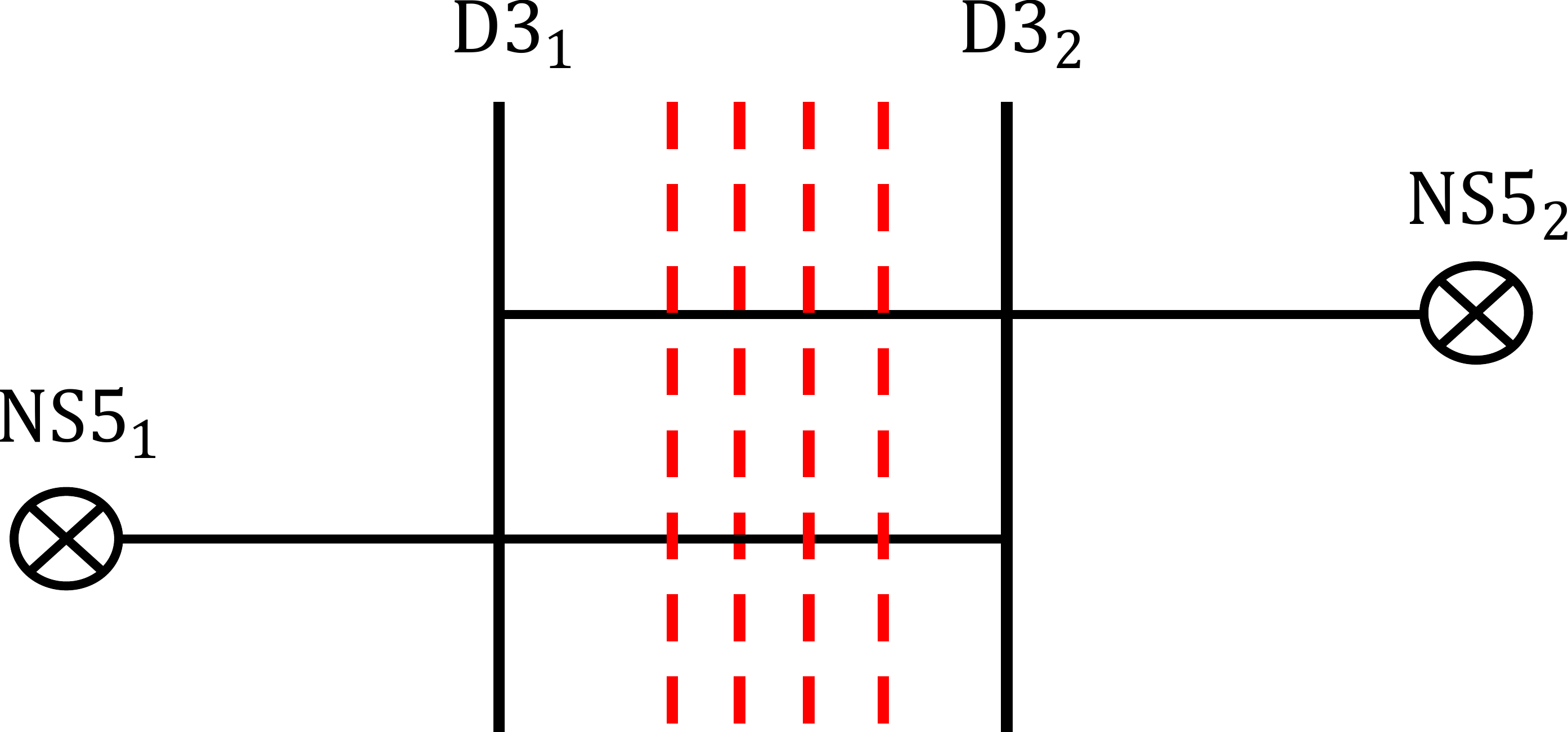}}
\subfigure[]{\label{fig:permute3}
\includegraphics[width=5cm]{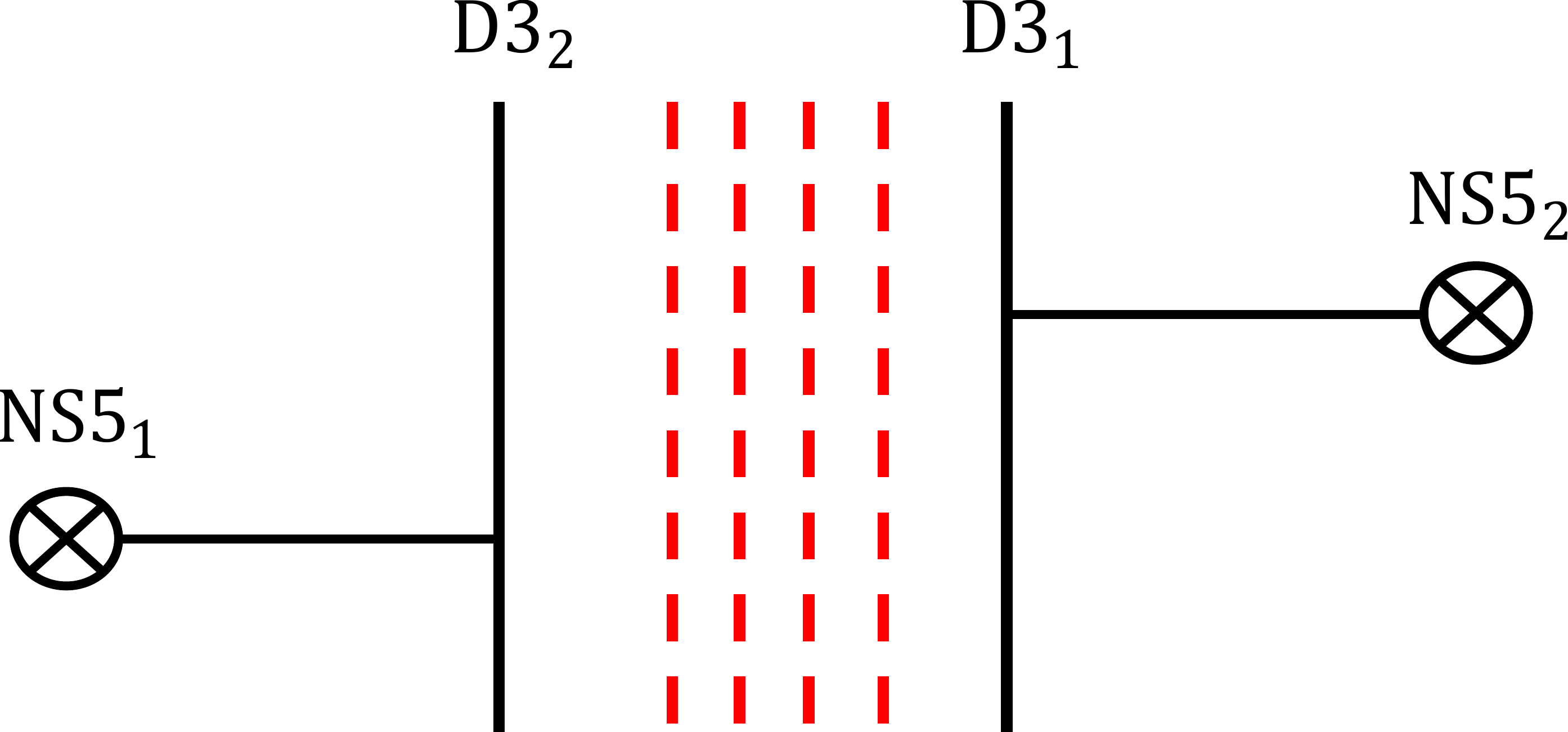}}
\caption{(a): A brane configuration for~${\bm B}=-{\bm e}_1+{\bm e}_2$ with finite D1-branes representing smooth monopoles. (b): The D1-branes reconnect to give charge ${\bm v}={\bm e}_1-{\bm e}_2$.  (c): After moving around in the~$(x^4,x^5)$-space the two D3-branes get permuted; the resulting magnetic charge is again $-{\bm e}_1+{\bm e}_2$ with non-decreasing coefficients.}
\label{fig:permute}
\end{figure}

Another step for the derivation of the SQM is a permutation of D3-branes.
After introducing the finite D1-branes, the coefficients $v_i$ in ${\bm v} = \sum_{i=1}^N v_i {\bm e}_i$ are not always non-decreasing.
We permute the $N$ D3-branes by moving them around in the~$(x^4,x^5)$-space so that after the permutation we have $v_1\leq v_2\leq \ldots \leq v_N$.
This step is illustrated in Figure~\ref{fig:permute}.

From the brane setup, it is now possible to read off the desired SQM by following the procedure in Section~3.3 of~\cite{Brennan:2018yuj}.
This involves moving NS5-branes via Hanany-Witten transitions across D3-branes as well as reconnecting D1-branes.
One can read off the matter content of the SQM from the brane configuration where D1-branes end only on NS5-branes and not on any D3-brane.

\begin{figure}
\centering
\hspace{2cm}
\subfigure[]{\label{fig:quiver1A1_N04}
\includegraphics[width=2.7cm]{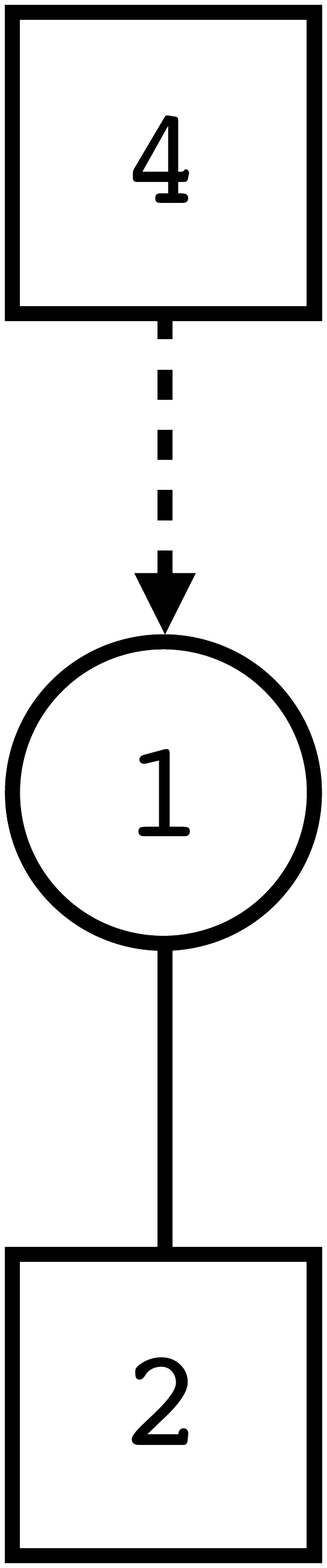}}
\subfigure[]{\label{fig:quiver1A1}
\includegraphics[width=7cm]{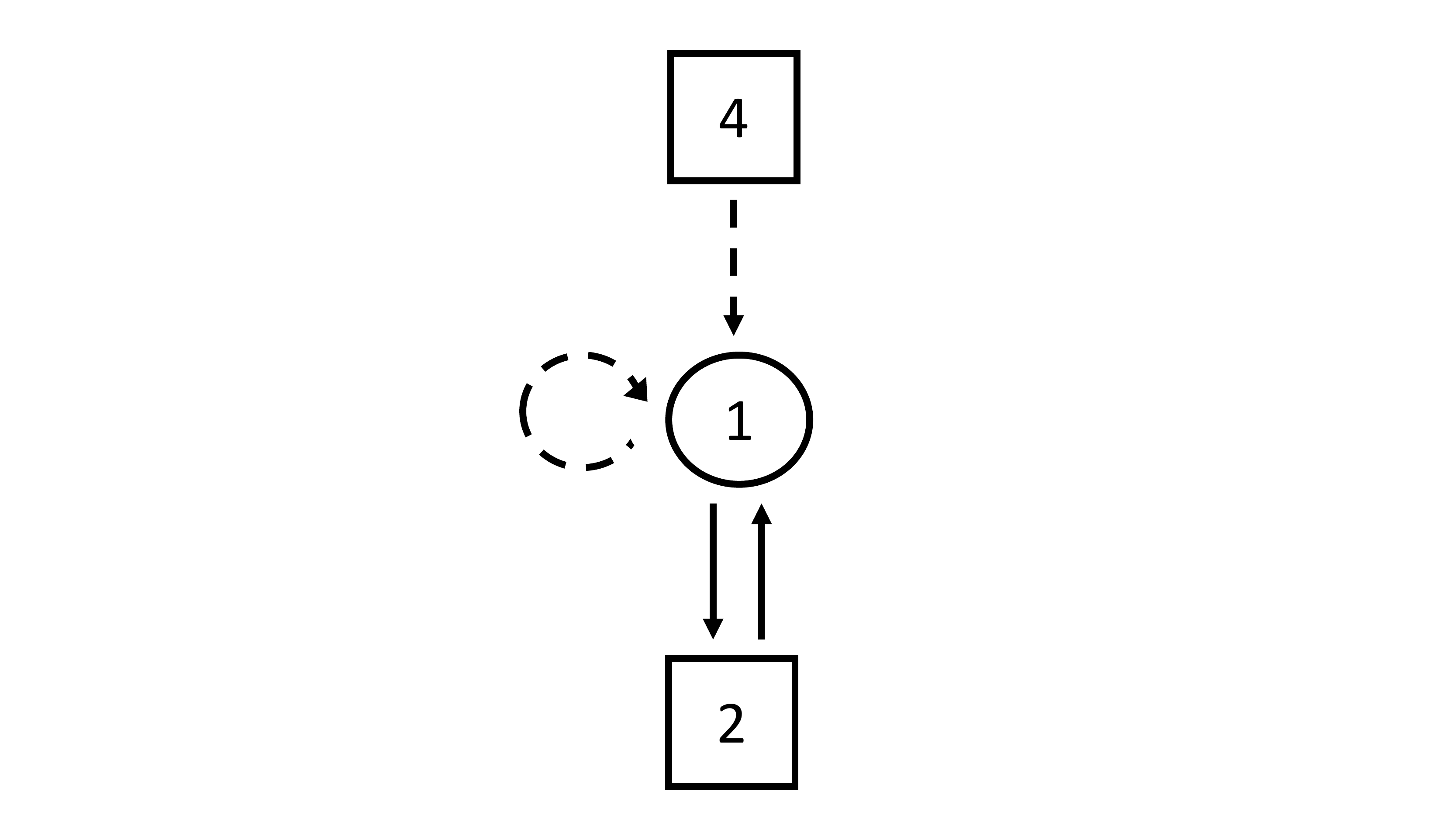}}
\caption{%
(a): A sample quiver diagram with nodes and arrows representing $\mathcal{N}=(0,4)$ multiplets.  
A circular node with a positive integer $k$ represents a $U(k)$ gauge group and the corresponding vector multiplet.  A square box with a positive integer $N$ represents a flavor symmetry group $U(N)$.  A solid line (or a dashed arrow) from a $U(n')$ node to a $U(n)$ node represents a hypermultiplet (or respectively a short Fermi multiplet) in the bifundamental representation $(\boldsymbol{n},  \overline{\boldsymbol{n'}})$ of $U(n)\times U(n')$. We used an arrow, not just a line, to indicate $\mathcal{N}=(0,4)$ a short Fermi multiplet because an orientation is needed to specify the representation.
(b): The equivalent quiver diagram with nodes and arrows representing $\mathcal{N}=(0,2)$ multiplets. In this case each line needs to be equipped with an arrow to specify the representation. 
}
\end{figure}

The SQM is the dimensional reduction of a two-dimensional $\mathcal{N}=(0, 4)$ supersymmetric gauge theory \cite{Tong:2014yna, Hwang:2014uwa}. 
D1-D1 strings on $n$ D1-branes give rise to a $\mathcal{N}=(0, 4)$ $U(n)$ vector multiplet. 
D1-D1' strings on $n$ D1-branes and $n'$ D1-branes which are separated by an NS5-brane yield an $\mathcal{N}=(0, 4)$ hypermultiplet in the bifundamental representation of $U(n) \times U(n')$. Furthermore, D1-D3 strings on $n$ D1-branes and a D3-brane give an $\mathcal{N}=(0, 4)$ hypermultiplet in the fundamental representation of $U(n)$. Finally, D1-D7 strings on $n$ D1-branes and a D7-brane yield a short $\mathcal{N}=(0, 4)$ Fermi multiplet in the fundamental representation of $U(n)$. 
Note that the matter multiplets from D1-D3 strings or D1-D7 strings exist when D1-branes intersect the D3-branes or D7-branes respectively.
For $N_F = 2N$ we assume that the Chern-Simons term is absent in the SQM.
For example, the SQM realized on the D1-brane in Figure~\ref{fig:sampletHooft4} is described by the quiver diagram in Figure~\ref{fig:quiver1A1_N04}.
In general we get a quiver with more than one gauge node. 
We emphasize that we can determine the gauge node that the short Fermi multiplets couple to; we begin with a configuration where every NS5-brane is either on the left or on the right of the stack of all the D3- and D7-branes and chase a sequence of Hanany-Witten transitions.  
We give concrete examples in Appendix~\ref{sec:branes}; these slightly generalize the brane construction of 't Hooft operators with monopole screening found in~\cite{Brennan:2018yuj,Brennan:2018moe,Brennan:2018rcn}.

The $\mathcal{N}=(0, 4)$ supersymmetry admits three real FI parameters for each $U(n)$ gauge node. 
The three FI parameters 
$(\zeta^1,\zeta^2,\zeta^3)$
are related to the positions of NS5-branes in the $(x^1, x^2, x^3)$-space \cite{Brennan:2018yuj, Brennan:2018moe, Brennan:2018rcn}. 
Let $\ell$ NS5-branes be located at $(x^1, x^2, x^3)=(s_a^1, s_a^2, s_a^3)$ with $a=1,\ldots,\ell$.
Let $n$ D1-branes stretch between the $a$-th and the $(a+1)$-th NS5-branes.
Then the FI parameters associated with the $U(n)$ gauge symmetry on the D1 worldvolume are given by $\zeta_a^i = s^i_{a+1} - s^i_a$ ($i=1, 2, 3$). 
In the presence of hypermultiplets non-zero FI parameters Higgs the gauge symmetry. 
To see this from the brane configuration, let us consider D3-branes between the two NS5-branes.
As the relative position of the two NS5-branes changes in the $(x^1, x^2, x^3)$-space, the D1-branes break and end on the D3-branes.
 This is the process from~Figure \ref{fig:sampletHooft4} to Figure~\ref{fig:sampletHooft3}. 
 
The $\Omega$-deformation breaks the 1d $\mathcal{N}=(0,4)$  supersymmetry to~$\mathcal{N}=(0,2)$ and demands that $\zeta^1_a=\zeta^2_a=0$; we write $\zeta_a:= \zeta_a^3$.
A quiver description of an SQM
allows one to compute its Witten index by localization~\cite{
Hwang:2014uwa, Cordova:2014oxa, Hori:2014tda}. 
The main formulae are summarized in Appendix~\ref{sec:locforSQM}.

For a given sector specified by $\bm{v}$, $\zeta_a$ does not arise as an FI parameter in the SQM if there is no D1-brane between the $a$-th and the $(a+1)$-th NS5-branes; in the extreme case $\bm{v}=\bm{B}$ the SQM is trivial and there is no FI parameter.
Even in such cases we will use the notations $\bm\zeta=(\zeta_1,\ldots,\zeta_{\ell-1})=\sum_{a=1}^{\ell-1} \zeta_a \bm{e}_a \in\mathbb{R}^{\ell-1}$.
We formally regard each $\zeta_a$ as the FI parameter for the $a$-th gauge node in the quiver while allowing the node to be absent.

\subsection{Relation between wall-crossing and operator ordering}\label{sec:proposal}


In Sections~\ref{sec:Moyal} and~\ref{sec:SQMbrane} we reviewed the two methods, the Moyal product and the SQMs, which we will employ in Sections~\ref{sec:twoops} and~\ref{sec:threeops} to compute $Z_\text{mono}$.
Both methods exhibit discrete changes in~$Z_\text{mono}$.
We now consider products of several $T_{\Box}$'s and $T_{\overline{\Box}}$'s in the 4d $\mathcal{N}=2$ $U(N)$ gauge theory with $2N$ flavors and see the relation between the discrete changes in the two methods. 

Because the Moyal product is non-commutative, the correlator of 't Hooft operators $T_a$ ($a=1,\ldots,\ell$) in general depends on their ordering. 
The $\Omega$-deformation requires the 't Hooft operators $T_a$ to be placed at $x^1 = x^2 = 0$, $x^3=s_a$.
The Euclidean time ordering 
\begin{equation} \label{eq:s-sigma-inequalities}
s_{\sigma(1)} > s_{\sigma(2)} >\ldots > s_{\sigma(\ell)}  \end{equation}
along the $x^3$-direction, where $\sigma\in S_\ell$, determines the ordering
\begin{equation}
\Braket{T_{\sigma(1)}} * \Braket{T_{\sigma(2)}} * \ldots * \Braket{T_{\sigma(\ell)}} 
\end{equation}
 in the Moyal product.
When two different operators pass each other there can be a discrete change%
\footnote{\label{footnote:Poynting}%
In an abelian gauge theory and for operators with electric and magnetic charges $(e_a,m_a)$ ($a=1,2$) satisfying $e_1 m_2 - e_2 m_1 \neq 0$, 
the jump occurs due to a change in the angular momentum $J_3$ induced by the electromagnetic fields.
See Section 3.6 of~\cite{Gaiotto:2010be}.
It is a non-trivial non-perturbative effect that the two purely magnetic operators do not commute. Indeed we will see that they are non-commutative only for restricted values of $N_F$. 
}.
Although the correlator is  
insensitive to infinitesimal changes in the positions, it may depend on the ordering; the correlators of half-BPS line operators in the $\Omega$-background form a 1d topological sector of the 4d theory~\cite{Yagi:2014toa}.

If we canonically quantize the theory on $S^1\times\mathbb{R}^3$ with an $\Omega$-deformation regarding $x^3$ as the Euclidean time, the operators $\widehat{T}_{\sigma(a)}$ that on the Hilbert space are time-ordered as
\begin{equation}\label{eq:operator-product-canonical}
 \widehat T_{\sigma(1)} \cdot \widehat T_{\sigma(2)} \cdot \ldots  \cdot \widehat T_{\sigma(\ell)}.
\end{equation}
One may interpret this as a product in the Heisenberg picture because it comes from a path integral, or as one in the Schr\"odinger picture for the 1d topological sector in which the effective Hamiltonian is zero.
In this case the operator product~(\ref{eq:operator-product-canonical}) realizes the non-commutativity.
In~\cite{Ito:2011ea} it was argued for class $\mathcal{S}$ theories that the so-called Weyl transform relates the line operator vev and the Verlinde operator~\cite{Verlinde:1988sn,Alday:2009fs,Drukker:2009id} that acts on conformal blocks.
We note that the Moyal product of the vevs is mapped to the product of Verlinde operators by the Weyl transform.  See, for example, \cite{Harvey:2001yn}.
It is then natural to identify the Verlinde operators with the operators in the 1d topological sector.

The discrete change in~$Z_\text{mono}$ should also be visible in the Witten index of the SQM that describes monopole screening.
Indeed the Witten index of an SQM, with $\mathcal{N}=(0, 4)$  supersymmetry broken to $\mathcal{N}=(0, 2)$ by a parameter $\epsilon_+$, 
is in general piecewise constant but can vary discretely in the space of FI parameters~\cite{Hwang:2014uwa, Cordova:2014oxa, Hori:2014tda}. 
The space of FI parameters is divided into FI-chambers~(\ref{def-FI-chamber}).
The Witten index may jump when one crosses an FI-wall separating two FI-chambers; this is wall-crossing.

We can see that the two origins (ordering and wall-crossing) of the discrete change in~$Z_\text{mono}$ are directly related to each other. 
This is because, in the brane picture considered in Section~\ref{sec:SQMbrane}, the locations $(x^1,x^2,x^3)=(0,0,s_a)$ of the NS5-branes and the minimal operators  ($a\leq \ell$) are related to the FI parameters $\zeta_a=s_{a+1}-s_a$ ($a\leq \ell-1$) of the SQMs.  
Different orderings correspond to different FI-chambers. 
A discrete change in the expectation value should match the discrete changes in the Witten indices. 

\begin{figure}[t]
\centering
\includegraphics[width=8cm]{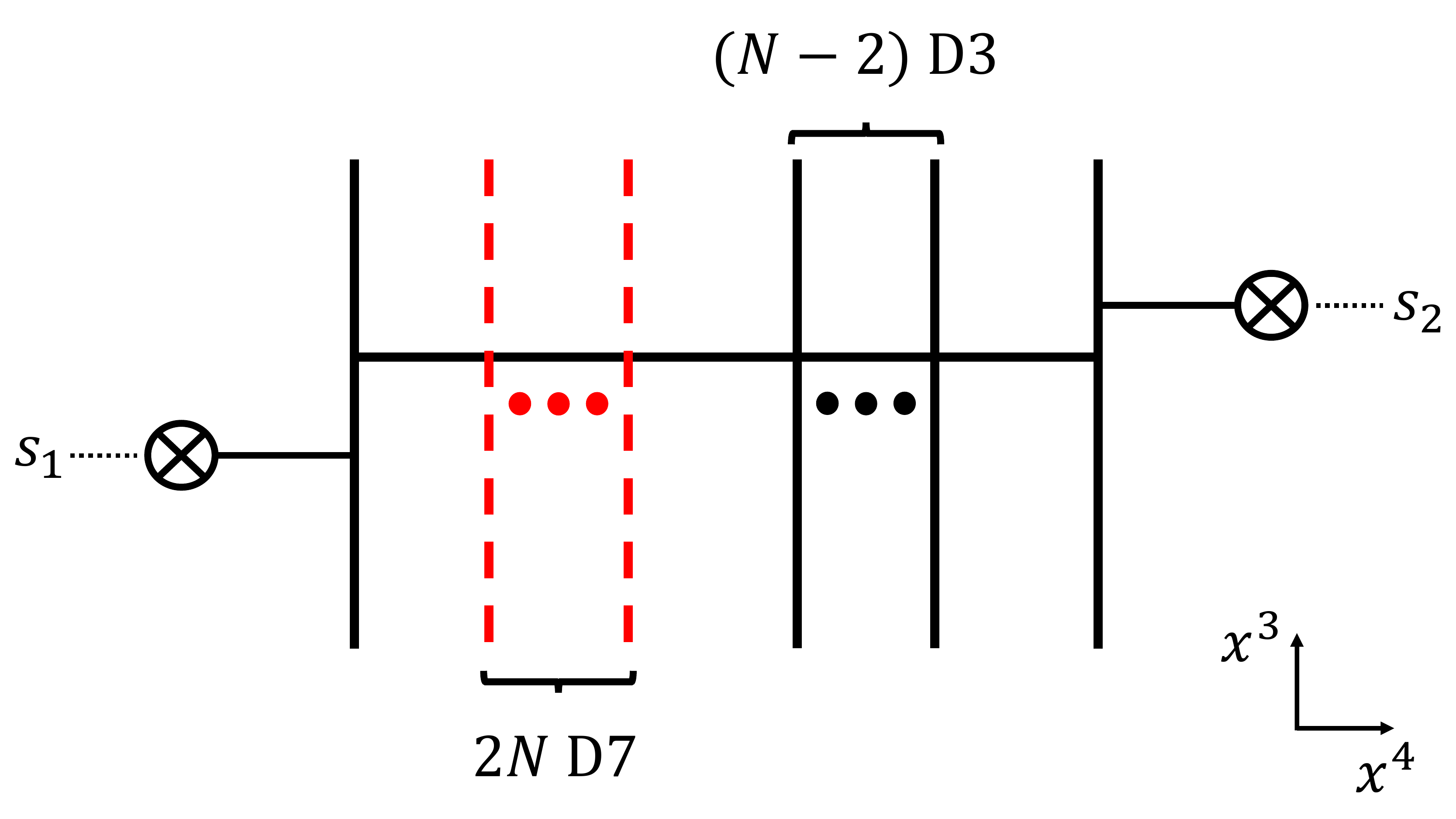}
\caption{A brane configuration for a product of two minimal 't Hooft line operator with the magnetic charge ${\bm B} = -{\bm e}_1 + {\bm e}_N$. A smooth monopole with magnetic charge $
{\bm e}_1 - {\bm e}_N$ is also introduced. $s_{1, 2}$ indicates the location of the NS5-branes in the $x^3$-space. }
\label{fig:relation1}
\end{figure}
To illustrate this we again consider the product of two minimal operators $T_{\Box}$ and $T_{\overline{\Box}}$ in the 4d $ \mathcal{N}=2$ $U(N)$ gauge theory with $2N$ flavors. %
In Figure~\ref{fig:relation1}  two D1-branes stretching between two NS5-branes and D3-branes inject the magnetic charge ${\bm B} = -{\bm e}_1 + {\bm e}_N$ of the product operator, and a finite D1-brane represents a smooth monopole with the opposite charge $ {\bm e}_1 - {\bm e}_N$ that screens the operator.
The locations of the NS5-branes can be interpreted in two ways. 
From the 4d viewpoint, the locations are the insertion points of the operators along the 3-axis;  $T_{\overline{\Box}}$ is inserted at $x^3 = s_1$ and $T_{\Box}$ at $x^3 = s_2$, where $s_1$ and $s_2$ are the locations of the corresponding NS5-branes.
The ordering $s_2 > s_1$ implies the ordering $ \Braket{T_{\Box}} \ast \Braket{T_{\overline{\Box}}}$ in the product. 
From the 1d viewpoint, the difference $s_2 - s_1$ is the FI parameter of the $U(1)$ gauge node on the D1-brane. 
When $\zeta = s_2 -s_1$ is non-zero and the gauge symmetry is spontaneously broken.
In Figure~\ref{fig:relation1} for example, $\zeta$ is positive. 
When we let the NS5-branes pass each other so that $s_1 > s_2$, the product in the 4d theory becomes $\Braket{T_{\overline{\Box}}}\ast\Braket{T_{\Box}}$  and the FI parameter becomes negative. Hence, the ordering in the Moyal product is directly related to the sign of the FI parameter, and the different expectation values due to the two different orderings need to give the same results as the two results of the Witten index related by the wall-crossing. 

It is also possible to consider the product of operators of the same type, for which a change in the ordering should not change the monopole screening contributions. 
Although the corresponding SQMs may still possess FI parameters, the relation above implies that there is no wall-crossing for the associated change in the parameters.

In some products of 't Hooft operators there can be several monopole screening sectors. 
In some sectors, the expectation value of the product operator does not change even when the ordering of $T_{\Box}$ and $T_{\overline{\Box}}$ is changed. 
This happens for example when smooth monopoles screen the magnetic charges of the $T_{\Box}$'s alone or the $T_{\overline{\Box}}$'s alone. 
We will encounter an example of this kind in Section~\ref{sec:threeops}. The ordering of minimal 't Hooft operators typically matters when smooth monopoles screen the magnetic charges of both $T_{\Box}$ and $T_{\overline{\Box}}$.


\section{$U(N)$ SQCD with $2N$ flavors: two minimal 't~Hooft operators}
\label{sec:twoops}

In the introduction and in Section~\ref{sec:proposal}, we explained a correspondence between the ordering of minimal 't~Hooft operators and the choice of an FI-chamber of the SQMs which describe monopole screening sectors. We will explicitly check the relation in the case of  
a product of minimal 't~Hooft operators in the 4d $\mathcal{N}=2$ $U(N)$ gauge theory with $2N$ hypermultiplets in the fundamental representation. Namely, we will compute monopole screening contributions in two ways, {\it i.e.}, by the Moyal product and by the SQMs, and show that the results match.

A minimal 't~Hooft operator in the 4d $U(N)$ gauge theory with $2N$ flavors has its magnetic charge corresponding to the weights of the fundamental or the anti-fundamental representation of the Langlands dual group $U(N)$.  
The expectation values of the minimal 't~Hooft operators have been given in \eqref{TF} and \eqref{TAF}. 
The expectation values of the minimal 't~Hooft operators themselves do not contain a monopole screening contribution. 
We start with the Moyal product of two minimal 't~Hooft operators.  

\subsection{Moyal product}

We first compute monopole screening contributions from the Moyal product of two minimal 't~Hooft line operators. There are three possible combinations for the product of the vevs of the two minimal 't~Hooft operators when we ignore the orderings. The three are $\Braket{T_{\Box}} \ast \Braket{T_{\Box}}$, $\Braket{T_{\Box}} \ast \Braket{T_{\overline{\Box}}}$ and $\Braket{T_{\overline{\Box}}} \ast \Braket{T_{\overline{\Box}}}$. Since $\Braket{T_{\overline{\Box}}} \ast \Braket{T_{\overline{\Box}}}$ is related to $\Braket{T_{\Box}} \ast \Braket{T_{\Box}}$ by the replacement $b_i \to -b_i\, (i=1, \cdots, N)$, we focus on the combinations $\Braket{T_{\Box}} \ast \Braket{T_{\Box}}$ and $\Braket{T_{\Box}} \ast \Braket{T_{\overline{\Box}}}$. 

The former combination is the Moyal product of the two same operators and hence the ordering should not matter. More concretely, first note that $\Braket{T_{\Box}} \ast \Braket{T_{\Box}}$ can be decomposed as
\begin{align}
\Braket{T_{\Box}} \ast \Braket{T_{\Box}} =& \sum_{i=1}^N e^{2b_i}Z_{\text{1-loop}}({\bm v} = 2{\bm e}_i)\nn\\
& + \sum_{1 \leq i < j \leq N}e^{b_i + b_j}Z_{\text{1-loop}}({\bm v} = {\bm e}_i + {\bm e}_j)Z_{\text{mono}}({\bm v} = {\bm e}_i + {\bm e}_j). \label{TFF}
\end{align}
We identify $e^{2b_i}$ and $e^{b_i + b_j}$ with $e^{\bm{v}\cdot\bm{b}}$ in~(\ref{eq:TB-vev}).
The one-loop parts are always unique and the ordering dependence only affects monopole screening parts. In this case, the sector with $e^{2b_i}$ does not have a monopole screening contribution whereas the sector with $e^{b_i + b_j}$ $(i \neq j)$ exhibits monopole screening. Since the ordering is unique, the monopole screening contribution $Z_{\text{mono}}({\bm v} = {\bm e}_i + {\bm e}_j)$ should not depend on the choice of an FI-chamber.

On the other hand, the latter combination $\Braket{T_{\Box}} \ast \Braket{T_{\overline{\Box}}}$ may possibly give different results depending on the ordering, namely whether the ordering is $\Braket{T_{\Box}}\ast \Braket{T_{\overline{\Box}}}$ or $\Braket{T_{\overline{\Box}}}\ast \Braket{T_{\Box}}$, since the Moyal product is in general non-commutative. In both cases, the product can be decomposed as
\begin{align}
\Braket{T_{\Box}}\ast \Braket{T_{\overline{\Box}}}\text{ or }\Braket{T_{\overline{\Box}}}\ast \Braket{T_{\Box}} = &\sum_{1 \leq i \neq j \leq N}e^{b_i-b_j}Z_{\text{1-loop}}({\bm v} = {\bm e}_i - {\bm e}_j) + Z_{\text{mono}}({\bm v} = {\bm 0}). \label{TTFAF}
\end{align}
Note that $Z_{\text{1-loop}}({\bm v} = {\bm 0}) = 1$.  
In this case, the ordering of the two minimal 't~Hooft operators may affect the monopole screening contribution $Z_{\text{mono}}({\bm v} = {\bm 0})$ in the sector ${\bm v} = {\bm 0}$.  

It is straightforward to explicitly see the monopole screening contributions from the application of the Moyal product defined in \eqref{moyal}. For example, when we apply the Moyal product \eqref{moyal} to $\Braket{T_{\Box}} \ast \Braket{T_{\Box}}$, we obtain
\begin{align}
&\Braket{T_{\Box}} \ast \Braket{T_{\Box}} = \sum_{1 \leq i, j \leq N}e^{b_i + b_j}Z_i({\bm a} + \epsilon_+{\bm e}_j)Z_j({\bm a} - \epsilon_+{\bm e}_i)\nn\\
=&\sum_{i=1}^Ne^{2b_i}Z_i({\bm a} + \epsilon_+{\bm e}_i)Z_i({\bm a} - \epsilon_+{\bm e}_i) 
+
 \sum_{1 \leq i \neq j \leq N}
e^{b_i + b_j}
Z_i({\bm a} + \epsilon_+{\bm e}_j)Z_j({\bm a} - \epsilon_+{\bm e}_i). \label{TFFresult0}
\end{align}
Inserting the explicit expression \eqref{Zi} into \eqref{TFFresult0} yields
\begin{align}
\Braket{T_{\Box}} \ast \Braket{T_{\Box}} 
&=\sum_{i=1}^Ne^{2b_i}Z_{\text{1-loop}}({\bm v} = 2{\bm e}_i) 
+ \sum_{1 \leq i < j \leq N}e^{b_i+b_j}Z_{\text{1-loop}}({\bm v} = {\bm e}_i + {\bm e}_j) \times
\nn \\
&\qquad\times
\Bigg(\frac{1}{2\sinh\frac{a_i - a_j}{2}2\sinh\frac{- a_i + a_j + 2\epsilon_+}{2}}
 + \frac{1}{2\sinh\frac{a_j - a_i}{2}2\sinh\frac{- a_j + a_i + 2\epsilon_+}{2}} \Bigg).\label{TFFresult}
\end{align}
Hence the contribution for the monopole screening in the sector ${\bm v} = {\bm e}_i + {\bm e}_j$ is given by
\begin{align}
\Braket{T_{\Box}} \ast \Braket{T_{\Box}}\Big|_{Z_{\text{mono}}({\bm v} = {\bm e}_i + {\bm e}_j)} &=\frac{1}{2\sinh\frac{a_i - a_j}{2}2\sinh\frac{- a_i + a_j + 2\epsilon_+}{2}} 
\nn\\
&\qquad
+ \frac{1}{2\sinh\frac{a_j - a_i}{2}2\sinh\frac{- a_j + a_i + 2\epsilon_+}{2}}.  \label{eq:box-box-Zmono}
\end{align}

Next we consider the products $\Braket{T_{\Box}}\ast \Braket{T_{\overline{\Box}}}$ and $\Braket{T_{\overline{\Box}}}\ast \Braket{T_{\Box}}$. Applying the Moyal product \eqref{moyal} to $\Braket{T_{\Box}}\ast \Braket{T_{\overline{\Box}}}$ and $\Braket{T_{\overline{\Box}}}\ast \Braket{T_{\Box}}$ yields the expressions,
\begin{align}
\Braket{T_{\Box}} \ast \Braket{T_{\overline{\Box}}} =& \sum_{1 \leq i, j \leq N}e^{b_i - b_j}Z_i({\bm a}-\epsilon_+{\bm e}_j)Z_j({\bm a}-\epsilon_+{\bm e}_i)\nn\\
=& \sum_{1 \leq i \neq j \leq N}e^{b_i - b_j}Z_i({\bm a} - \epsilon_+{\bm e}_j)Z_j({\bm a}-\epsilon_+{\bm e}_i) + \sum_{i=1}^{N}Z_i({\bm a}- \epsilon_+{\bm e}_i)^2\label{TFAFresult01},
\end{align}
and 
\begin{align}
\Braket{T_{\overline{\Box}}}\ast \Braket{T_{\Box}} =& \sum_{1 \leq i, j \leq N}e^{-b_i + b_j}Z_i({\bm a}+\epsilon_+{\bm e}_j)Z_j({\bm a}+\epsilon_+{\bm e}_i)\nn\\
=& \sum_{1 \leq i \neq j \leq N }e^{-b_i + b_j}Z_i({\bm a} + \epsilon_+{\bm e}_j)Z_j({\bm a} + \epsilon_+{\bm e}_i) + \sum_{i=1}^{N}Z_i({\bm a} + \epsilon_+{\bm e}_i)^2. \label{TFAFresult02}
\end{align}
Using \eqref{Zi}, \eqref{TFAFresult01} and \eqref{TFAFresult02} can be expressed as
\begin{align}
\Braket{T_{\Box}} \ast \Braket{T_{\overline{\Box}}} =&\sum_{1 \leq i \neq j \leq N}e^{b_i - b_j}Z_{\text{1-loop}}({\bm v} = {\bm e}_i - {\bm e}_j) 
\nn\\
&\qquad
+  \sum_{i=1}^{N}\frac{\prod_{f=1}^{2N}2\sinh\frac{a_i - m_f - \epsilon_+}{2}}{\prod_{1 \leq j (\neq \text{ fixed }i) \leq N}2\sinh\frac{a_i - a_j}{2}2\sinh\frac{-a_i + a_j + 2\epsilon_+}{2}},\\
\Braket{T_{\overline{\Box}}}\ast \Braket{T_{\Box}}  =&\sum_{1 \leq i \neq j \leq N}e^{-b_i + b_j}Z_{\text{1-loop}}({\bm v} = -{\bm e}_i + {\bm e}_j) 
\nn\\
&\qquad
+\sum_{i=1}^{N}\frac{\prod_{f=1}^{2N}2\sinh\frac{a_i - m_f + \epsilon_+}{2}}{\prod_{1 \leq j (\neq \text{ fixed }i) \leq N}2\sinh\frac{a_i - a_j + 2\epsilon_+}{2}2\sinh\frac{-a_i + a_j}{2}}.
\end{align}
Let us introduce the notation
\begin{equation}
\braket{ T_{\bm B}}\Big|_{Z_\text{mono}({\bm v})} := Z_\text{mono}({\bm B},{\bm v}) \text{ in } (\ref{eq:TB-vev}).
\end{equation}
The contributions to the monopole screening in the sector ${\bm v} = {\bm 0}$ for the two orderings are
\begin{align}
\Braket{T_{\Box}} \ast \Braket{T_{\overline{\Box}}}\Big|_{Z_{\text{mono}}({\bm v} = {\bm 0})} =&  \sum_{i=1}^{N}\frac{\prod_{f=1}^{2N}2\sinh\frac{a_i - m_f - \epsilon_+}{2}}{\prod_{1 \leq j (\neq \text{ fixed }i) \leq N}2\sinh\frac{a_i - a_j}{2}2\sinh\frac{-a_i + a_j + 2\epsilon_+}{2}}, \label{TFAFv0}\\
\Braket{T_{\overline{\Box}}}\ast \Braket{T_{\Box}}\Big|_{Z_{\text{mono}}({\bm v} = {\bm 0})} = &  \sum_{i=1}^{N}\frac{\prod_{f=1}^{2N}2\sinh\frac{a_i - m_f + \epsilon_+}{2}}{\prod_{1 \leq j (\neq \text{ fixed }i) \leq N}2\sinh\frac{a_i - a_j + 2\epsilon_+}{2}2\sinh\frac{-a_i + a_j}{2}}.\label{TAFFv0}
\end{align}
We obtained different expressions for the monopole screening in the ${\bm v} = {\bm 0}$ sector from the different orderings $\Braket{T_{\Box}} \ast \Braket{T_{\overline{\Box}}}$ and $\Braket{T_{\overline{\Box}}}\ast \Braket{T_{\Box}}$. 
They indeed have different values as we will see in~(\ref{TFTAFDiff}).

\subsection{SQMs}

We then turn to the computations using the Witten index of SQMs for the monopole screening contributions.

\subsubsection{'t~Hooft operator with ${\bm B} = 2{\bm e}_N$}
\begin{figure}[t]
\centering
\subfigure[]{\label{fig:B02no1}
\includegraphics[width=8cm]{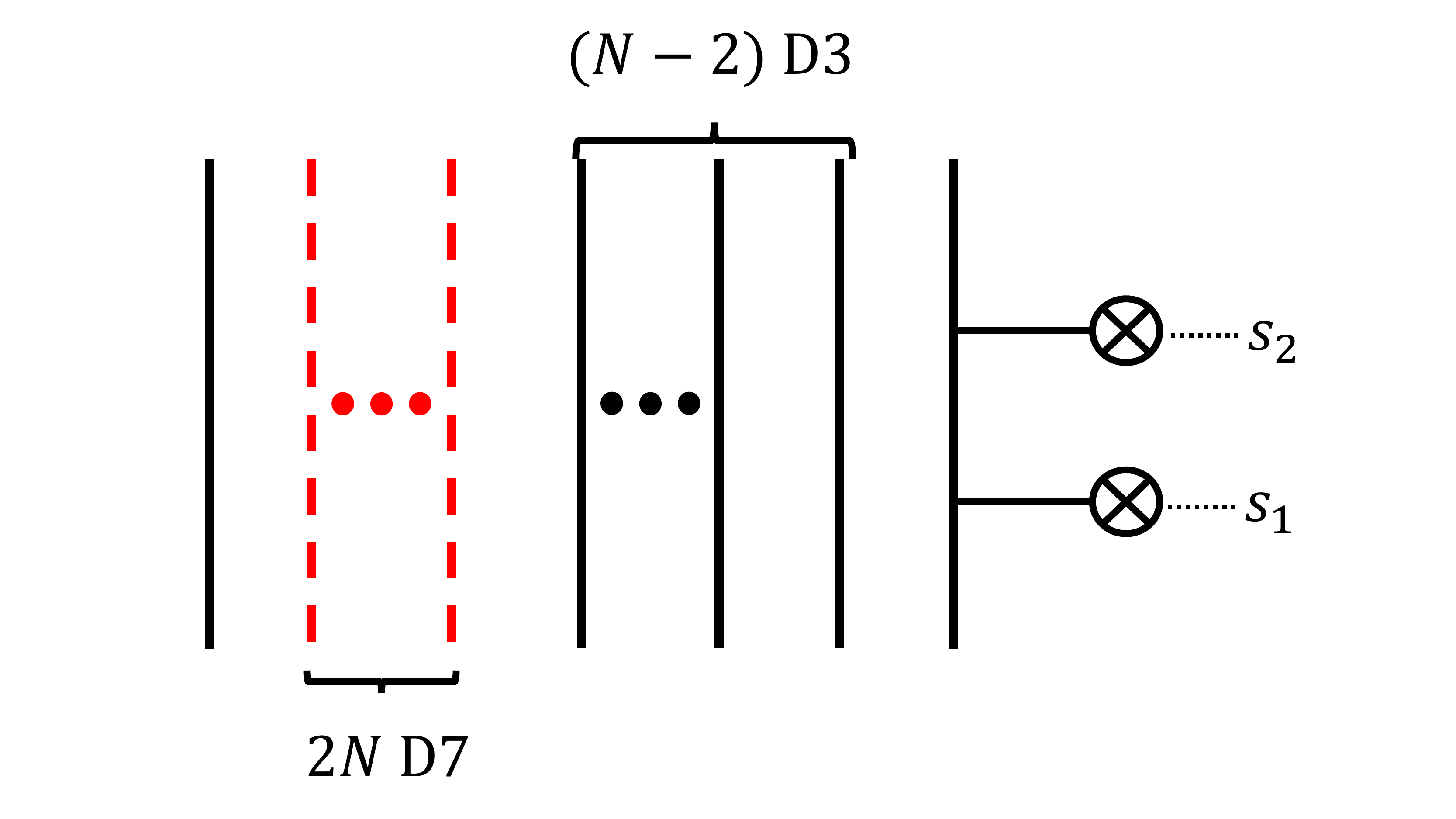}}
\subfigure[]{\label{fig:B02no2}
\includegraphics[width=8cm]{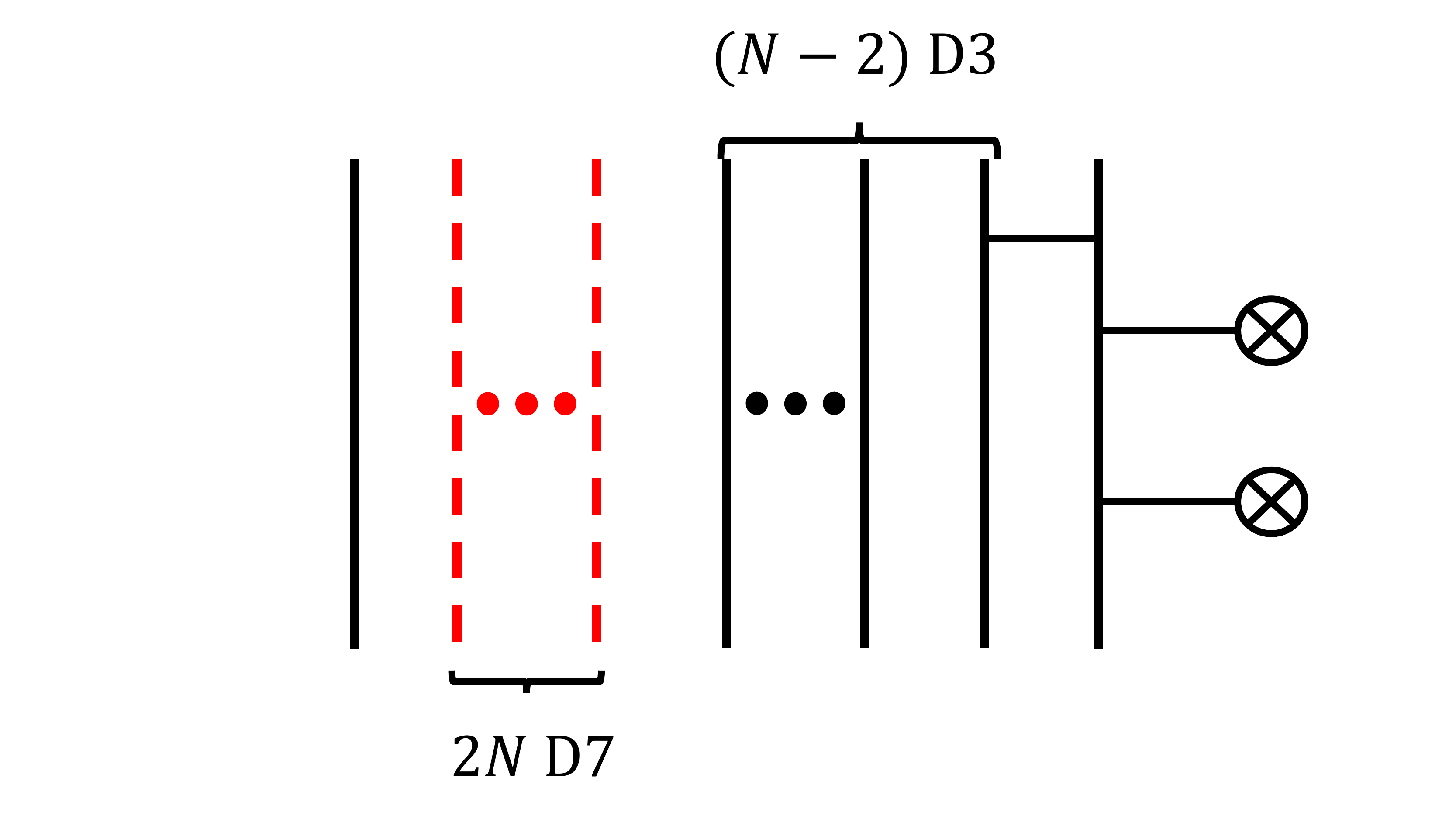}}
\subfigure[]{\label{fig:B02no2to3}
\includegraphics[width=8cm]{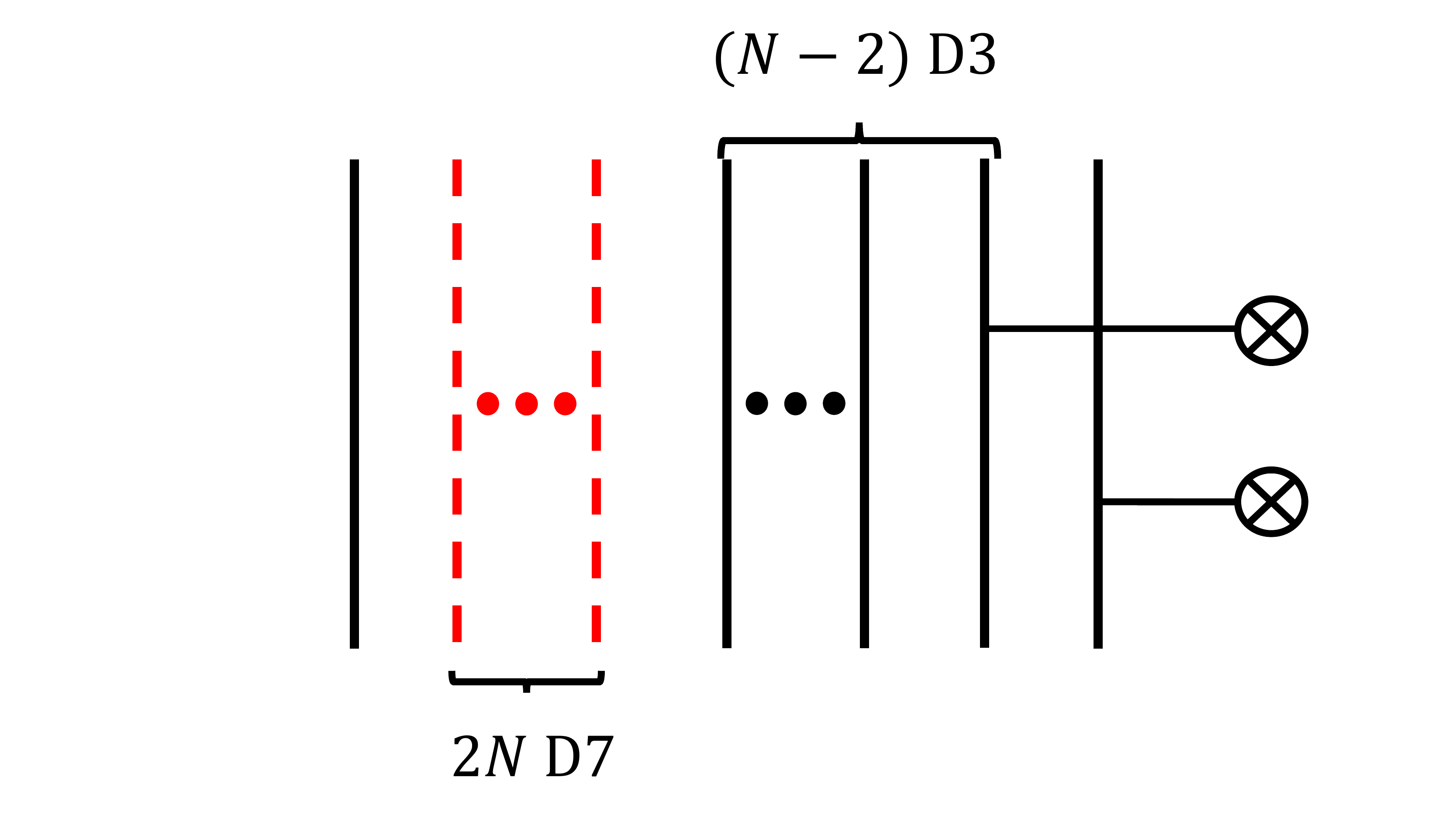}}
\subfigure[]{\label{fig:B02no3}
\includegraphics[width=8cm]{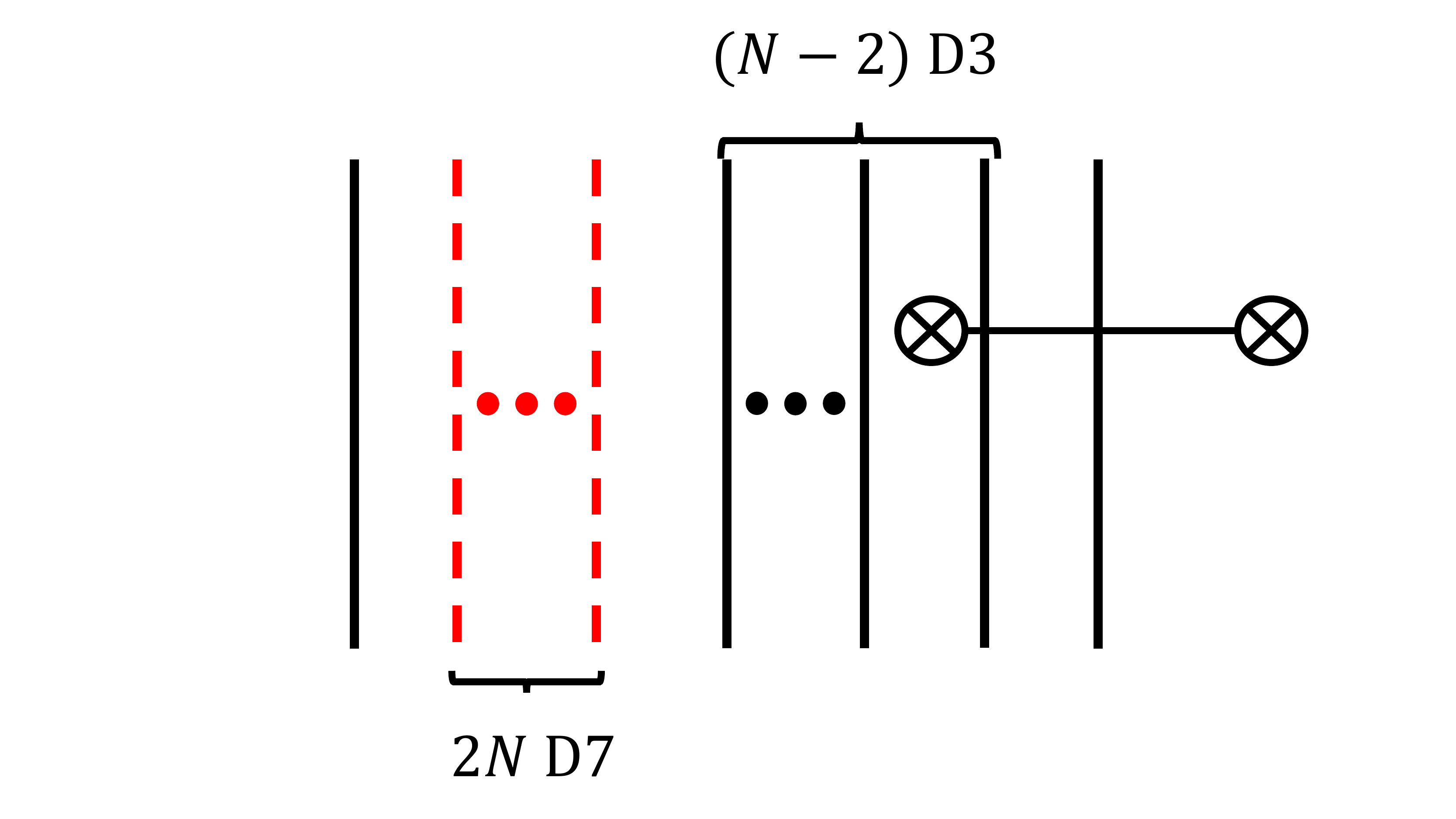}}
\caption{(a): A brane configuration realizing the 't~Hooft operator with the magnetic charge ${\bm B}=2{\bm e}_N$.  
(b): Introducing a smooth monopole with the magnetic charge $
{\bm e}_{N-1} - {\bm e}_N$  
by a D1-brane between two D3-branes to Figure \ref{fig:B02no1}. (c): Tuning the position of the D1-brane between D3-branes until it coincides with that for the upper D1-brane between an NS5-brane and the $N$th D3-brane, corresponding to the screening sector ${\bm v} = {\bm e}_{N-1} + {\bm e}_N$. (d): The brane configuration for reading off the SQM for the monopole screening contribution in the sector ${\bm v} = {\bm e}_{N-1} + {\bm e}_N$.
}
\label{fig:B02}
\end{figure}
First we consider the monopole screening contribution $Z_{\text{mono}}({\bm v} = {\bm e}_i + {\bm e}_j)$ which appears in the product $\Braket{T_{\Box}}\ast\Braket{T_{\Box}}$ in \eqref{TFF}. For that we focus on the case $i=N-1, j = N$. The 't~Hooft operator with the magnetic charge ${\bm B} = 2{\bm e}_N$  
is realized by a brane configuration in Figure \ref{fig:B02no1}. The coordinate in the $x^3$-direction for the lower NS5-brane is denoted by $s_1$ and that for the upper NS5-branes is denoetd by $s_2$. The $N$ D3-branes are labelled from left to right. The 't~Hooft operator has a screening sector ${\bm v} = {\bm e}_{N-1} + {\bm e}_N$. In order to see it, we introduce a smooth monopole with magnetic charge $
{\bm e}_{N-1} - {\bm e}_N$  
as in Figure~\ref{fig:B02no2}.
The smooth monopole is expressed as a D1-brane between two D3-branes. Then it is possible to screen the magnetic charge by tuning the position of the D1-brane between two D3-branes until the position coincides with that for one of the other D1-branes between an NS5-brane and the $N$-th D3-brane. The diagram in Figure \ref{fig:B02no2to3} shows the case when the position of the D1-brane between the D3-branes is set to the position of the upper D1-brane between an NS5-brane and the $N$-th D3-brane. It is possible to see that the 't~Hooft operator realized by the D1-branes has magnetic charge $
{\bm e}_{N-1} + {\bm e}_N$. In order to read off an SQM describing the degrees of freedom for the monopole screening, we further move the lower NS5-brane in Figure \ref{fig:B02no2to3}  
until we obtain the configuration in Figure \ref{fig:B02no3}. The configuration in Figure \ref{fig:B02no3} has a D1-brane between two NS5-branes and the effective field theory on the D1-brane gives an SQM. 

\begin{figure}[t]
\centering
\includegraphics[width=2.3cm]{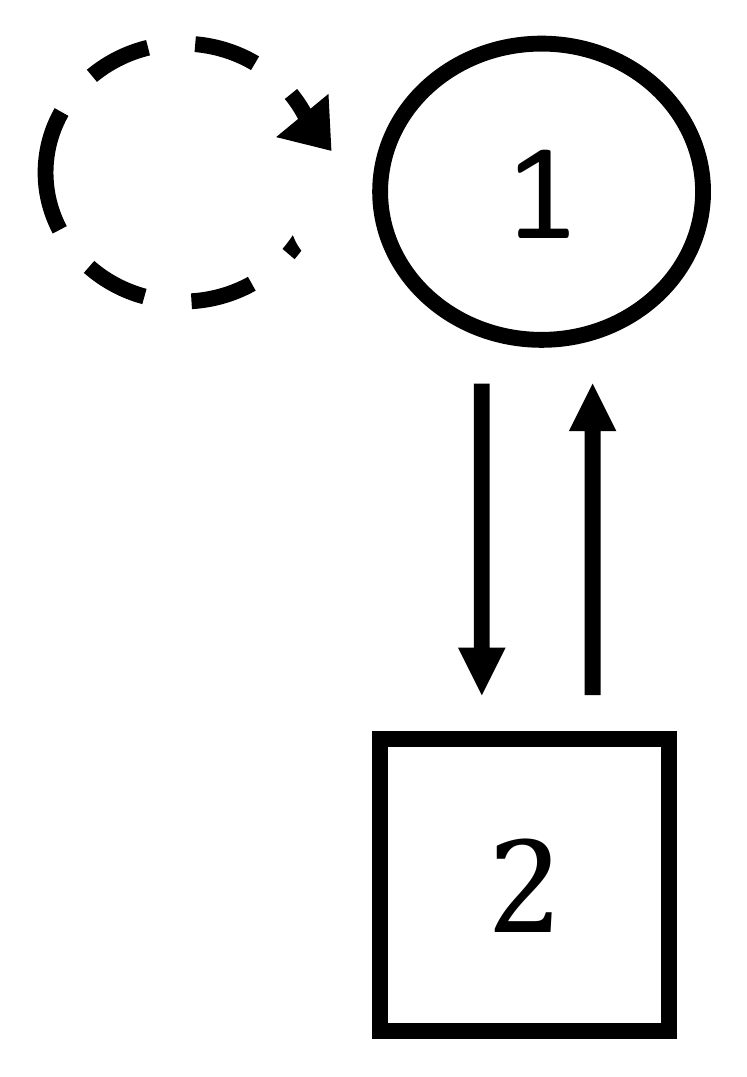}
\caption{The quiver diagram for the worldvolume theory on the D1-branes in the configuration in Figure \ref{fig:B02no3}. 
}
\label{fig:quiver02}
\end{figure}
From the brane configuration in Figure~\ref{fig:B02no3}, we can read off the field content of the SQM as we explained in Section~\ref{sec:SQMbrane};
we summarize it as a quiver diagram in Figure~~\ref{fig:quiver02}.
Each arrow represents an $\mathcal{N}=(0, 2)$ multiplet. 
Associated to the $U(1)$ gauge node we have an FI parameter $\zeta$, related to the positions $s_{1,2}$ of the two NS5-branes as $\zeta = s_2 - s_1$.
As reviewed in Appendix~\ref{sec:locforSQM}, the Witten index is given by the integral of the product of the one-loop determinants for all the multiplets.
In this case it is
\begin{align}
Z({\bm B} =2{\bm e}_N, {\bm v}= {\bm e}_{N-1} + {\bm e}_N; \zeta) = \oint_{JK(\zeta)}\frac{d\phi}{2\pi i}\frac{2\sinh\epsilon_+}{\prod_{i=N-1}^{N}2\sinh\frac{\phi - a_i + \epsilon_+}{2}2\sinh\frac{-\phi + a_i + \epsilon_+}{2}}, \label{Z011}
\end{align}
where $a_{N-1}$ and $a_N$ are  the chemical potentials for the flavor symmetry group $U(2)$, parameterizing the position of the $(N-1)$-th and the $N$-th D3-branes. The result of the integral generically depends on the FI parameter for the $U(1)$ gauge group in the quiver theory in Figure \ref{fig:quiver02}. It turns out that we can use the Jeffrey-Kirwan (JK) residue  prescription if we choose the FI parameter  as the JK parameter $\eta$ \cite{Hwang:2014uwa, Cordova:2014oxa, Hori:2014tda}.  
A brief review of the JK residue is given in Appendix \ref{sec:locforSQM}.  The subscript $JK(\zeta)$ in \eqref{Z011}  indicates that we evaluate the integral following the JK residue prescription  with the JK parameter $\eta$ set to $\zeta$.

 A choice of the sign of~$\zeta$ determines
the poles that contribute to the integral~(\ref{Z011}).
For the choice $\zeta > 0$, the poles contributing to the JK residue are  $\phi = a_{N-1} - \epsilon_+$ and $\phi = a_N - \epsilon_+$. 
 An explicit calculation of the residues of \eqref{Z011} gives
\begin{align}
&Z({\bm B} =2{\bm e}_N, {\bm v}= {\bm e}_{N-1} + {\bm e}_N; \zeta > 0) \nn\\
&= \frac{1}{2\sinh\frac{a_{N-1} - a_N}{2}2\sinh\frac{ - a_{N-1} + a_N + 2\epsilon_+}{2}} +  \frac{1}{2\sinh\frac{a_{N} - a_{N -1}}{2}2\sinh\frac{ - a_{N} + a_{N-1} + 2\epsilon_+}{2}}. \label{B011result1} 
\end{align}
 For the other choice $\zeta < 0$ we take the poles $\phi = a_{N-1} + \epsilon_+$ and $\phi = a_N + \epsilon_+$ to obtain
\begin{align}
&Z({\bm B} =2{\bm e}_N, {\bm v}= {\bm e}_{N-1} + {\bm e}_N; \zeta < 0)\nn\\
& = \frac{1}{2\sinh\frac{a_{N-1} - a_N + 2\epsilon_+}{2}2\sinh\frac{ - a_{N-1} + a_N}{2}} +  \frac{1}{2\sinh\frac{a_{N} - a_{N-1} + 2\epsilon_+}{2}2\sinh\frac{ - a_{N} + a_{N-1}}{2}}. \label{B011result2}
\end{align}
 The resulting expressions \eqref{B011result1} and \eqref{B011result2} turn out to be equal, {\it i.e.},
\begin{align}
Z({\bm B} =2{\bm e}_N, {\bm v}= {\bm e}_{N-1} + {\bm e}_N; \zeta > 0) = Z({\bm B} =2{\bm e}_N, {\bm v}= {\bm e}_{N-1} + {\bm e}_N; \zeta < 0),
\end{align}
and there is no wall-crossing. 
This perfectly agrees with the fact that the monopole screening contribution  for ${\bm v} = {\bm e}_{N-1} + {\bm e}_N$ appears in the product $\Braket{T_{\Box}}\ast \Braket{T_{\Box}}$  that has a unique ordering.
 One also sees that~\eqref{B011result1} and~\eqref{B011result2} both agree with~\eqref{eq:box-box-Zmono} when~$i= N-1, j= N$.

\subsubsection{'t~Hooft operator with ${\bm B} = -{\bm e}_1 + {\bm e}_N$}
\begin{figure}[t]
\centering
\subfigure[]{\label{fig:B101no1}
\includegraphics[width=8cm]{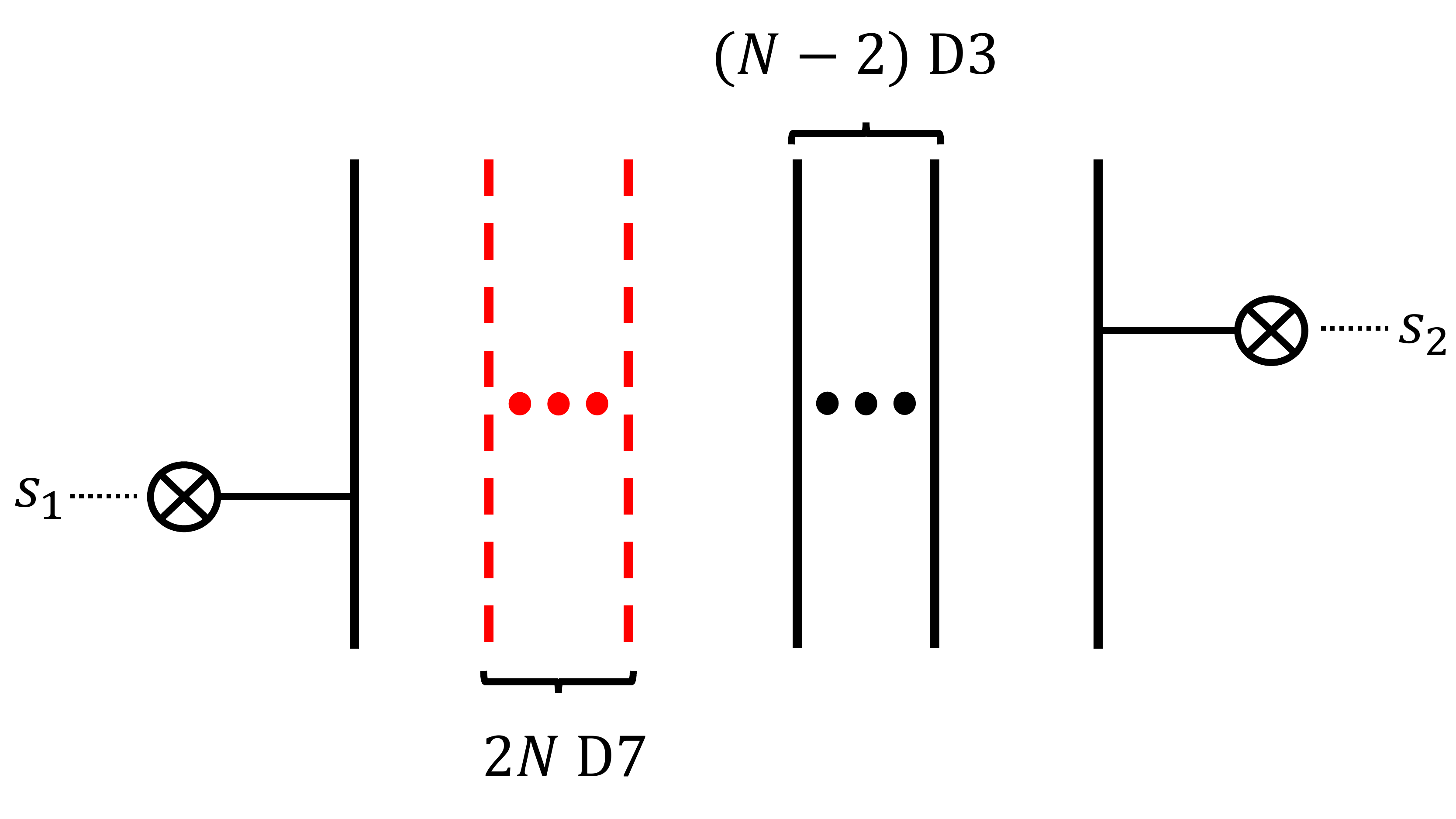}}
\subfigure[]{\label{fig:B101no2}
\includegraphics[width=8cm]{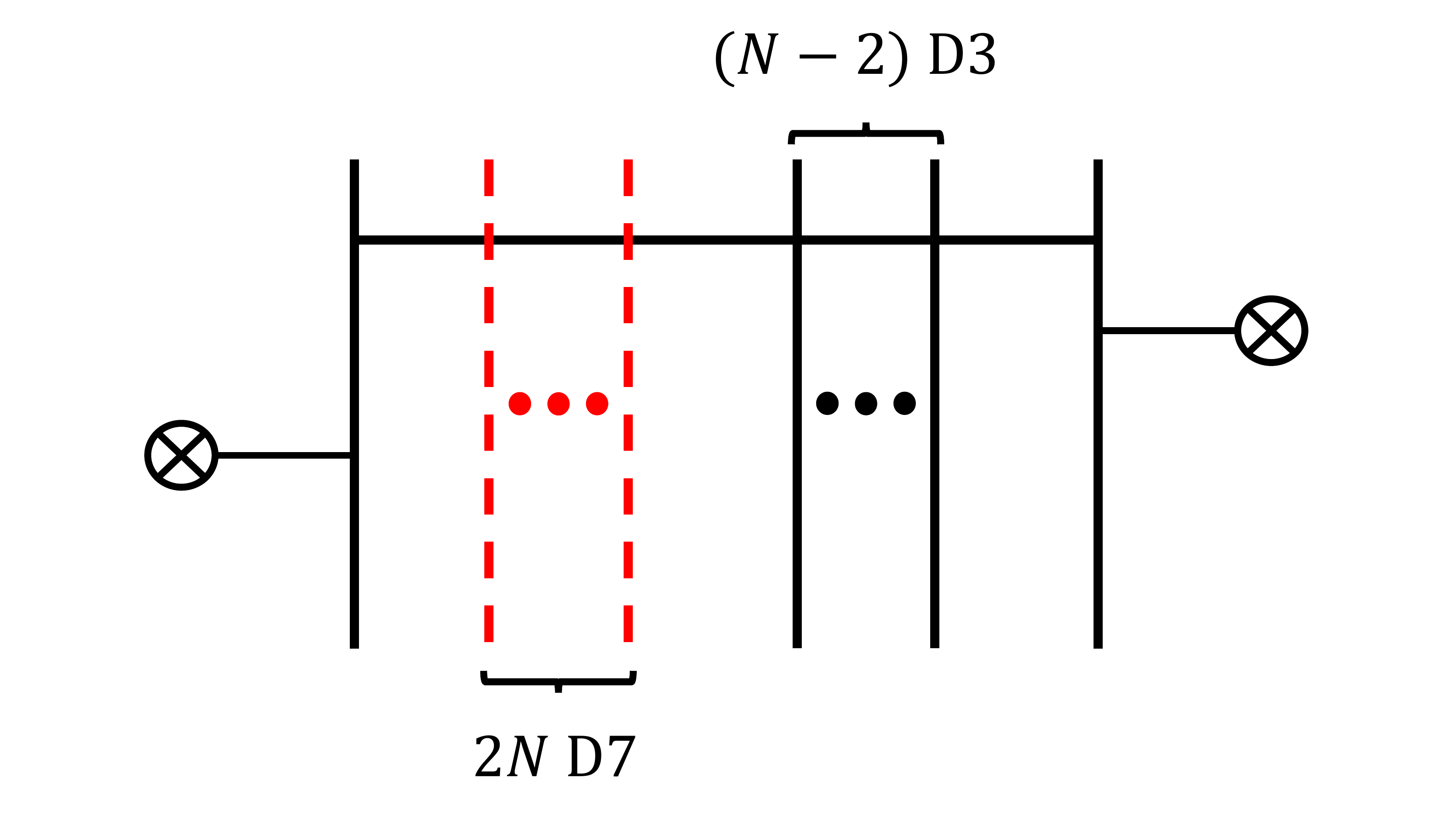}}
\subfigure[]{\label{fig:B101no3}
\includegraphics[width=8cm]{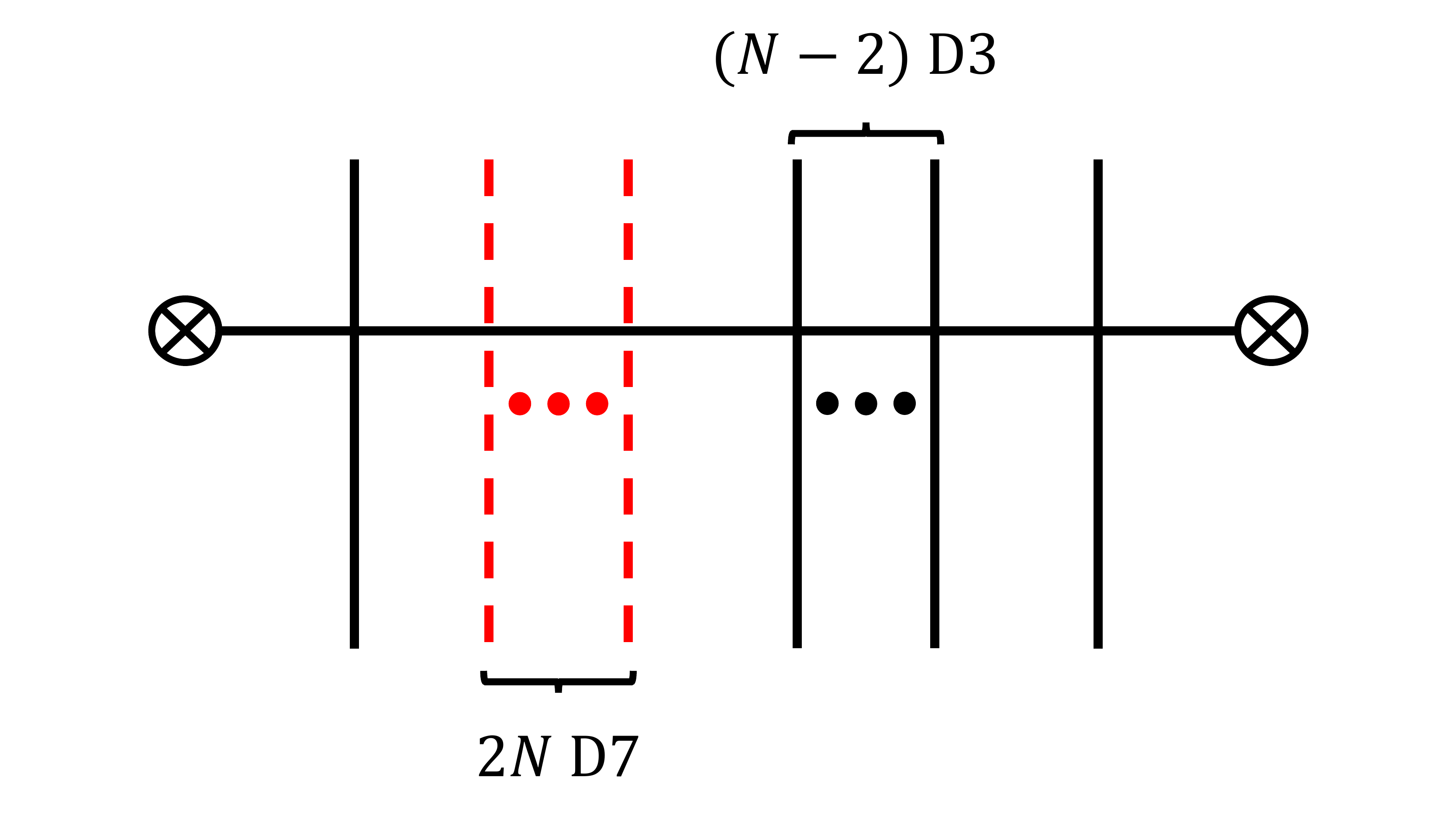}}
\subfigure[]{\label{fig:quiver101}
\includegraphics[width=8cm]{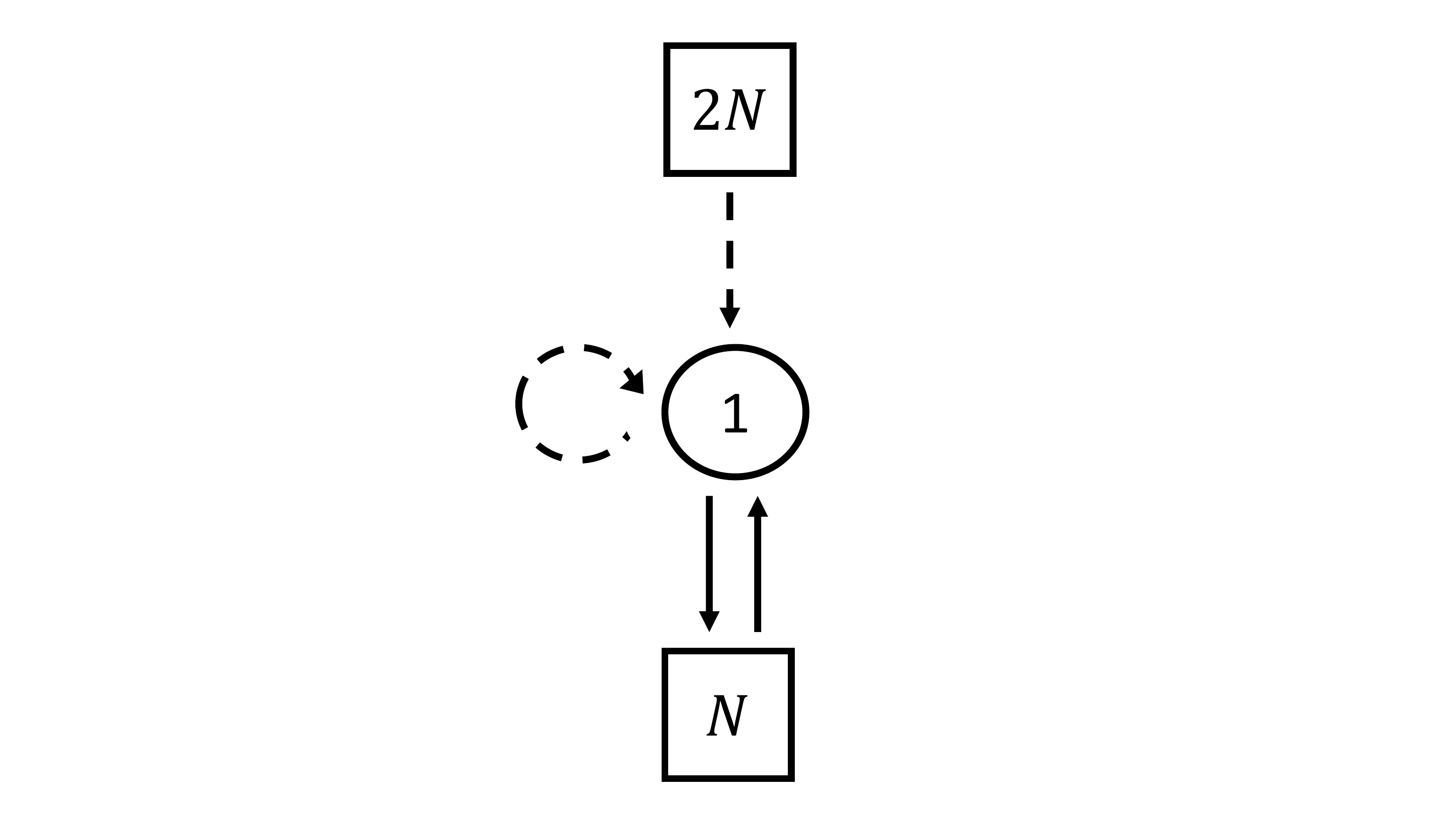}}
\caption{(a): A brane configuration realizing the 't~Hooft operator with the magnetic charge ${\bm B}=-{\bm e}_1 + {\bm e}_N$.  
(b): Introducing a smooth monopole with magnetic charge $
 {\bm e}_1 - {\bm e}_N$ 
to Figure~\ref{fig:B101no1}. (c): The brane configuration for reading off the SQM describing the contribution of the monopole screening of the 't~Hooft operator in the sector ${\bm v} = {\bm 0}$. (d): The quiver diagram for the worldvolume theory on the D1-branes in the configuration in Figure~\ref{fig:B101no3}. }
\label{fig:B101}
\end{figure}
Monopole screening also occurs in the sector ${\bm v} = {\bm 0}$  for the product $\Braket{T_{\Box}}\ast\Braket{T_{\overline{\Box}}}$ and $\Braket{T_{\overline{\Box}}}\ast\Braket{T_{\Box}}$.
 To study it from the viewpoint of an SQM, we consider the sector ${\bm v} = {\bm 0}$  for the 't~Hooft operator with the magnetic charge ${\bm B}= - {\bm e}_1 + {\bm e}_N$.  
The SQM for the monopole screening can be read off from the corresponding brane configuration. The 't~Hooft operator with the magnetic charge ${\bm B}=-{\bm e}_1 + {\bm e}_N$  
can be realized by a brane configuration in Figure \ref{fig:B101no1}. 
 The values of the coordinate~$x^3$ for the positions of the NS5-branes on the left and on the right are denoted by $s_1$ and $s_2$, respectively.
In order to realize the ${\bm v} = {\bm 0}$ sector, we introduce a D1-brane corresponding to a smooth monopole with magnetic charge $
 {\bm e}_1 - {\bm e}_N$  
as in Figure \ref{fig:B101no2}. Then the monopole screening occurs when the position of the D1-branes becomes equal to each other and we have a single D1-brane connecting the two NS5-branes, which is depicted in Figure \ref{fig:B101no3}. From the brane configuration in Figure \ref{fig:B101no3}, it is possible to read off the content of the effective field theory of the worldvolume theory on the D1-branes. The field theory content can be summarized by the quiver diagram in Figure \ref{fig:quiver101} and FI parameter of the $U(1)$ gauge node is given by $\zeta = s_2 - s_1$. 

From the quiver diagram in Figure \ref{fig:quiver101}, we can compute the Witten index of the SQM using \eqref{partitionfunction}. The partition function is given by
\begin{align}
Z({\bm B}=-{\bm e}_1 + {\bm e}_N, {\bm v}={\bm 0}; \zeta)= \oint_{JK(\zeta)}\frac{d\phi}{2\pi i}\frac{2\sinh\epsilon_+\prod_{f=1}^{2N}2\sinh\frac{\phi - m_f}{2}}{\prod_{i=1}^{N}2\sinh\frac{\phi - a_i + \epsilon_+}{2}2\sinh\frac{-\phi + a_i + \epsilon_+}{2}}, \label{Z101}
\end{align}
where $m_f\, (f=1, \cdots, 2N)$ are chemical potentials for the flavor symmetry group $U(2N)$, which correspond to the mass parameters of the $2N$ flavors, and $a_i\, (i=1, \cdots, N)$ are chemical potentials for the flavor symmetry group $U(N)$. 

\begin{figure}[t]
\centering
\includegraphics[width=8cm]{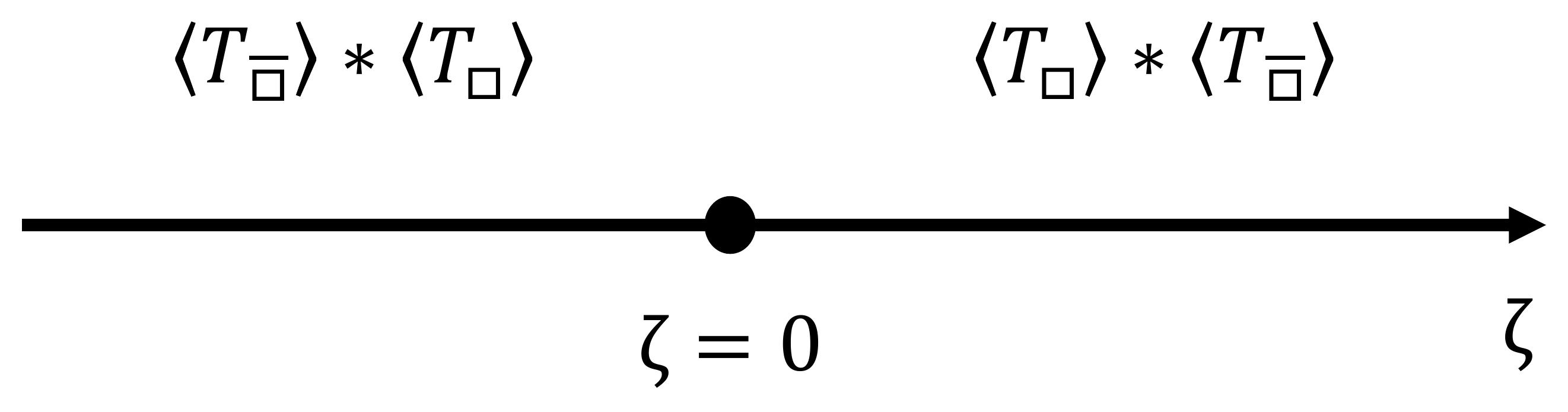}
\caption{The relation between the order of the Moyal product and the sign of the FI parameter in the case of the ${\bm v} = {\bm 0}$ sector.
}
\label{fig:phase1}
\end{figure}

In a similar way to the evaluation for the integral \eqref{Z011}, there are two possible choices of poles depending on the sign of the FI parameter. For $\zeta > 0$, we choose the poles $\phi = a_i - \epsilon_+\, (i=1, \cdots, N)$. Then the evaluation of the integral \eqref{Z101} yields
\begin{align}
&Z({\bm B}=-{\bm e}_1 + {\bm e}_N, {\bm v}={\bm 0}; \zeta > 0)=\sum_{i=1}^{N}\frac{\prod_{f=1}^{2N}2\sinh\frac{a_i - m_f - \epsilon_+}{2}}{\prod_{1 \leq j (\neq \text{ fixed }i) \leq N}2\sinh\frac{a_i - a_j}{2}2\sinh\frac{-a_i + a_j + 2\epsilon_+}{2}}.\label{Z101result1}
\end{align}
On the other hand, the choice $\zeta < 0$ requires poles $\phi = a_i + \epsilon_+\, (i=1, \cdots, N)$. Evaluating the integral \eqref{Z101} then gives
\begin{align}
&Z({\bm B}=-{\bm e}_1 + {\bm e}_N, {\bm v}={\bm 0}; \zeta < 0) = \sum_{i=1}^{N}\frac{\prod_{f=1}^{2N}2\sinh\frac{a_i - m_f + \epsilon_+}{2}}{\prod_{1 \leq j (\neq \text{ fixed }i) \leq N}2\sinh\frac{a_i - a_j + 2\epsilon_+}{2}2\sinh\frac{-a_i + a_j}{2}}.\label{Z101result2}
\end{align}
In this case the two results \eqref{Z101result1} and \eqref{Z101result2} are different from each other and they show the wall-crossing. This agrees with the fact that the ${\bm v} = {\bm 0}$ sector appears in the product $\Braket{T_{\Box}}\ast\Braket{T_{\overline{\Box}}}$ or $\Braket{T_{\overline{\Box}}}\ast\Braket{T_{\Box}}$ and the different orderings indeed gave different results as in \eqref{TFAFv0} and \eqref{TAFFv0}. 
The explicit comparison of \eqref{Z101result1} and \eqref{Z101result2} with \eqref{TFAFv0} and \eqref{TAFFv0} gives the relations
\begin{align}
\Braket{T_{\Box}} \ast \Braket{T_{\overline{\Box}}}\Big|_{Z_{\text{mono}}({\bm v} = {\bm 0})}  =& Z({\bm B}=-{\bm e}_1 + {\bm e}_N, {\bm v}={\bm 0}; \zeta > 0),\\
\Braket{T_{\overline{\Box}}}\ast \Braket{T_{\Box}}\Big|_{Z_{\text{mono}}({\bm v} = {\bm 0})} = &Z({\bm B}=-{\bm e}_1 + {\bm e}_N, {\bm v}={\bm 0}; \zeta < 0),
\end{align}
which shows that the difference of the orderings is precisely related to the sign of the FI parameter which is related to the wall-crossing. Namely $\zeta > 0$ implies $s_1 < s_2$ and the 't~Hooft operator $T_{\Box}$ is to the left of the operator $T_{\overline{\Box}}$ in the Moyal product. On the other hand, $\zeta < 0$ means $s_1 > s_2$ and $T_{\Box}$ is to the right of $T_{\overline{\Box}}$ in the product. A schematic picture of the relation between the FI parameter space and the ordering is given in Figure \ref{fig:phase1}. 

The difference between \eqref{Z101result1} and \eqref{Z101result2}, which is also the difference between $\Braket{T_{\Box}}\ast\Braket{T_{\overline{\Box}}}$ and $\Braket{T_{\overline{\Box}}}\ast\Braket{T_{\Box}}$,  can be written as a contour integral with an integration variable $u=e^\phi$
\begin{equation} \label{Z101-difference}
\oint_\gamma \frac{du}{2\pi i u}
 \frac{
\prod_{f=1}^{2N}(u-e^{m_f})
}{
\prod_{i=1}^N (u-e^{a_i-\epsilon_+}) (u-e^{a_i+\epsilon_+})
} 
\times (-1)^{N-1} e^{\sum_i a_i - \frac{1}{2} \sum_f m_f} 2\sinh \epsilon_+
,
\end{equation}
where the contour $\gamma$ encloses $u=0$ and $u=\infty$ but no poles at finite non-zero $u$.
Evaluating the integral by residues we obtain
\begin{equation} \label{TFTAFDiff}
\Braket{T_{\Box}}\ast\Braket{T_{\overline{\Box}}} -  \Braket{T_{\overline{\Box}}}\ast\Braket{T_{\Box}}
=
4 (-1)^{N} \sinh\left(\sum_{i=1}^N a_i - \frac{1}{2} \sum_{f=1}^{2N} m_f \right) \sinh \epsilon_+ .
\end{equation}

\section{$U(N)$ SQCD with $2N$ flavors: three minimal 't~Hooft operators}
\label{sec:threeops}

In Section~\ref{sec:twoops}, we considered the product of two minimal 't~Hooft operators. In this section we expand the analysis by introducing one more minimal 't~Hooft operator. Namely we consider the product of three minimal 't~Hooft operators in the 4d $U(N)$ gauge theory with $2N$ flavors. 
Although the chamber structure becomes more involved, we will see that the computations of the Moyal products and the Witten indices give identical results, and also that wall-crossing occurs only across those FI-walls where the ordering of distinct operators changes.

\subsection{Moyal product}

We consider the Moyal product of the expectation values of three minimal 't~Hooft operators. There are four possible combinations when we ignore the ordering. The four combinations are $\Braket{T_{\Box}}\ast\Braket{T_{\Box}}\ast\Braket{T_{\Box}}$,  $\Braket{T_{\Box}}\ast\Braket{T_{\Box}}\ast\Braket{T_{\overline{\Box}}}$,  $\Braket{T_{\Box}}\ast\Braket{T_{\overline{\Box}}}\ast\Braket{T_{\overline{\Box}}}$ and $\Braket{T_{\overline{\Box}}}\ast\Braket{T_{\overline{\Box}}}\ast\Braket{T_{\overline{\Box}}}$. Since  $\Braket{T_{\Box}}\ast\Braket{T_{\overline{\Box}}}\ast\Braket{T_{\overline{\Box}}}$ is related to $\Braket{T_{\Box}}\ast\Braket{T_{\Box}}\ast\Braket{T_{\overline{\Box}}}$ and also  $\Braket{T_{\overline{\Box}}}\ast\Braket{T_{\overline{\Box}}}\ast\Braket{T_{\overline{\Box}}}$ is related to $\Braket{T_{\Box}}\ast\Braket{T_{\Box}}\ast\Braket{T_{\Box}}$ by the replacement $b_i \to -b_i$ for $i=1, \cdots, N$ up to ordering, it is enough to focus on the former two cases, $\Braket{T_{\Box}}\ast\Braket{T_{\Box}}\ast\Braket{T_{\Box}}$ and $\Braket{T_{\Box}}\ast\Braket{T_{\Box}}\ast\Braket{T_{\overline{\Box}}}$. 

For the product of three $\braket{T_\Box}$'s we have the decomposition
\begin{align}
& \Braket{T_{\Box}}\ast\Braket{T_{\Box}}\ast\Braket{T_{\Box}} 
\nn\\
=& \sum_{i=1}^Ne^{3b_i}Z_{\text{1-loop}}({\bm v} = 3{\bm e}_i) +
 \sum_{1 \leq i < j \leq N}e^{2b_i + b_j}Z_{\text{1-loop}}({\bm v} = 2{\bm e}_i + {\bm e}_j)Z_{\text{mono}}({\bm v} = 2{\bm e}_i + {\bm e}_j)\nn\\
&+  \sum_{1 \leq i < j < k\leq N}e^{b_i + b_j + b_k}Z_{\text{1-loop}}({\bm v} = {\bm e}_i + {\bm e}_j + {\bm e}_k)Z_{\text{mono}}({\bm v} = {\bm e}_i + {\bm e}_j + {\bm e}_k). \label{TFFFmono}
\end{align}
Since the ordering is unique $Z_{\text{mono}}({\bm v} = 2{\bm e}_i + {\bm e}_j)$ and $Z_{\text{mono}}({\bm v} = {\bm e}_i + {\bm e}_j + {\bm e}_k)$ should not exhibit wall-crossing.

For the product of two $\braket{T_\Box}$'s and one $\braket{T_{\overline\Box}}$, there are three orderings: $\Braket{T_{\Box}}\ast\Braket{T_{\Box}}\ast\Braket{T_{\overline{\Box}}}$, $\Braket{T_{\Box}}\ast\Braket{T_{\overline{\Box}}}\ast\Braket{T_{\Box}}$ and $\Braket{T_{\overline{\Box}}}\ast\Braket{T_{\Box}}\ast\Braket{T_{\Box}}$. In each case the product can be decomposed as
\begin{align}
&\Braket{T_{\Box}}\ast\Braket{T_{\Box}}\ast\Braket{T_{\overline{\Box}}} \text{ or }\Braket{T_{\Box}}\ast\Braket{T_{\overline{\Box}}}\ast\Braket{T_{\Box}} \text{ or }\Braket{T_{\overline{\Box}}}\ast\Braket{T_{\Box}}\ast\Braket{T_{\Box}}\nn\\
=&\sum_{1 \leq i \neq j \leq N}e^{2b_i - b_j}Z_{\text{1-loop}}({\bm v} = 2{\bm e}_i - {\bm e}_j)\nn\\
&+\sum_{1 \leq i < j \leq N, 1 \leq k (\neq i, j) \leq N}e^{b_i + b_j - b_k}Z_{\text{1-loop}}({\bm v} ={\bm e}_i + {\bm e}_j - {\bm e}_k)Z_{\text{mono}}({\bm v} ={\bm e}_i + {\bm e}_j - {\bm e}_k)\nn\\
&+\sum_{i=1}^{N}e^{b_i}Z_{\text{1-loop}}({\bm v} =  {\bm e}_i)Z_{\text{mono}}({\bm v} =  {\bm e}_i). \label{TFFAFmono}
\end{align}
The contributions $Z_{\text{mono}}({\bm v} ={\bm e}_i + {\bm e}_j - {\bm e}_k)$ and $Z_{\text{mono}}({\bm v} =  {\bm e}_i)$ computed for different orderings may have different values.

\subsubsection{Three $\Braket{T_{\Box}}$'s}

The definition~\eqref{moyal} of the Moyal product applied to \eqref{TF} gives
\begin{align}
&\quad
\Braket{T_{\Box}}\ast\Braket{T_{\Box}}\ast\Braket{T_{\Box}}
\nn\\
 = &\sum_{1 \leq i, j, k \leq N}e^{b_i + b_j + b_k}Z_i({\bm a} + \epsilon_+{\bm e}_j + \epsilon_+{\bm e}_k)Z_j({\bm a} - \epsilon_+{\bm e}_i + \epsilon_+{\bm e}_k)Z_k({\bm a} - \epsilon_+{\bm e}_i - \epsilon_+{\bm e}_j) .
 \label{TFFFresult0}
\end{align}
We identify $e^{b_i + b_j + b_k}$ with $e^{\bm{v}\cdot\bm{b}}$ in~(\ref{eq:TB-vev}).
We restrict to terms with ${\bm v} = {\bm e}_{N-1} + 2{\bm e}_N$ and ${\bm v} = {\bm e}_{N-2} + {\bm e}_{N-1} + {\bm e}_N$;
the others can be obtained by the Weyl group action.

The sum of the coefficients of $e^{b_{N-2} + 2b_N}$ in \eqref{TFFFresult0} is 
\begin{align}
&
Z_{N-1}({\bm a} + 2\epsilon_+{\bm e}_N)Z_N({\bm a}-\epsilon_+{\bm e}_{N-1} + \epsilon_+{\bm e}_N)Z_N({\bm a}-\epsilon_+{\bm e}_{N-1} - \epsilon_+{\bm e}_N)\nn\\
&
+Z_{N-1}({\bm a})Z_N({\bm a} + \epsilon_+{\bm e}_{N-1} + \epsilon_+{\bm e}_N)Z_N({\bm a} - \epsilon_+{\bm e}_{N-1} - \epsilon_+{\bm e}_N)\nn\\
&
\quad
+ Z_{N-1}({\bm a}-2\epsilon_+{\bm e}_N)Z_N({\bm a} + \epsilon_+{\bm e}_{N-1} + \epsilon_+{\bm e}_N)Z_N({\bm a} + \epsilon_+{\bm e}_{N-1} + \epsilon_+{\bm e}_N).
\label{TFFFresult01}
\end{align} 
Substituting \eqref{Zi} to \eqref{TFFFresult01} we obtain
\begin{align}
\Braket{T_{\Box}}\ast\Braket{T_{\Box}}\ast\Braket{T_{\Box}}\Big|_{Z_{\text{mono}}({\bm v} = {\bm e}_{N-1} + 2{\bm e}_N)}
=&
\sum_{k=0}^1\frac{1}{2\sinh\frac{a_{N-1}-a_N-(-1)^k\epsilon_+}{2}2\sinh\frac{-a_{N-1}+a_N+(-1)^k3\epsilon_+}{2}} 
\nn\\
&
+ \frac{1}{2\sinh\frac{a_{N}-a_{N-1}+\epsilon_+}{2}2\sinh\frac{-a_{N}+a_{N-1}+\epsilon_+}{2}}
.\label{TFFFv12}
\end{align}

Similarly the sum of the coefficients of $e^{b_{N-2} + b_{N-1} + b_N}$ in \eqref{TFFFresult0} is
\begin{align}
&
\quad\
Z_{N-2}({\bm a}+\epsilon_+{\bm e}_{N-1} + \epsilon_+{\bm e}_N)Z_{N-1}({\bm a} - \epsilon_+{\bm e}_{N-2} + \epsilon_+{\bm e}_N)Z_N({\bm a} - \epsilon_+{\bm e}_{N-2} - \epsilon_+{\bm e}_{N-1})\nn\\
&
+ Z_{N-2}({\bm a} + \epsilon_+{\bm e}_{N-1} + \epsilon_+{\bm e}_N)Z_{N-1}({\bm a} - \epsilon_+{\bm e}_{N-2} - \epsilon_+{\bm e}_N)Z_N({\bm a} - \epsilon_+{\bm e}_{N-2} + \epsilon_+{\bm e}_{N-1})\nn\\
&
+ Z_{N-2}({\bm a}-\epsilon_+{\bm e}_{N-1}+\epsilon_+{\bm e}_N)Z_{N-1}({\bm a}+\epsilon_+{\bm e}_{N-2}+\epsilon_+{\bm e}_N)Z_N({\bm a}-\epsilon_+{\bm e}_{N-2}-\epsilon_+{\bm e}_{N-1})\nn\\
&
+ Z_{N-2}({\bm a}+\epsilon_+{\bm e}_{N-1}-\epsilon_+{\bm e}_N)Z_{N-1}({\bm a}-\epsilon_+{\bm e}_{N-2} - \epsilon_+{\bm e}_N)Z_N({\bm a}+\epsilon_+{\bm e}_{N-2}+ \epsilon_+{\bm e}_{N-1})\nn\\
&
+ Z_{N-2}({\bm a}-\epsilon_+{\bm e}_{N-1}-\epsilon_+{\bm e}_N)Z_{N-1}({\bm a}+\epsilon_+{\bm e}_{N-2}+\epsilon_+{\bm e}_N)Z_{N}({\bm a}+\epsilon_+{\bm e}_{N-2} - \epsilon_+{\bm e}_{N-1})\nn\\
&
+ Z_{N-2}({\bm a}-\epsilon_+{\bm e}_{N-1}-\epsilon_+{\bm e}_N)Z_{N-1}({\bm a}+\epsilon_+{\bm e}_{N-1} - \epsilon_+{\bm e}_N)Z_{N}({\bm a} + \epsilon_+{\bm e}_{N-2} + \epsilon_+{\bm e}_{N-1})
.
\label{TFFFv111result0}
\end{align}
This leads to
\begin{align}
&\Braket{T_{\Box}}\ast\Braket{T_{\Box}}\ast\Braket{T_{\Box}}\Big|_{Z_{\text{mono}}({\bm v} = {\bm e}_{N-2} + {\bm e}_{N-1} + {\bm e}_N)} =
\nn\\
&\sum_{k=0}^1\left(\frac{1}{{\displaystyle \prod_{N-2 \leq i < j \leq N}}2\sinh\frac{a_{i} - a_j}{2}2\sinh\frac{-a_i + a_j +(-1)^k 2\epsilon_+}{2}}\right.\nn\\
&+\frac{1}{{\displaystyle\prod_{N-1 \leq  j \leq N}}2\sinh\frac{a_{N-2} - a_j}{2}2\sinh\frac{-a_{N-2} + a_j + (-1)^k2\epsilon_+}{2}2\sinh\frac{-a_{N-1} + a_N}{2}2\sinh\frac{a_{N-1} - a_N + (-1)^k2\epsilon_+}{2}}\nn\\
&+
\left.\frac{1}{2\sinh\frac{-a_{N-2} + a_{N-1}}{2}2\sinh\frac{a_{N-2} - a_{N-1} + (-1)^k2\epsilon_+}{2}
{\displaystyle \prod_{N-2 \leq  i \leq N-1}}
2\sinh\frac{a_i - a_N}{2}2\sinh\frac{-a_i + a_N + (-1)^k2\epsilon_+}{2}}\right)
. \label{TFFFv111}
\end{align}

\subsubsection{Two $\Braket{T_{\Box}}$'s and one $\Braket{T_{\overline{\Box}}}$}

We can also apply the Moyal product \eqref{moyal} to  \eqref{TF} and \eqref{TAF} 
and get
\begin{align}
&\Braket{T_{\Box}} \ast \Braket{T_{\Box}} \ast \Braket{T_{\overline{\Box}}} 
\nn\\
&= \sum_{1 \leq i, j, k \leq N}e^{b_i + b_j - b_k}Z_i({\bm a} + \epsilon_+{\bm e}_j - \epsilon_+{\bm e}_k)Z_j({\bm a}-\epsilon_+{\bm e}_i -\epsilon_+{\bm e}_k)Z_k({\bm a} - \epsilon_+{\bm e}_i - \epsilon_+{\bm e}_j), \label{TFTFTAF}\\
&\Braket{T_{\Box}} \ast \Braket{T_{\overline{\Box}}} \ast \Braket{T_{\Box}} 
\nn\\
&= \sum_{1 \leq i, j, k \leq N}e^{b_i - b_j + b_k}Z_i({\bm a}- \epsilon_+{\bm e}_j+ \epsilon_+{\bm e}_k)Z_j({\bm a}- \epsilon_+{\bm e}_i+ \epsilon_+{\bm e}_k)Z_k({\bm a}- \epsilon_+{\bm e}_i+ \epsilon_+{\bm e}_j), \label{TFTAFTF}\\
&\Braket{T_{\overline{\Box}}} \ast \Braket{T_{\Box}} \ast \Braket{T_{\Box}}
\nn\\
 &= \sum_{1 \leq i, j, k \leq N}e^{-b_i + b_j + b_k}Z_i({\bm a}+ \epsilon_+{\bm e}_j+ \epsilon_+{\bm e}_k)Z_j({\bm a}+ \epsilon_+{\bm e}_i+ \epsilon_+{\bm e}_k)Z_k({\bm a}+ \epsilon_+{\bm e}_i -  \epsilon_+{\bm e}_j). \label{TAFTFTF}
\end{align}
 By the Weyl group action it is enough to consider ${\bm v} = -{\bm e}_1 + {\bm e}_{N-1} + {\bm e}_N$ and ${\bm v} = {\bm e}_{N}$. 
The sum of the coefficients
of $e^{-b_1 + b_{N-1} + b_N}$ in each of \eqref{TFTFTAF}, \eqref{TFTAFTF} and \eqref{TAFTFTF} is 
\begin{align}
&
Z_1({\bm a}- \epsilon_+{\bm e}_{N-1} -  \epsilon_+{\bm e}_N)Z_{N-1}({\bm a}- \epsilon_+{\bm e}_1+ \epsilon_+{\bm e}_N)Z_N({\bm a}- \epsilon_+{\bm e}_1 -  \epsilon_+{\bm e}_{N-1})\nn\\
&\hspace{1cm} + Z_1({\bm a}- \epsilon_+{\bm e}_{N-1} -  \epsilon_+{\bm e}_N)Z_{N-1}({\bm a}- \epsilon_+{\bm e}_1- \epsilon_+{\bm e}_N)Z_N({\bm a}- \epsilon_+{\bm e}_1+ \epsilon_+{\bm e}_{N-1})
\label{TFTFTAFresult0}
\end{align}
for \eqref{TFTFTAF} $=\Braket{T_{\Box}} \ast \Braket{T_{\Box}} \ast \Braket{T_{\overline{\Box}}}$,
\begin{align}
&
Z_1({\bm a}- \epsilon_+{\bm e}_{N-1}+ \epsilon_+{\bm e}_N)Z_{N-1}({\bm a}- \epsilon_+{\bm e}_1 +  \epsilon_+{\bm e}_N)Z_N({\bm a}+ \epsilon_+{\bm e}_1- \epsilon_+{\bm e}_{N-1})
\nn\\
&\hspace{1cm} +
Z_1({\bm a}+ \epsilon_+{\bm e}_{N-1}- \epsilon_+{\bm e}_N)Z_{N-1}({\bm a}+ \epsilon_+{\bm e}_1 -  \epsilon_+{\bm e}_N)Z_N({\bm a}- \epsilon_+{\bm e}_1+ \epsilon_+{\bm e}_{N-1})
 \label{TFTAFTFresult0}
\end{align}
for \eqref{TFTAFTF} $=\Braket{T_{\Box}} \ast \Braket{T_{\overline{\Box}}} \ast \Braket{T_{\Box}}$, and
\begin{align}
&
Z_1({\bm a}+ \epsilon_+{\bm e}_{N-1}+ \epsilon_+{\bm e}_N)Z_{N-1}({\bm a} +  \epsilon_+{\bm e}_1 +  \epsilon_+{\bm e}_N)Z_N({\bm a}+ \epsilon_+{\bm e}_1- \epsilon_+{\bm e}_{N-1})
\nn\\
&\hspace{1cm} + 
Z_1({\bm a}+ \epsilon_+{\bm e}_{N-1}+ \epsilon_+{\bm e}_N)Z_{N-1}({\bm a}+ \epsilon_+{\bm e}_1 -  \epsilon_+{\bm e}_N)Z_N({\bm a}+ \epsilon_+{\bm e}_1+ \epsilon_+{\bm e}_{N-1})
\label{TAFTFTFresult0}
\end{align}
for \eqref{TAFTFTF} $=\Braket{T_{\overline{\Box}}} \ast \Braket{T_{\Box}} \ast \Braket{T_{\Box}}$, respectively.
Substituting  \eqref{Zi} into \eqref{TFTFTAFresult0}-\eqref{TAFTFTFresult0}
we find that the three orderings give the same result for the monopole screening contribution
\begin{align}
&\Braket{T_{\Box}} \ast \Braket{T_{\Box}} \ast \Braket{T_{\overline{\Box}}}\Big|_{Z_{\text{mono}}({\bm v} =  -{\bm e}_1 + {\bm e}_{N-1} + {\bm e}_N)} = \Braket{T_{\Box}} \ast \Braket{T_{\overline{\Box}}} \ast \Braket{T_{\Box}}\Big|_{Z_{\text{mono}}({\bm v} =  -{\bm e}_1 + {\bm e}_{N-1} + {\bm e}_N)}\nn\\
=& \Braket{T_{\overline{\Box}}} \ast \Braket{T_{\Box}} \ast \Braket{T_{\Box}}\Big|_{Z_{\text{mono}}({\bm v} = -{\bm e}_1 + {\bm e}_{N-1} + {\bm e}_N)} \nn\\
=&\sum_{k=0}^1\frac{1}{2\sinh\frac{a_{N-1} - a_N}{2}2\sinh\frac{-a_{N-1} + a_N + (-1)^k2\epsilon_+}{2}}
.
\label{TFFAFvm111}
\end{align}

On the other hand, the sum of the coefficients of $e^{b_N}$ in each of \eqref{TFTFTAF}, \eqref{TFTAFTF} and \eqref{TAFTFTF} is
\begin{align}
&
Z_N({\bm a})Z_N({\bm a}-2 \epsilon_+{\bm e}_N)^2
+\sum_{i=1}^{N-1}\Big(Z_N({\bm a})Z_i({\bm a}- \epsilon_+{\bm e}_i- \epsilon_+{\bm e}_N)^2
\nn\\
&\hspace{1cm} 
+ Z_i({\bm a}- \epsilon_+{\bm e}_i +  \epsilon_+{\bm e}_N)Z_N({\bm a}-2 \epsilon_+{\bm e}_i)Z_i({\bm a}- \epsilon_+{\bm e}_i- \epsilon_+{\bm e}_N)\Big)
\label{TFFAFv01result0}
\end{align}
for  \eqref{TFTFTAF} $= \Braket{T_{\Box}}\ast \Braket{T_{\Box}} \ast \Braket{T_{\overline{\Box}}}$, 
\begin{align}
&
\sum_{i=1}^{N-1}\left(Z_N({\bm a})Z_i({\bm a}+ \epsilon_+{\bm e}_i- \epsilon_+{\bm e}_N)^2  +Z_i({\bm a}- \epsilon_+{\bm e}_i+ \epsilon_+{\bm e}_N)^2Z_N({\bm a})\right)
+ Z_N({\bm a})^3 
\label{TFAFFv01result0}
\end{align}
for \eqref{TFTAFTF} $=\Braket{T_{\Box}}\ast \Braket{T_{\overline{\Box}}} \ast \Braket{T_{\Box}}$, and 
\begin{align}
& 
Z_N({\bm a})Z_N({\bm a} + 2 \epsilon_+{\bm e}_N)^2
+\sum_{i=1}^{N-1}\Big(Z_i({\bm a}+ \epsilon_+{\bm e}_i+ \epsilon_+{\bm e}_N)Z_N({\bm a}+ 2\epsilon_+{\bm e}_i)Z_i({\bm a}+ \epsilon_+{\bm e}_i- \epsilon_+{\bm e}_N)
\nn\\
&\hspace{2cm}  + Z_i({\bm a}+ \epsilon_+{\bm e}_i+ \epsilon_+{\bm e}_N)^2Z_N({\bm a})
\Big)
.\label{TAFFFv01result0}
\end{align}
for \eqref{TAFTFTF} $=\Braket{T_{\overline{\Box}}}\ast \Braket{T_{\Box}} \ast \Braket{T_{\Box}}$. 
Rewriting \eqref{TFFAFv01result0}-\eqref{TAFFFv01result0} using \eqref{Zi} yields
\begin{align}
&\quad
\Braket{T_{\Box}}\ast \Braket{T_{\Box}} \ast \Braket{T_{\overline{\Box}}}\Big|_{Z_{\text{mono}}({\bm v} =
 {\bm e}_N)}  \nn
 \\
= &
\sum_{i=1}^{N-1}\left(\frac{\prod_{f=1}^{2N}2\sinh\frac{a_i - m_f - \epsilon_+}{2}}{\prod_{1\leq j\neq i \leq N-1}2\sinh\frac{a_i - a_j}{2}2\sinh\frac{-a_i + a_j + 2\epsilon_+}{2}}\frac{1}{2\sinh\frac{a_i - a_N - \epsilon_+}{2}2\sinh\frac{-a_i + a_N + 3\epsilon_+}{2}}\right.\nn\\
&+\left.\frac{1}{2\sinh\frac{a_i - a_N + \epsilon_+}{2}2\sinh\frac{-a_i + a_N + \epsilon_+}{2}}\frac{\prod_{f=1}^{2N}2\sinh\frac{a_i - m_f - \epsilon_+}{2}}{\prod_{1 \leq j \neq i \leq N-1}2\sinh\frac{a_i - a_j}{2}2\sinh\frac{-a_i + a_j + 2\epsilon_+}{2}}\right)\nn\\
&+\frac{\prod_{f=1}^{2N}2\sinh\frac{a_N - m_f - 2\epsilon_+}{2}}{\prod_{i=1}^{N-1}2\sinh\frac{a_N - a_i - \epsilon_+}{2}2\sinh\frac{-a_N + a_i + 3\epsilon_+}{2}},
\label{TFFAFv01result1}
\end{align}
\begin{align}
&\quad
\Braket{T_{\Box}}\ast \Braket{T_{\overline{\Box}}} \ast \Braket{T_{\Box}}\Big|_{Z_{\text{mono}}({\bm v} 
={\bm e}_N)}
 \nn\\
=&
\sum_{i=1}^{N-1}\left(\frac{1}{2\sinh\frac{a_i - a_N - \epsilon_+}{2}2\sinh\frac{-a_i + a_N + 3\epsilon_+}{2}}\frac{\prod_{f=1}^{2N}2\sinh\frac{a_i - m_f - \epsilon_+}{2}}{\prod_{1 \leq j \neq i \leq N-1}2\sinh\frac{a_i - a_j}{2}2\sinh\frac{-a_i + a_j + 2\epsilon_+}{2}}\right.\nn\\
&+\left.\frac{\prod_{f=1}^{2N}2\sinh\frac{a_i - m_f + \epsilon_+}{2}}{\prod_{1\leq j\neq i \leq N-1}2\sinh\frac{a_i - a_j + 2\epsilon_+}{2}2\sinh\frac{-a_i + a_j}{2}}\frac{1}{2\sinh\frac{a_i - a_N + 3\epsilon_+}{2}2\sinh\frac{-a_i + a_N - \epsilon_+}{2}}\right)\nn\\
&+ \frac{\prod_{f=1}^{2N}2\sinh\frac{a_N - m_f}{2}}{\prod_{i=1}^{N-1}2\sinh\frac{a_N - a_i + \epsilon_+}{2}2\sinh\frac{-a_N + a_i + \epsilon_+}{2}},\label{TFAFFv01result1}
\end{align}
\begin{align}
 &\quad\Braket{T_{\overline{\Box}}}\ast \Braket{T_{\Box}} \ast \Braket{T_{\Box}}\Big|_{Z_{\text{mono}}({\bm v} = {\bm e}_N)}
 \nn\\
=&
\sum_{i=1}^{N-1}\left(\frac{\prod_{f=1}^{2N}2\sinh\frac{a_i - m_f + \epsilon_+}{2}}{\prod_{1\leq j\neq i \leq N-1}2\sinh\frac{a_i - a_j + 2\epsilon_+}{2}2\sinh\frac{-a_i + a_j}{2}}\frac{1}{2\sinh\frac{a_i - a_N +  3\epsilon_+}{2}2\sinh\frac{-a_i + a_N - \epsilon_+}{2}}\right.\nn\\
&+\left.\frac{1}{2\sinh\frac{a_i - a_N + \epsilon_+}{2}2\sinh\frac{-a_i + a_N + \epsilon_+}{2}}\frac{\prod_{f=1}^{2N}2\sinh\frac{a_i - m_f + \epsilon_+}{2}}{\prod_{1 \leq j \neq i \leq N-1}2\sinh\frac{a_i - a_j + 2\epsilon_+}{2}2\sinh\frac{-a_i + a_j}{2}}\right) \nn\\
&+\frac{\prod_{f=1}^{2N}2\sinh\frac{a_N - m_f + 2\epsilon_+}{2}}{\prod_{i=1}^{N-1}2\sinh\frac{a_N - a_i + 3\epsilon_+}{2}2\sinh\frac{-a_N + a_i - \epsilon_+}{2}}.\label{TAFFFv01result1}
\end{align}
In this case, the three different orderings give different monopole screening contributions.

\subsection{SQMs}

We now turn to the computations from SQMs for the monopole screening contributions. 

\subsubsection{'t~Hooft operator with ${\bm B} = 3{\bm e}_N$}
\begin{figure}[t]
\centering
\includegraphics[width=10cm]{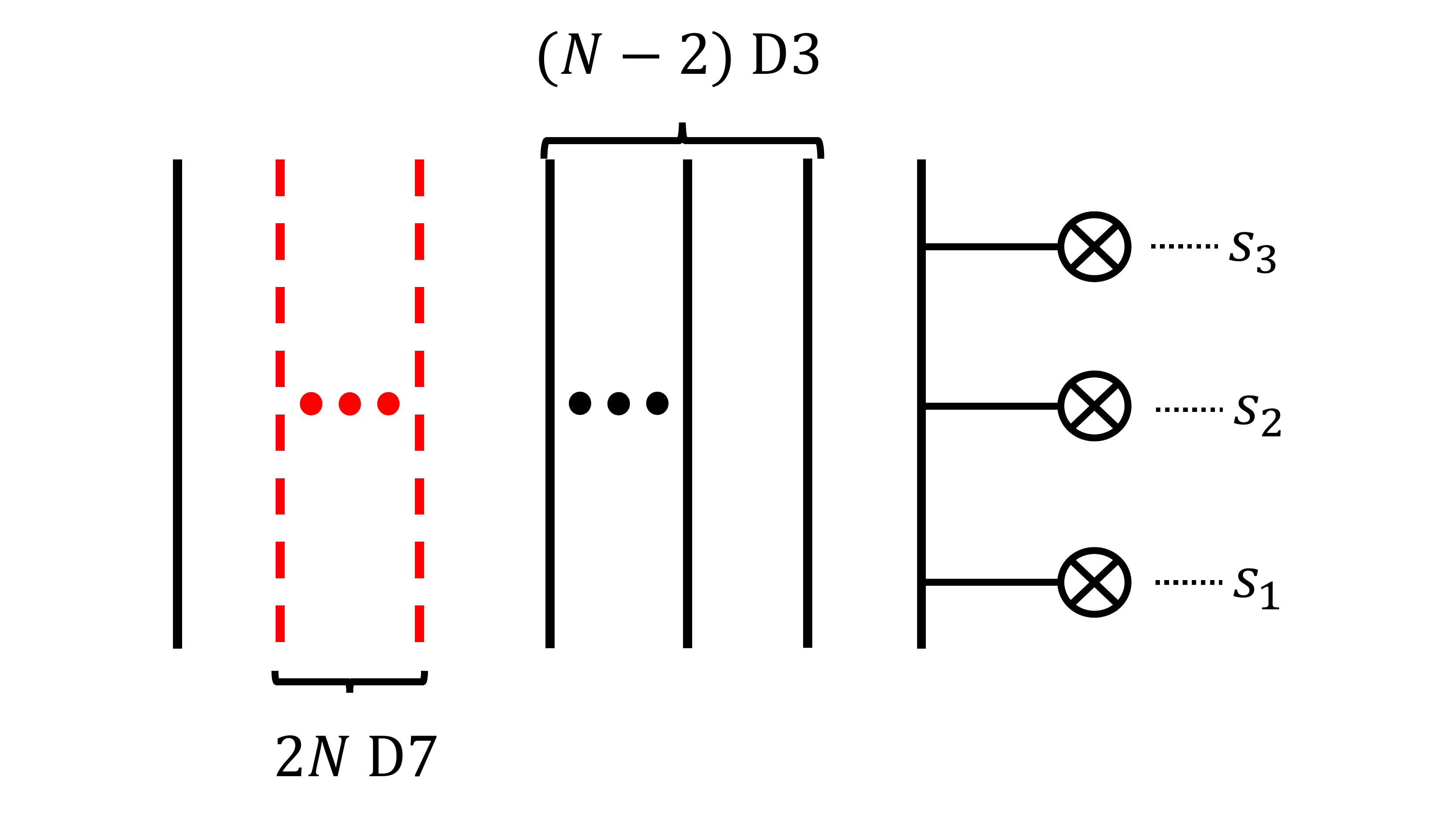}
\caption{A brane configuration for the 't~Hooft operator with the magnetic charge ${\bm B}=3{\bm e}_N$.}
\label{fig:B03no1}
\end{figure}
We begin by analyzing monopole screening in $\Braket{T_{\bm B}}=\Braket{T_{\Box}}\ast\Braket{T_{\Box}}\ast\Braket{T_{\Box}}$ with $B=3{\bm e}_N$.
The operator $T_{\bm B}$ is realized by the  brane diagram in Figure \ref{fig:B03no1}. We denote the values of the coordinate $x^3$ for the three NS5-branes by $s_1, s_2, s_3$ respectively from bottom to top as in Figure \ref{fig:B03no1}.  Up to the Weyl group action there are two sectors with monopole screening as can be seen in~\eqref{TFFFmono}.

\begin{figure}[t]
\centering
\subfigure[]{\label{fig:B03v12no2}
\includegraphics[width=8cm]{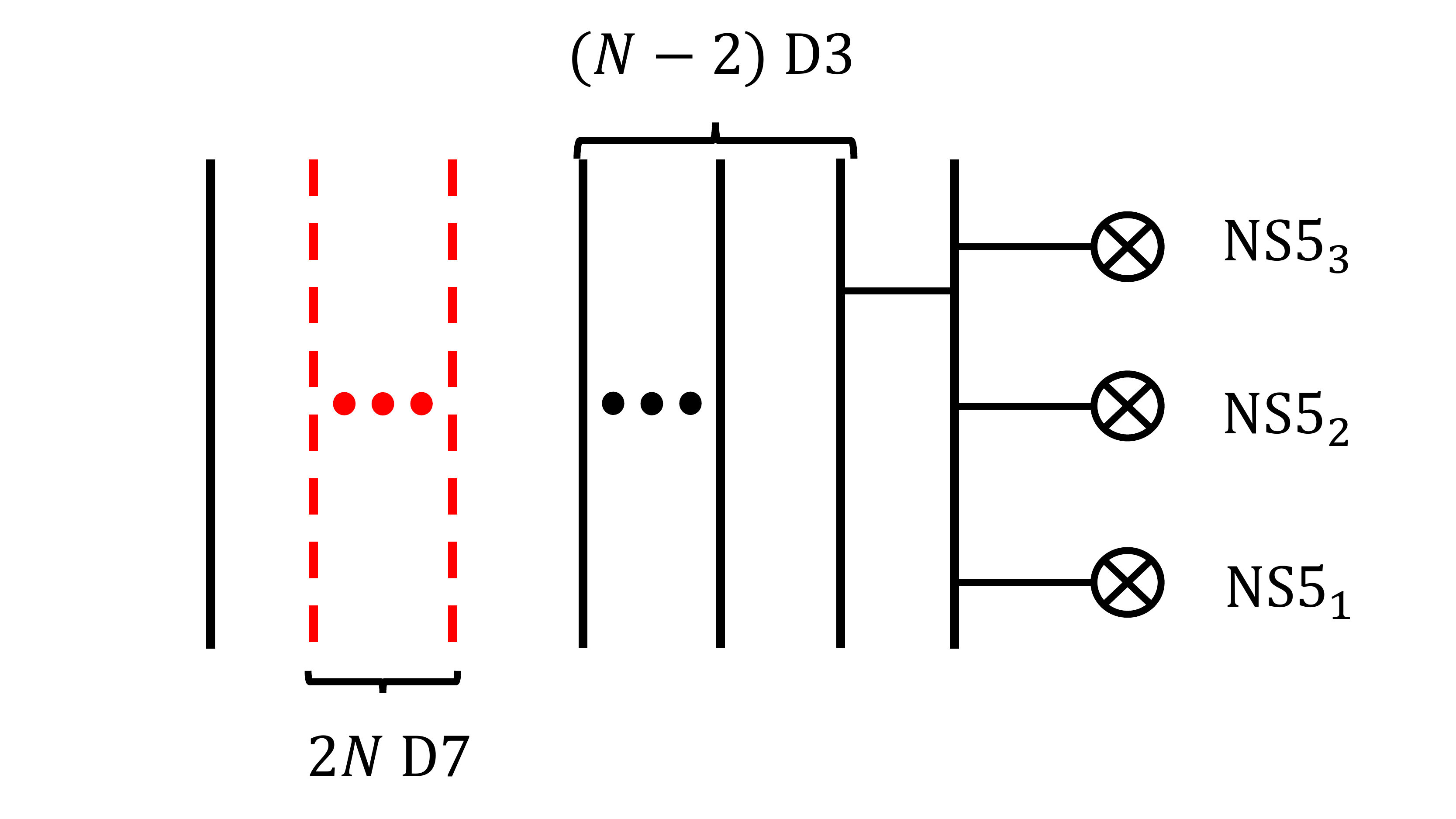}}
\subfigure[]{\label{fig:B03v12no2to3}
\includegraphics[width=8cm]{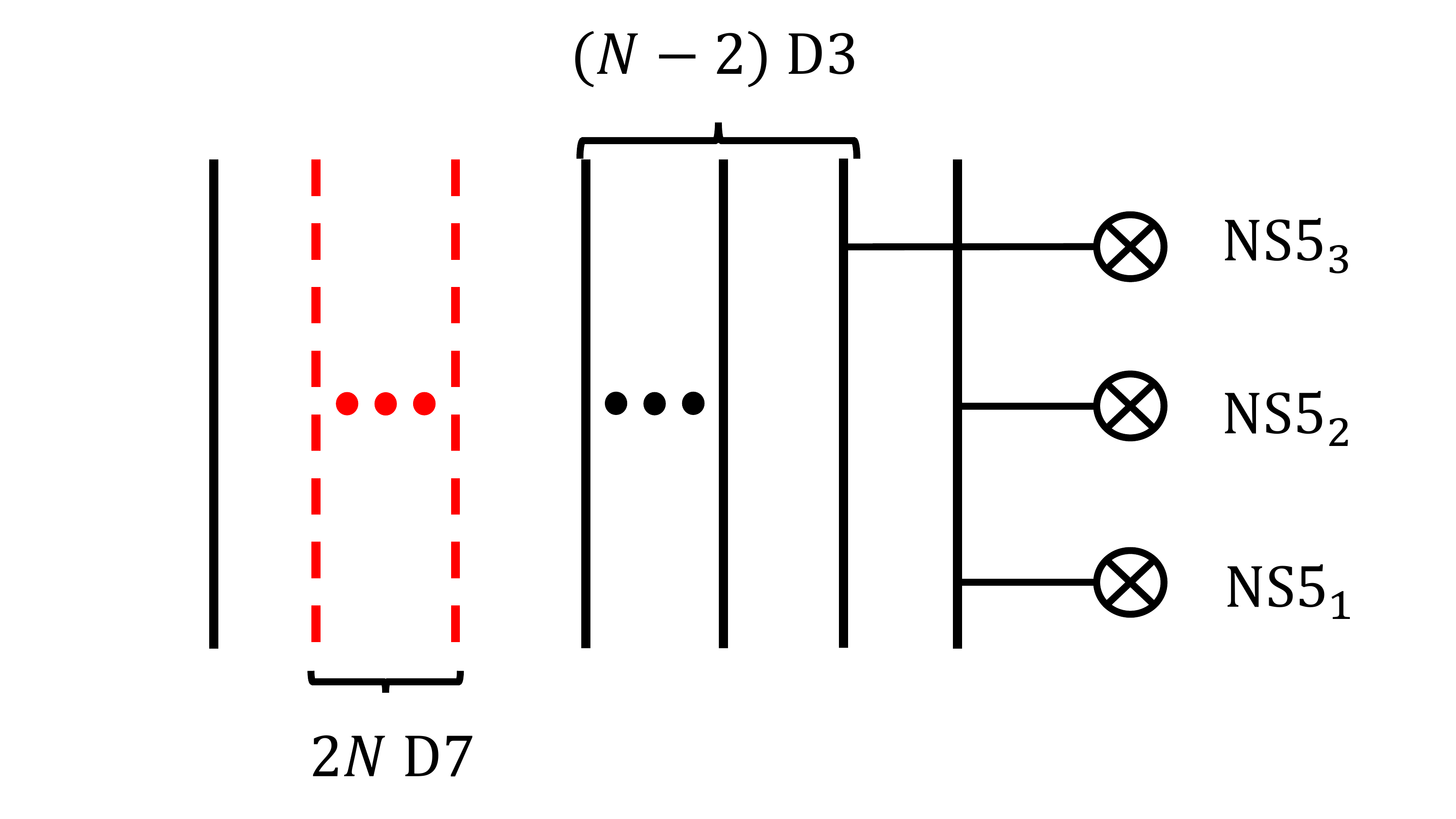}}
\subfigure[]{\label{fig:B03v12no3}
\includegraphics[width=8cm]{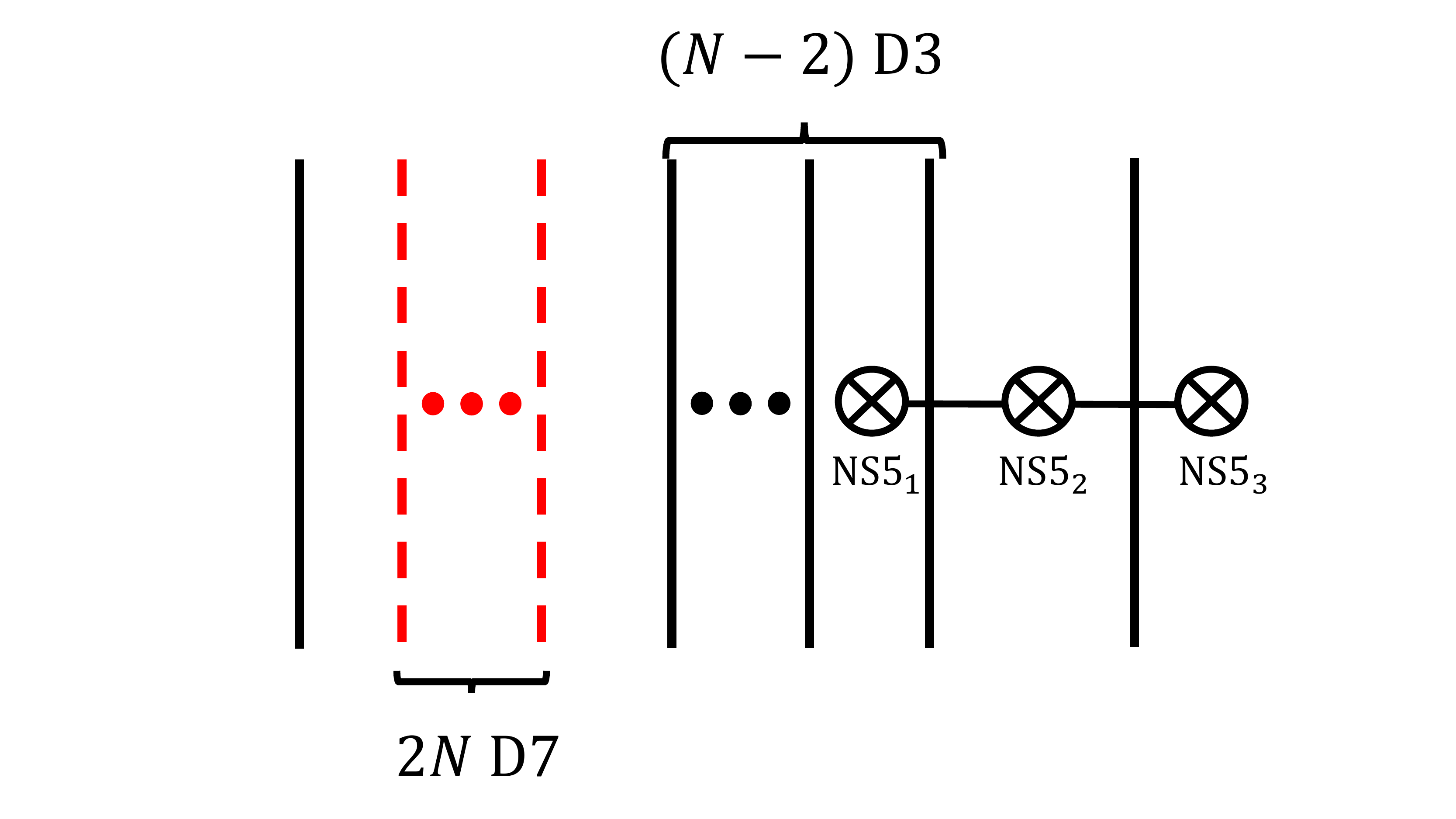}}
\subfigure[]{\label{fig:quiver12}
\includegraphics[width=8cm]{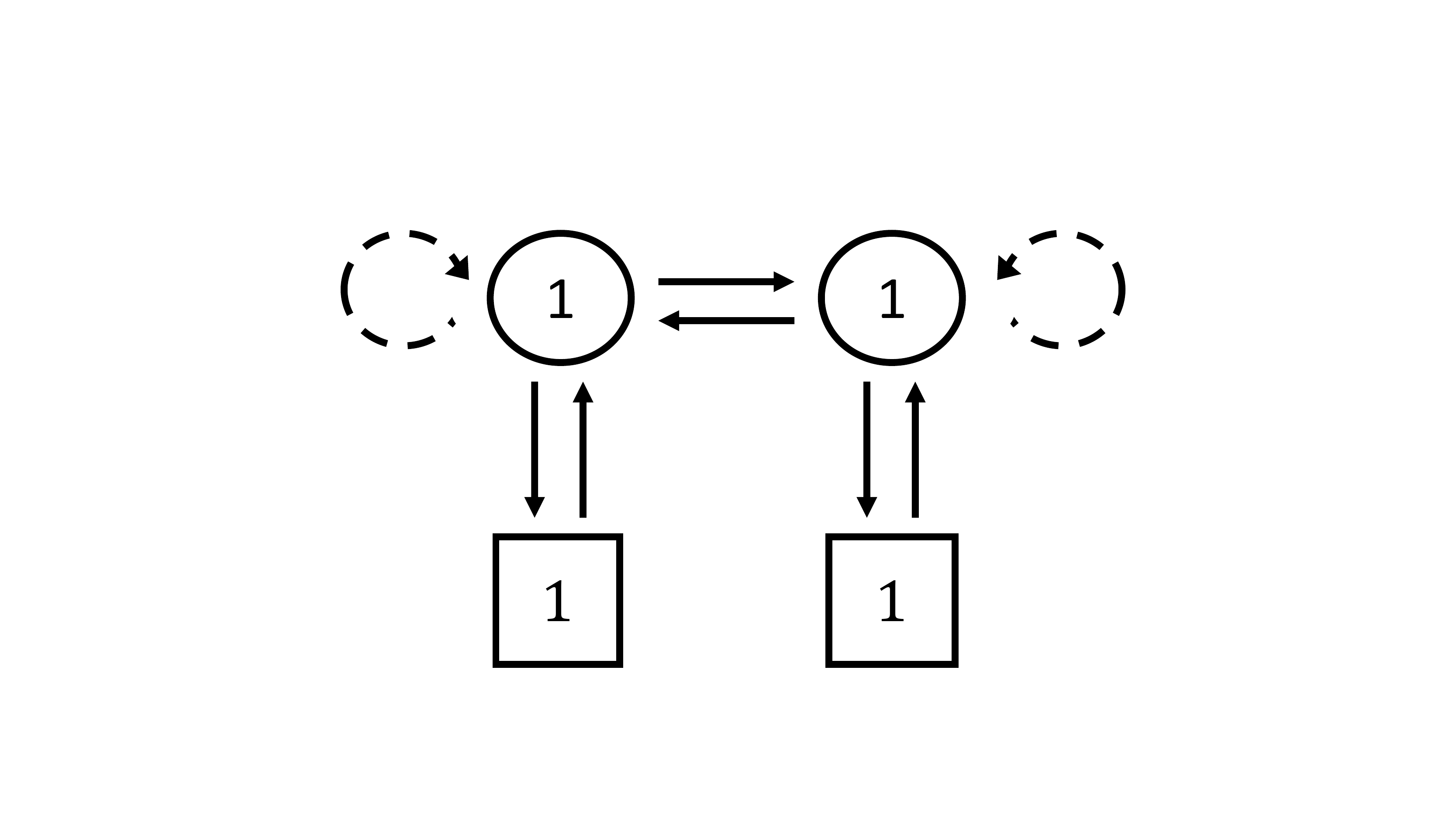}}
\caption{(a):  An introduction of a smooth monopole with magnetic charge $
{\bm e}_{N-1} - {\bm e}_N$ to Figure~\ref{fig:B03no1}. (b):  We adjust the position of the D1-brane  to the position of the top D1-brane between the $N$-th D3-brane and an NS5-brane.  This configuration corresponds to the sector ${\bm v} = {\bm e}_{N-1} + 2{\bm e}_N$. (c): The brane configuration for reading off the SQM for the monopole screening contribution in the sector ${\bm v} = {\bm e}_{N-1} + 2{\bm e}_N$. (d): The quiver diagram for the worldvolume theory on the D1-branes in the configuration in Figure \ref{fig:B03v12no3}. }
\label{fig:B03v12}
\end{figure}
\paragraph{Sector ${\bm v} = {\bm e}_{N-1} + 2{\bm e}_N$.}
For this sector, we introduce a smooth monopole with the magnetic charge ${\bm B} = {\bm e}_{N-1} - {\bm e}_N$ as in Figure \ref{fig:B03v12no2}.  A rearrangement of the positions of the D1- and NS5-branes in Figure~\ref{fig:B03v12no2} through Figure~\ref{fig:B03v12no2to3} leads to the configuration in Figure~\ref{fig:B03v12no3}, from which we can read off the SQM living on the D1-branes. The quiver diagram for the SQM is depicted in Figure \ref{fig:quiver12}. The quiver has two $U(1)$ gauge nodes, which we call $U(1)_1$ and $U(1)_2$. The $U(1)_1$ comes from a D1-brane between the NS5$_1$-brane and the NS5$_2$-brane and the $U(1)_2$ comes from a D1-brane between the NS5$_2$-brane and the NS5$_3$-brane. 
We write $\zeta_1$ and $\zeta_2$ for the FI parameters of the gauge nodes $U(1)_1$ and $U(1)_2$, respectively. Then the FI parameters are related to the positions of the three NS5-branes as $\zeta_1 = s_2 - s_1, \zeta_2 = s_3 - s_2$. 

It is straightforward to compute the Witten index of the quiver theory in Figure \ref{fig:quiver12}. Using the formulae in Appendix~\ref{sec:locforSQM}, the Witten index is given by
\begin{align}
&Z({\bm B}=3{\bm e}_N, {\bm v} = {\bm e}_{N-1} + 2{\bm e}_N; {\bm \zeta})
=\oint_{JK({\bm \zeta})}\frac{d\phi_1}{2\pi i}\frac{d\phi_2 }{2\pi i}\frac{\left(2\sinh\epsilon_+\right)^2}{2\sinh\frac{\phi_1 - \phi_2 + \epsilon_+}{2}2\sinh\frac{\phi_2 - \phi_1 + \epsilon_+}{2}}
\nn\\
&
\hspace{6.5cm}
 \times \frac{1}{\prod_{i=1}^22\sinh\frac{\phi_i - a_{N-2+i} + \epsilon_+}{2}2\sinh\frac{-\phi_i + a_{N-2+i} + \epsilon_+}{2}}.\label{Z03v12}
\end{align}
In this case, we have two FI parameters $\zeta_1, \zeta_2$.  
We set ${\bm \zeta}=\zeta_1{\bm e}_1 + \zeta_2{\bm e}_2 \in \mathfrak{h}^{\ast}_{U(1)_1} \oplus \mathfrak{h}^{\ast}_{U(1)_2}$ and use this vector as the JK parameter ${\bm \eta}$ for the JK residue prescription. 
According to the prescription summarized in Appendix~\ref{sec:JK}, the poles that contribute depend on ${\bm \eta}={\bm \zeta}$.
In the present case the relevant poles turn out to be non-degenerate; for each pole to contribute the charge vectors associated with the pole have to satisfy the condition in~(\ref{JK-def-non-degenerate}).
The condition on the poles then determines six JK-chambers, defined in~(\ref{def-JK-chamber}), for which different sets of poles contribute.
\begin{figure}[t]
\centering
\includegraphics[width=4cm]{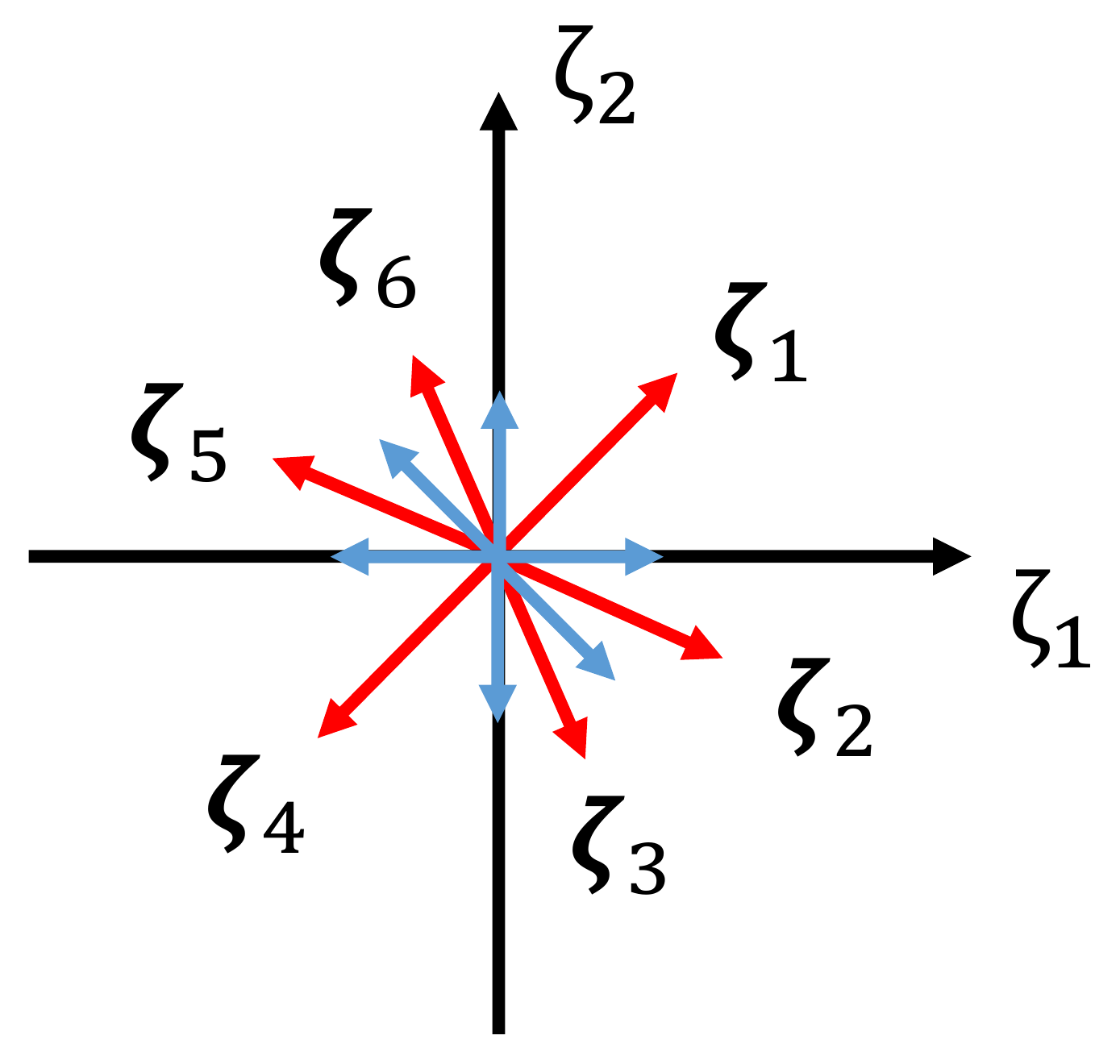}
\caption{
Possible choices of the FI parameter~$\bm{\zeta}$.
Each choice corresponds to a cone generated by two neighboring short (blue) arrows.
Representatives from the cones are the long (red) arrows and are denoted by~${\bm \zeta}_i$ ($i=1, \ldots, 6$).}
\label{fig:JK1}
\end{figure}
In Figure~\ref{fig:JK1} we display the cones and their representatives~${\bm \zeta}_i$ $(i=1, \ldots, 6)$ characterized by
\begin{align}
{\bm \zeta}_1 & \in  \mathbb{R}_{> 0}{\bm e}_1 + \mathbb{R}_{> 0}{\bm e}_2 , \label{B102FI1}\\
{\bm \zeta}_2 & \in  \mathbb{R}_{> 0}{\bm e}_1 + \mathbb{R}_{> 0}({\bm e}_1 - {\bm e}_2) ,\\
{\bm \zeta}_3 & \in  \mathbb{R}_{> 0}({\bm e}_1 - {\bm e}_2) + \mathbb{R}_{> 0}(-{\bm e}_2) ,\\
{\bm \zeta}_4 & \in  \mathbb{R}_{> 0}(-{\bm e}_1) + \mathbb{R}_{> 0}(-{\bm e}_2) ,\\
{\bm \zeta}_5 & \in  \mathbb{R}_{> 0}(-{\bm e}_1) + \mathbb{R}_{> 0}(-{\bm e}_1 + {\bm e}_2) ,\\
{\bm \zeta}_6 & \in  \mathbb{R}_{> 0}(-{\bm e}_1 + {\bm e}_2) + \mathbb{R}_{> 0}{\bm e}_2 . \label{B102FI6}
\end{align}
In this case the JK-chambers turn out to coincide with the FI-chambers defined in~(\ref{def-FI-chamber}).
To compute~(\ref{Z03v12}) for each vector ${\bm \zeta}_i\, (i=1, \cdots, 6)$, we include, as dictated by~(\ref{JK-def-non-degenerate}), those poles whose associated charge vectors form a cone that contains ${\bm \zeta}_i$.

For example, if we choose ${\bm \zeta}={\bm \zeta}_1$ there are three cones spanned by charge vectors%
\footnote{%
For a cone spanned by charge vectors
we write
$\text{Cone}[{\bm Q}_1, \ldots, {\bm Q}_n]$ for $\mathbb{R}_{>0}{\bm Q}_1 + \ldots + \mathbb{R}_{>0}{\bm Q}_n$.
}
\begin{align}
C_{1a} = \text{Cone}\left[(1, 0), (0, 1)\right],
\
C_{1b} = \text{Cone}\left[(-1, 1), (1, 0)\right], 
\
 C_{1c} = \text{Cone}\left[(0, 1), (1, -1)\right] \label{cone1}
\end{align}
in $\mathfrak{h}^{\ast}_{U(1)_1} \oplus \mathfrak{h}^{\ast}_{U(1)_2}$.
 We let $\widetilde Z({\bm B}, {\bm v}; C)$ express the sum of the JK residues of the poles inside the charge cone $C$ for the Witten index $ Z({\bm B}, {\bm v}; {\bm \zeta})$.
Then the contribution from each cone is evaluated as
\begin{align}
\widetilde Z({\bm B}=3{\bm e}_N, {\bm v} = {\bm e}_{N-1} + 2{\bm e}_N; C_{1a})&= \frac{1}{2\sinh\frac{a_{N-1} - a_{N} + \epsilon_+}{2}2\sinh\frac{-a_{N-1} + a_N + \epsilon_+}{2}},\\
\widetilde Z({\bm B}=3{\bm e}_N, {\bm v} = {\bm e}_{N-1} + 2{\bm e}_N; C_{1b}) &= \frac{1}{2\sinh\frac{a_{N-1} - a_N - \epsilon_+}{2}2\sinh\frac{-a_{N-1} + a_N + 3\epsilon_+}{2}},\\
\widetilde Z({\bm B}=3{\bm e}_N, {\bm v} = {\bm e}_{N-1} + 2{\bm e}_N; C_{1c}) &= \frac{1}{2\sinh\frac{a_{N} - a_{N-1} - \epsilon_+}{2}2\sinh\frac{-a_N + a_{N-1} + 3\epsilon_+}{2}}.
\end{align}
Therefore the JK residue prescription with the vector ${\bm \zeta}_1$ in Figure \ref{fig:JK1} for the integral of~\eqref{Z03v12} yields
\begin{align}
&Z({\bm B}=3{\bm e}_N, {\bm v} = {\bm e}_{N-1} + 2{\bm e}_N; {\bm \zeta}_1)=\sum_{\alpha=a,b,c}\widetilde Z({\bm B}=3{\bm e}_N, {\bm v} = {\bm e}_{N-1} + 2{\bm e}_N;C_{1\alpha})\nn\\
=&\sum_{k=0}^1\frac{1}{2\sinh\frac{a_{N-1}-a_N-(-1)^k\epsilon_+}{2}2\sinh\frac{-a_{N-1}+a_N+(-1)^k3\epsilon_+}{2}} + \frac{1}{2\sinh\frac{a_{N}-a_{N-1}+\epsilon_+}{2}2\sinh\frac{-a_{N}+a_{N-1}+\epsilon_+}{2}}, \label{Z03v12result1}
\end{align}
which precisely equals the monopole screening contribution in \eqref{TFFFv12}. 

For the other choices~${\bm \zeta}={\bm \zeta}_i$ $(i=2, \cdots, 6)$ we can compute the integral \eqref{Z03v12} in a similar way.
The results turn out to be exactly the same as~\eqref{Z03v12result1} although different sets of poles contribute; this is as expected because the ordering is unique for the product $\Braket{T_{\Box}}\ast\Braket{T_{\Box}}\ast\Braket{T_{\Box}}$ in which the ${\bm v} = {\bm e}_{N-1} + 2{\bm e}_N$ sector appears.
The SQM exhibits no wall-crossing.

\begin{figure}[t]
\centering
\subfigure[]{\label{fig:B03v111no2}
\includegraphics[width=8cm]{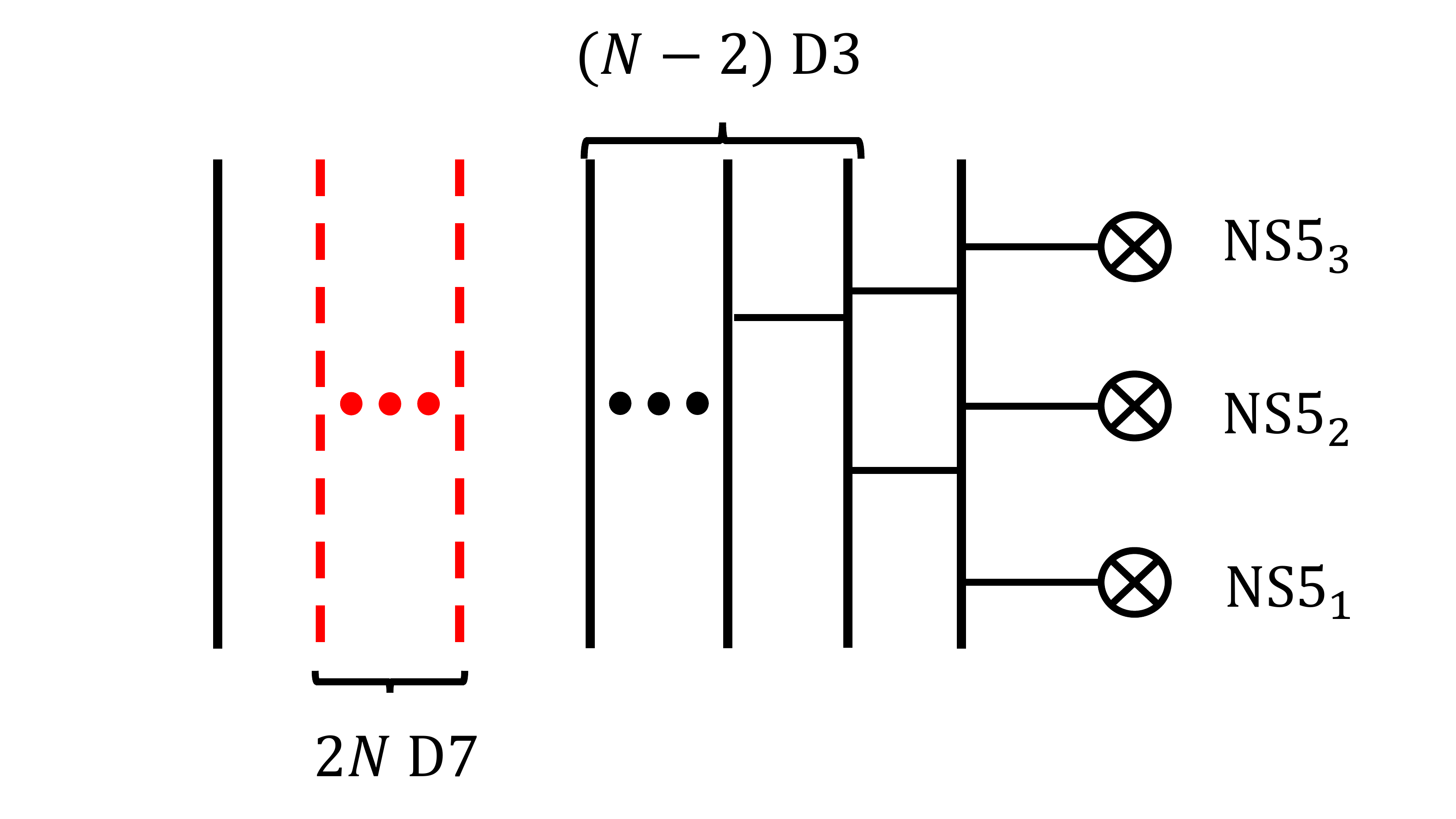}}
\subfigure[]{\label{fig:B03v111no2to3}
\includegraphics[width=8cm]{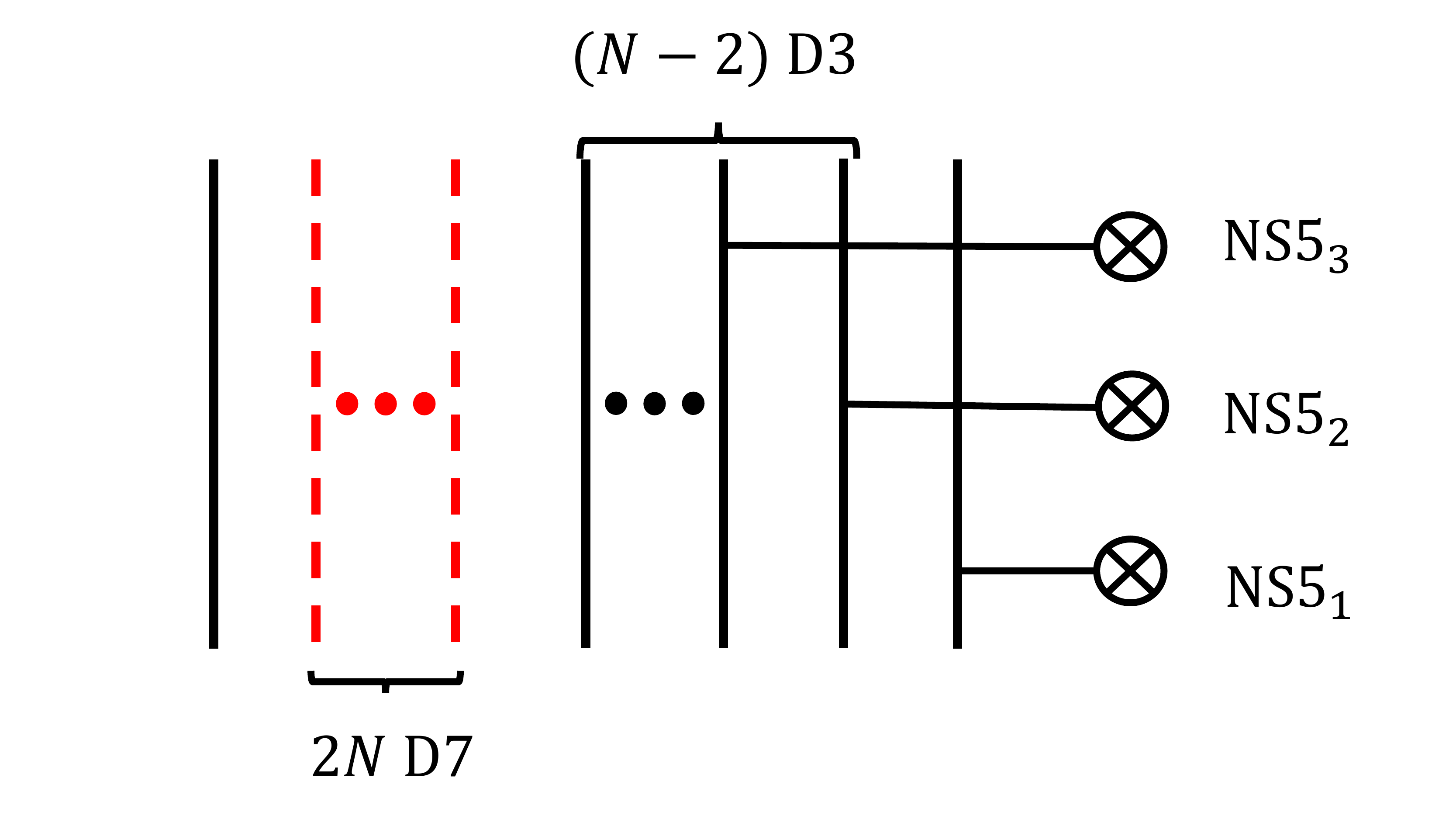}}
\subfigure[]{\label{fig:B03v111no3}
\includegraphics[width=8cm]{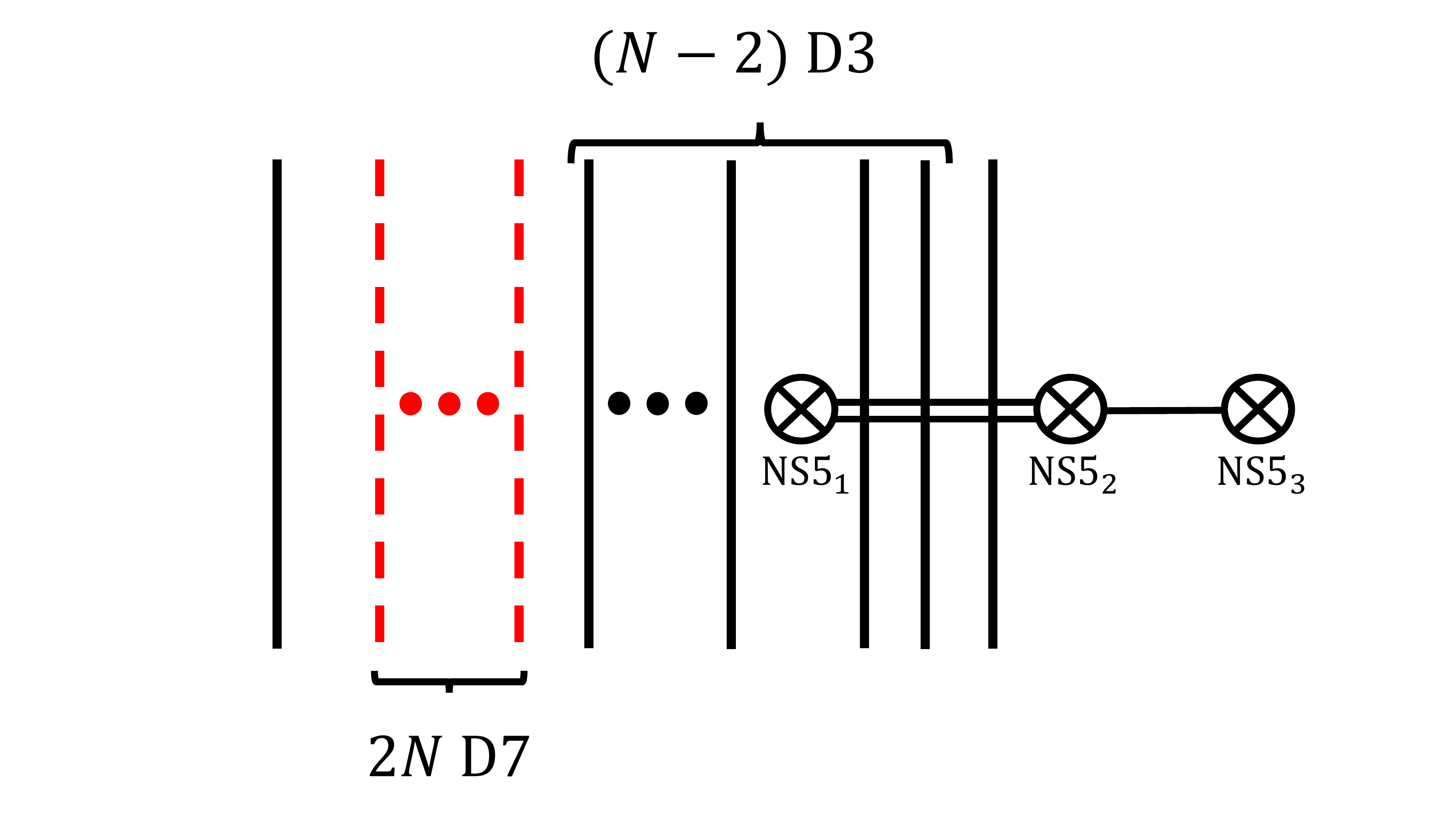}}
\subfigure[]{\label{fig:quiver111}
\includegraphics[width=8cm]{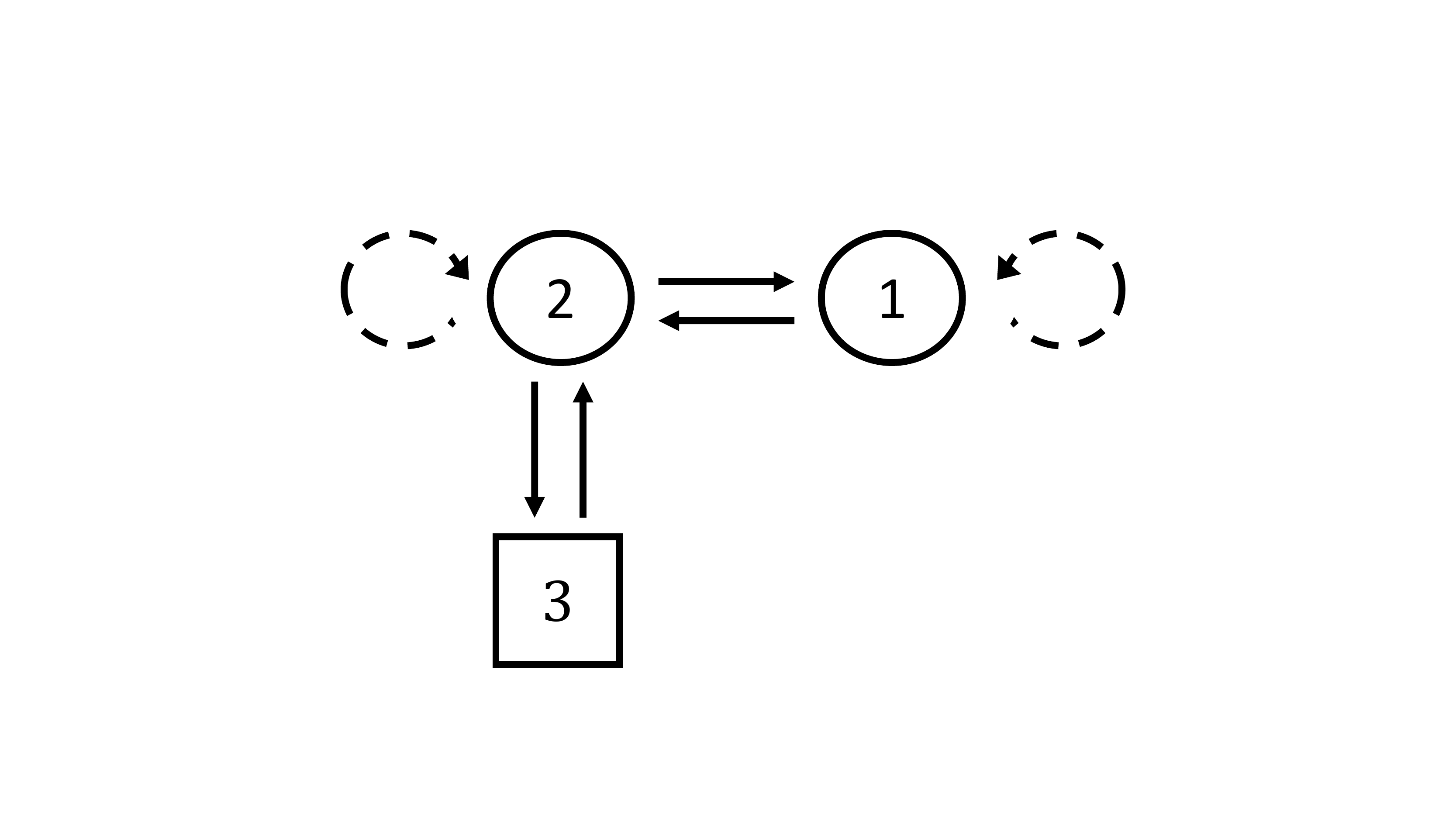}}
\caption{(a):  An introduction of smooth monopoles to Figure \ref{fig:B03no1}. (b):  We adjust the positions of the D1-branes to realize the screening sector ${\bm v} = {\bm e}_{N-2} + {\bm e}_{N-1} + {\bm e}_N$. (c): The brane configuration  that allows us to read off the SQM for the sector. (d): The  corresponding quiver diagram.
 }
\label{fig:B03v111}
\end{figure}
\paragraph{Sector ${\bm v} = {\bm e}_{N-2} + {\bm e}_{N-1} + {\bm e}_N$.}
Again we do not expect wall-crossing in the SQM result as 
the ordering is unique.
The ${\bm v} = {\bm e}_{N-2} + {\bm e}_{N-1} + {\bm e}_N$ sector may be realized by introducing smooth monopoles as in Figure \ref{fig:B03v111no2}. Then moving NS5-branes leads to a configuration in Figure \ref{fig:B03v111no3}, which yields the quiver theory depicted in Figure \ref{fig:quiver111}. Hence the Witten index of the worldvolume theory on the D1-branes in Figure \ref{fig:B03v111no3} gives the monopole screening contribution in the sector ${\bm v} = {\bm e}_{N-2} + {\bm e}_{N-1} + {\bm e}_N$. We again have two FI parameters $\zeta_1, \zeta_2$ from the gauge nodes $U(2)$ and $U(1)$ respectively in the quiver theory in Figure \ref{fig:quiver111}. The relation between the FI parameters and the position of the NS5-branes is given by $\zeta_1 = s_2 - s_1, \zeta_2 = s_3 - s_2$. 

The Witten index of the quiver theory in Figure \ref{fig:quiver111} is given by the contour integral
\begin{align}
&Z({\bm B}=3{\bm e}_N, {\bm v} = {\bm e}_{N-2} + {\bm e}_{N-1} + {\bm e}_{N}; \widetilde{{\bm \zeta}}) \nn\\
=& \frac{1}{2}\oint_{JK(\widetilde{\bm \zeta})} \frac{d\phi_1 }{2\pi i} \frac{d\phi_2 }{2\pi i} \frac{d\varphi}{2\pi i} \left(\frac{\left(2\sinh\epsilon_+\right)^3\prod_{1\leq i\neq j \leq 2}2\sinh\frac{\phi_i -\phi_j + 2\epsilon_+}{2}2\sinh\frac{\phi_i - \phi_j}{2}}{\prod_{i=1}^22\sinh\frac{\phi_i - \varphi + \epsilon_+}{2}2\sinh\frac{\varphi - \phi_i + \epsilon_+}{2}}\right.\nn\\
&\hspace{5cm}\times\left.\frac{1}{\prod_{i=N-2}^{N}\prod_{k=1}^22\sinh\frac{\phi_k - a_i + \epsilon_+}{2}2\sinh\frac{-\phi_k + a_i + \epsilon_+}{2}}\right). \label{Z03v111}
\end{align}
In order to evaluate the integral by the JK residue prescription we set
\begin{align}
\widetilde{{\bm \zeta}} := \zeta_1({\bm e}_1 + {\bm e}_2) + \zeta_2{\bm e}_3 \in \mathfrak{h}^{\ast}_{U(2)} \oplus \mathfrak{h}_{U(1)}^{\ast}, \label{zeta-tilde-U2U1}
\end{align}%
where ${\bm e}_1$ and ${\bm e}_2$ are the orthonormal basis of $\mathfrak{h}^{\ast}_{U(2)}$ for the $U(2)$ and ${\bm e}_3$ is the normalized basis in $\mathfrak{h}_{U(1)}^{\ast}$. 
We denote by $\widetilde{{\bm \zeta}}_i$ ($i=1,\ldots,6$) representatives (to be specified below) from the six FI-chambers corresponding to the orderings of the three 't Hooft operators:
\begin{align}
&\widetilde{{\bm \zeta}}_1: \zeta _1 >0, \, \zeta_2 >0 \, \, \longleftrightarrow  \, \,  s_3 > s_2 > s_1, \label{eq:FIregion1} \\
&\widetilde{{\bm \zeta}}_2: \zeta _2 < 0, \, \zeta_1+\zeta_2 >0 \, \, \longleftrightarrow \, \,   s_2 > s_3 > s_1, \label{eq:FIregion2} \\
&\widetilde{{\bm \zeta}}_3: \zeta _1 > 0, \, \zeta_1+\zeta_2 < 0 \, \, \longleftrightarrow \, \,    s_2 > s_1 > s_3, \\
&\widetilde{{\bm \zeta}}_4: \zeta _1 < 0, \, \zeta_2 < 0 \, \, \longleftrightarrow  \, \,  s_1 > s_2 > s_3, \\
&\widetilde{{\bm \zeta}}_5: \zeta _2 > 0, \, \zeta_1+\zeta_2 < 0 \, \, \longleftrightarrow  \, \,  s_1 > s_3 > s_2, \\
&\widetilde{{\bm \zeta}}_6: \zeta _1 < 0, \, \zeta_1+\zeta_2 < 0 \, \, \longleftrightarrow  \, \,  s_3 > s_1 > s_2.  \label{eq:FIregion6}
\end{align}
Unlike the case of~${\bm v} = {\bm e}_{N-1} + 2{\bm e}_N$ where the JK-chambers~(\ref{B102FI1})-(\ref{B102FI6}) coincide with the FI-chambers, in the current case the embedding, by the map $\bm{\zeta}\mapsto \widetilde{\bm{\zeta}}$, of the FI-chambers~(\ref{eq:FIregion1})-(\ref{eq:FIregion6}) to the dual of the Cartan subalgebra gives only subsets of JK-chambers.

Let us explain the details of the computation for the FI-chamber~(\ref{eq:FIregion1}).
We choose the representative to be $\widetilde{{\bm \zeta}}_1 = {\bm e}_1 + {\bm e}_2 +   {\bm e}_3$.
The intersections of the singular hyperplanes (see Appendix~\ref{sec:locforSQM} for the definition) 
 of \eqref{Z03v111}  include non-degenerate and degenerate poles.
First we evaluate the contributions from the degenerate poles using the constructive definition of the JK residue. 
The definition is given in~(\ref{eq:consJKres}) and the relevant details are summarized in Appendix~\ref{sec:JK}.
There are three degenerate poles
 \begin{align}
{\bm \phi}_{\ast}=(a_i- \epsilon_+, a_i- \epsilon_+, a_i -2 \epsilon_+),
\quad  i \in \{N-2 ,N-1, N \}, 
\label{eq:degept}
\end{align}
each of which is the intersection point of four singular hyperplanes
\begin{align}
&\{\phi_1 -a_i +\epsilon_+=0\} \cap \{\phi_2 -a_i +\epsilon_+=0\} 
\nn\\
& \qquad\qquad
\cap \{ -\phi_1 +\varphi +\epsilon_+=0\} \cap \{ -\phi_2 +\varphi +\epsilon_+=0 \}.
\label{eq:degint}
\end{align}
We define the ordered set of charge vectors associated with the four singular hyperplanes
\begin{align}
{\bm Q}_*=\{{\bm Q}_{1}, {\bm Q}_{2}, {\bm Q}_{3}, {\bm Q}_{4}  \}:= \{(1,0,0), (0,1,0), (-1,0,1) , (0,-1,1)\}.
\end{align}
Though $\widetilde{{\bm \zeta}}_1 $ belongs to a JK-chamber it does not satisfy the strong regularity condition \eqref{eq:strongref}.

When the strong regularity condition is violated, we compute the JK residues according to the procedure%
\footnote{A similar procedure was applied in the evaluation of a Witten index in Section 7.2 of~\cite{Hori:2014tda}.} 
we summarize in Appendix~\ref{sec:JK}.
For example, the possible  small shifts of the representative
$\widetilde{{\bm \zeta}}_1 = {\bm e}_1 + {\bm e}_2 +   {\bm e}_3$ are given by
 \begin{align}
\widetilde{\bm \zeta}_1' = {\bm e}_1 + (1 \pm \delta_1) {\bm e}_2 + (1 \pm \delta_2) {\bm e}_3 \quad \text{with} \, \, 0 < \delta_i  \ll 1. 
\label{eq:shiftFI}
\end{align}
The flags that contribute to JK residues depend on the signs and the relative magnitudes of~$\delta_1$ and $\delta_2$. 
If we choose $\widetilde{\bm \zeta}_1' = {\bm e}_1 + (1 - \delta_1) {\bm e}_2 + (1 - \delta_2) {\bm e}_3$ with $\delta_1 < \delta_2$,  the two flags  
\begin{align}
F^{(1)}&=[\{0\} \subset \langle {\bm Q}_1 \rangle \subset \langle {\bm Q}_1,  {\bm Q}_2 \rangle \subset \langle {\bm Q}_1,  {\bm Q}_2, {\bm Q}_3\rangle =\mathbb{R}^3], \\
F^{(2)}&=[\{0\} \subset \langle {\bm Q}_2  \rangle \subset \langle {\bm Q}_2,  {\bm Q}_1 \rangle \subset \langle {\bm Q}_2,  {\bm Q}_1, {\bm Q}_4 \rangle=\mathbb{R}^3]
\end{align}
satisfy the condition $\widetilde{\bm \zeta}_1' \in \mathfrak{s}^+(F^{(a)}, {\bm Q}_*), a=1, 2$. 
Here $\langle {\bm Q}_{1}, {\bm Q}_2, \cdots \rangle$ denotes the vector space spanned by the basis $\{ {\bm Q}_{1}, {\bm Q}_{2}, \cdots \}$.
Both of the iterated residues for the two flags are zero.  
In fact all the iterated residues computed in a similar manner vanish for the other choices of signs in~\eqref{eq:shiftFI} and the relative magnitudes of $\delta_1$ and $\delta_2$.
We interpret these results as the vanishing of the contributions from the degenerate poles to the Witten index~(\ref{Z03v111}).

Next we consider the non-degenerate poles.
There are four possible cones which contain $\widetilde{{\bm \zeta}}_1= {\bm e}_1 +  {\bm e}_2+ {\bm e}_3 $;
\begin{align}
\hspace{-3mm}
C_{a}^{(1)}\hspace{-1mm}&=\text{Cone}\left[(1, 0, 0), (0, 1, 0), (-1, 0, 1)\right], \ C_{b}^{(1)}\hspace{-1mm}=\text{Cone}\left[(1, 0, 0), (0, 1, 0), (0, -1, 1)\right], \nn\\
\hspace{-3mm}
C_{c}^{(1)}\hspace{-1mm}&=\text{Cone}\left[(1, 0, 0), (0, 1, -1), (-1, 0, 1)\right], \ C_{d}^{(1)}\hspace{-1mm}=\text{Cone}\left[(0, 1, 0), (1, 0, -1), (0, -1, 1)\right].\hspace{-1mm} \label{cone3dv1}
\end{align}
For example  there are the  six non-degenerate poles in
$C_{a}^{(1)}
$, each the intersection of  three hyperplanes:
\begin{align}
&\{\phi_1 -a_1 +\epsilon_+=0 \}\cap \{ \phi_2 -a_2 +\epsilon_+=0\} \cap \{ -\phi_1 +\varphi +\epsilon_+=0  \}, \\
&\{\phi_1 -a_1 +\epsilon_+=0 \}\cap\{ \phi_2 -a_3 +\epsilon_+=0\} \cap \{ -\phi_1 +\varphi +\epsilon_+=0  \}, \\
&\{\phi_1 -a_2 +\epsilon_+=0 \}\cap\{ \phi_2 -a_1 +\epsilon_+=0 \} \cap \{ -\phi_1 +\varphi +\epsilon_+=0  \}, \\
&\{\phi_1 -a_2 +\epsilon_+=0 \} \cap \{ \phi_2 -a_3 +\epsilon_+=0 \} \cap \{ -\phi_1 +\varphi +\epsilon_+=0  \}, \\
&\{\phi_1 -a_3 +\epsilon_+=0 \} \cap \{ \phi_2 -a_1 +\epsilon_+=0 \} \cap \{-\phi_1 +\varphi +\epsilon_+=0  \}, \\
&\{\phi_1 -a_3 +\epsilon_+=0 \} \cap \{\phi_2 -a_2 +\epsilon_+=0 \} \cap \{ -\phi_1 +\varphi +\epsilon_+=0  \},
\end{align}
 and the JK residue is given by
\begin{align}
&\widetilde Z({\bm B}=3{\bm e}_N, {\bm v} = {\bm e}_{N-2} + {\bm e}_{N-1} + {\bm e}_{N}; C_{a}^{(1)})\nn\\
&=\frac{1}{2} \sum_{k=0}^1\left(\frac{1}{{\displaystyle \prod_{N-2 \leq i < j \leq N}}2\sinh\frac{a_{i} - a_j}{2}2\sinh\frac{-a_i + a_j +(-1)^k 2\epsilon_+}{2}}\right.\nn\\
&+\frac{1}{\prod_{N-1 \leq  j \leq N}2\sinh\frac{a_{N-2} - a_j}{2}2\sinh\frac{-a_{N-2} + a_j + (-1)^k2\epsilon_+}{2}2\sinh\frac{-a_{N-1} + a_N}{2}2\sinh\frac{a_{N-1} - a_N + (-1)^k2\epsilon_+}{2}}\nn\\
&\left.+\frac{1}{2\sinh\frac{-a_{N-2} + a_{N-1}}{2}2\sinh\frac{a_{N-2} - a_{N-1} + (-1)^k2\epsilon_+}{2}
{\displaystyle \prod_{N-2 \leq  i \leq N-1}}
2\sinh\frac{a_i - a_N}{2}2\sinh\frac{-a_i + a_N + (-1)^k2\epsilon_+}{2}}\right),
\end{align}
In fact the explicit evaluation shows that the poles corresponding to the cones $C_{a}^{(1)}$ and $C_{b}^{(1)}$ lead to the same result whereas the poles corresponding to the cones $C^{(1)}_c$ and $C_{d}^{(1)}$ yield zero. The end result is 
\begin{align}
&Z({\bm B}=3{\bm e}_N, {\bm v} = {\bm e}_{N-2} + {\bm e}_{N-1} + {\bm e}_{N}; \widetilde{{\bm \zeta}}_1)\nn\\
=&\sum_{k=0}^1\left(\frac{1}{{\displaystyle \prod_{N-2 \leq i < j \leq N}}2\sinh\frac{a_{i} - a_j}{2}2\sinh\frac{-a_i + a_j +(-1)^k 2\epsilon_+}{2}}\right.\nn\\
&+\frac{1}{{\displaystyle \prod_{N-1 \leq  j \leq N}}2\sinh\frac{a_{N-2} - a_j}{2}2\sinh\frac{-a_{N-2} + a_j + (-1)^k2\epsilon_+}{2}2\sinh\frac{-a_{N-1} + a_N}{2}2\sinh\frac{a_{N-1} - a_N + (-1)^k2\epsilon_+}{2}}\nn\\
&+\left.\frac{1}{2\sinh\frac{-a_{N-2} + a_{N-1}}{2}2\sinh\frac{a_{N-2} - a_{N-1} + (-1)^k2\epsilon_+}{2}
\hspace{-3mm}
{\displaystyle \prod_{N-2 \leq  i \leq N-1}}
\hspace{-3mm}
2\sinh\frac{a_i - a_N}{2}2\sinh\frac{-a_i + a_N + (-1)^k2\epsilon_+}{2}}\right),
\label{eq:jkreszeta1}
\end{align}
which precisely equals the monopole screening contribution in \eqref{TFFFv111}. 

Similarly the contour integral~(\ref{Z03v111}) for the other FI-chambers \eqref{eq:FIregion2}-\eqref{eq:FIregion6} can be evaluated according to the procedure summarized in Appendix~\ref{sec:JK}.
As representatives we can choose
 $\widetilde{{\bm \zeta}}_2 = 3({\bm e}_1 + {\bm e}_2) -  {\bm e}_3$, $\widetilde{{\bm \zeta}}_3 = {\bm e}_1 + {\bm e}_2 -3  {\bm e}_3$, $\widetilde{{\bm \zeta}}_4 = -({\bm e}_1 + {\bm e}_2) -  {\bm e}_3$, $\widetilde{{\bm \zeta}}_5 = -3({\bm e}_1 + {\bm e}_2) +  {\bm e}_3$, 
  $\widetilde{{\bm \zeta}}_6 =- ({\bm e}_1 +{\bm e}_2) +3  {\bm e}_3$.
For each $\widetilde{{\bm \zeta}}_i$ let us list those cones containing~$\widetilde{{\bm \zeta}}_i$ which are spanned by the charge vectors at the non-degenerate points.
\begin{itemize}
\item The cones that contain $\widetilde{{\bm \zeta}}_2 = 3({\bm e}_1 + {\bm e}_2) -  {\bm e}_3$:
\begin{align}
&C_{a}^{(2)}=\text{Cone}\left[(1, 0, 0), (0, 1, 0), (1, 0, -1)\right], \nn\\
& C_{b}^{(2)}=\text{Cone}\left[(1, 0, 0), (0, 1, 0), (0, 1, -1)\right], \nn\\
&C_{c}^{(2)}=\text{Cone}\left[(1, 0, 0), (0, 1, -1), (-1, 0,1)\right], \nn\\
& C_{d}^{(2)}=\text{Cone}\left[(0, 1, 0), (1, 0, -1), (0, -1, 1)\right]. \label{cone3dv3}
\end{align}
\item The cones that contain $\widetilde{{\bm \zeta}}_3 = {\bm e}_1 + {\bm e}_2 -3  {\bm e}_3$:
\begin{align}
&C_{a}^{(3)}=\text{Cone}\left[(1, 0, 0), (0, -1, 0), (0, 1, -1)\right], \nn\\
& C_{b}^{(3)}=\text{Cone}\left[(-1, 0, 0), (0, 1, 0), (1, 0, -1)\right]. 
\end{align}
\item The cones that contain $\widetilde{{\bm \zeta}}_4= -({\bm e}_1 + {\bm e}_2) -  {\bm e}_3$:
\begin{align}
&C_{a}^{(4)}=\text{Cone}\left[(-1, 0, 0), (0, -1, 0), (1, 0, -1)\right], \nn\\
& C_{b}^{(4)}=\text{Cone}\left[(-1, 0, 0), (0, -1, 0), (0, 1, -1)\right], \nn\\
&C_{c}^{(4)}=\text{Cone}\left[(-1, 0, 0), (0, -1, 1), (1, 0,-1)\right], \nn\\
& C_{d}^{(4)}=\text{Cone}\left[(0, -1, 0), (-1, 0, 1), (0, 1, -1)\right]. \label{cone3dv2}
\end{align}
\item The cones that contain $\widetilde{{\bm \zeta}}_5 = -3({\bm e}_1 + {\bm e}_2) +  {\bm e}_3$:
\begin{align}
&C_{a}^{(5)}=\text{Cone}\left[(-1, 0, 0), (0, -1, 0), (-1, 0, 1)\right], \nn\\
& C_{b}^{(5)}=\text{Cone}\left[(-1, 0, 0), (0, -1, 0), (0, -1, 1)\right], \nn\\
&C_{c}^{(5)}=\text{Cone}\left[(-1, 0, 0), (0, -1, 1), (1, 0,-1)\right], \nn\\
& C_{d}^{(5)}=\text{Cone}\left[(0, -1, 0), (-1, 0, 1), (0, 1, -1)\right]. \label{cone3dv4}
\end{align}
\item The cones that contain $\widetilde{{\bm \zeta}}_6 =-( {\bm e}_1 + {\bm e}_2) +3  {\bm e}_3$:
\begin{align}
&C_{a}^{(6)}=\text{Cone}\left[(1, 0, 0), (0, -1, 0), (-1, 0, 1)\right], \nn\\
& C_{b}^{(6)}=\text{Cone}\left[(-1, 0, 0), (0, 1, 0), (0, -1, 1)\right]. 
\end{align}
\end{itemize}
For every~$\widetilde{\zeta}_i$ we checked that summing the JK residues of the poles in these charge cones gives~\eqref{eq:jkreszeta1}.
Again, the JK residues at  the degenerate  poles vanish for any choice of infinitesimal shifts similar to~\eqref{eq:shiftFI}.
 Therefore the SQM exhibits no wall-crossing, as expected from the uniqueness of the ordering in~$\Braket{T_{\Box}}\ast \Braket{T_{\Box}} \ast \Braket{T_{\Box}}$.

\subsubsection{'t~Hooft operator with ${\bm B} = -{\bm e}_1 + 2{\bm e}_N$}
\begin{figure}[t]
\centering
\includegraphics[width=8cm]{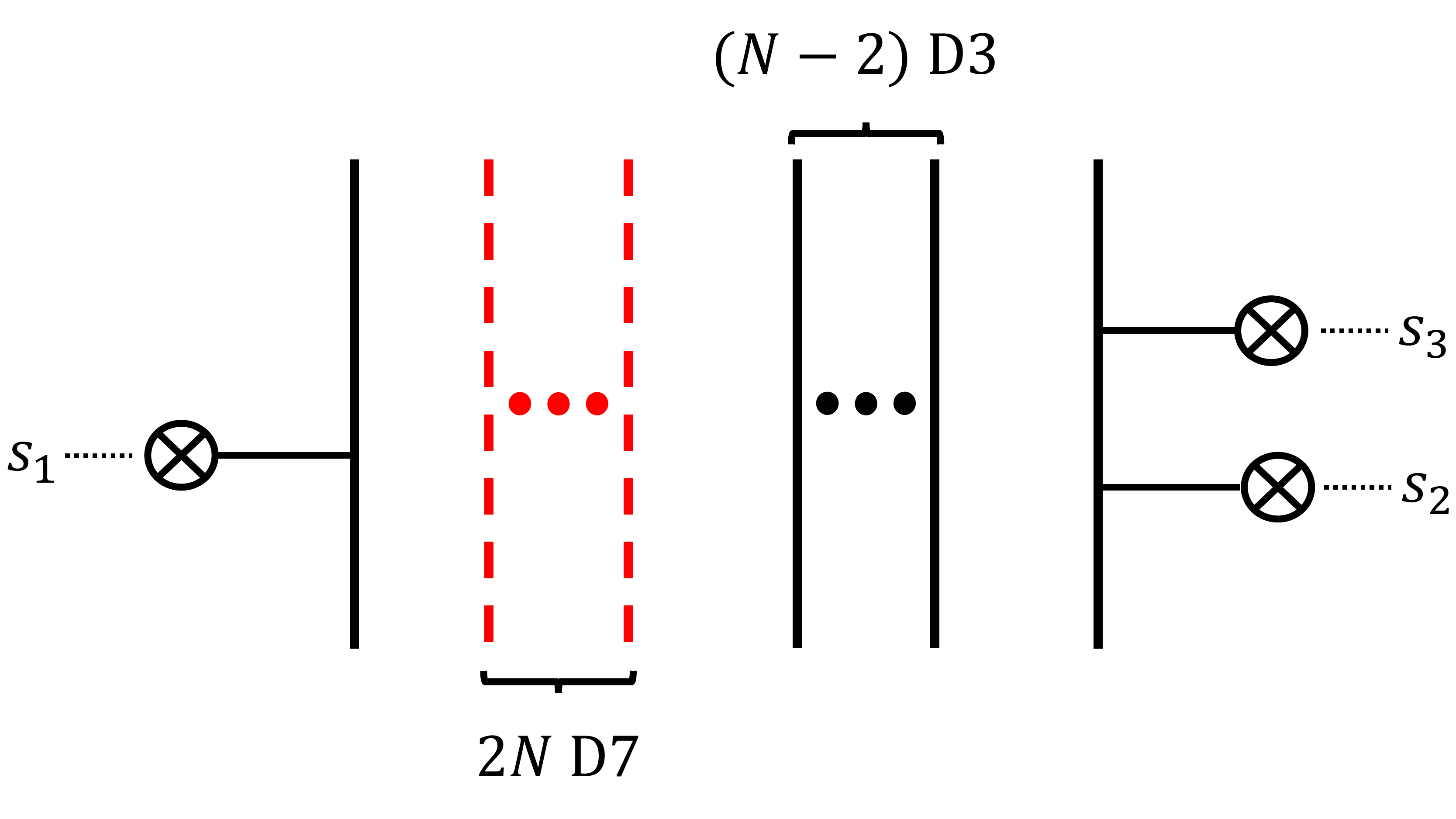}
\caption{A brane configuration for the 't~Hooft operator with the magnetic charge ${\bm B}=-{\bm e}_1 + 2{\bm e}_N$.
}
\label{fig:B102no1}
\end{figure}
Next we consider the products $\Braket{T_{\Box}}\ast\Braket{T_{\Box}}\ast\Braket{T_{\overline{\Box}}}$, $\Braket{T_{\Box}}\ast\Braket{T_{\overline{\Box}}}\ast\Braket{T_{\Box}}$ and $\Braket{T_{\overline{\Box}}}\ast\Braket{T_{\Box}}\ast\Braket{T_{\Box}}$. The relevant 't~Hooft operator with the magnetic charge ${\bm B} = -{\bm e}_1 + 2{\bm e}_N$ is realized by a brane configuration in Figure \ref{fig:B102no1}. $s_1, s_2, s_3$ in Figure \ref{fig:B102no1} are the coordinates in the $x^3$-direction of the three NS5-branes. As in \eqref{TFFAFmono}, the products contain two sectors for the monopole screening. One is characterized by ${\bm v} = -{\bm e}_1 + {\bm e}_{N-1} + {\bm e}_N$ and the other is characterized by ${\bm v} = {\bm e}_N$.

\begin{figure}[t]
\centering
\subfigure[]{\label{fig:B102v111no2}
\includegraphics[width=8cm]{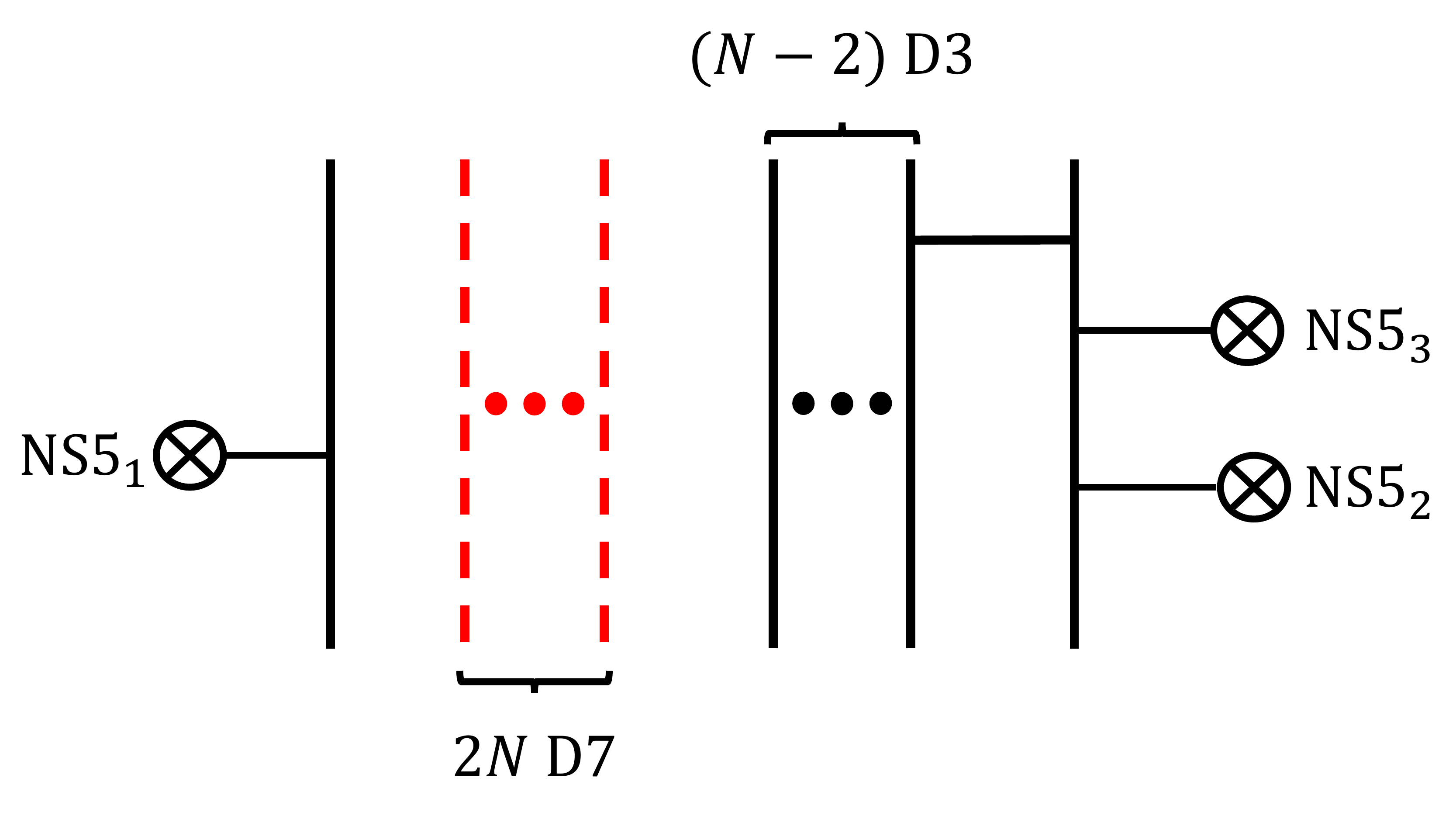}}
\subfigure[]{\label{fig:B102v111no2to3}
\includegraphics[width=8cm]{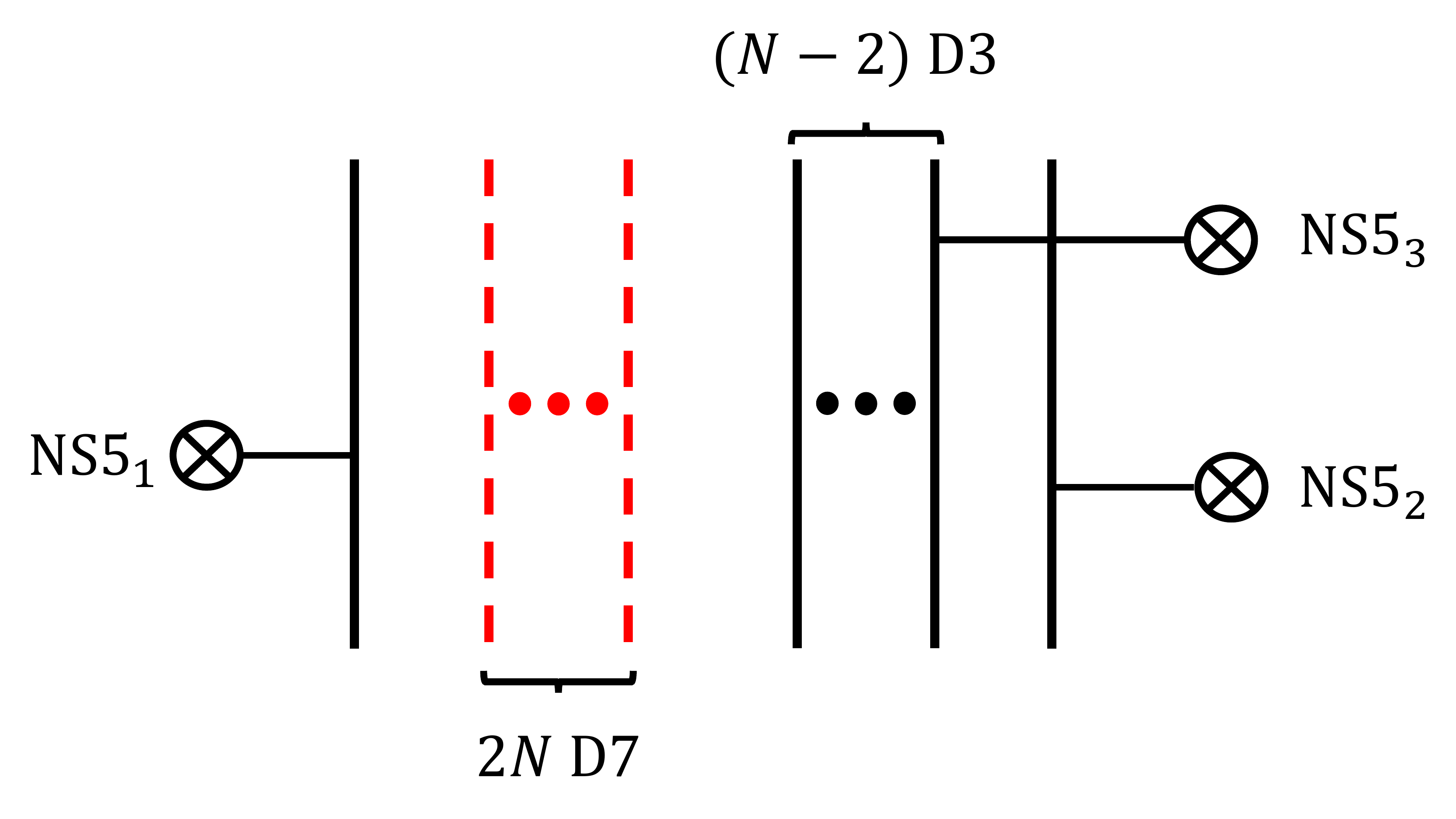}}
\subfigure[]{\label{fig:B102v111no3}
\includegraphics[width=8cm]{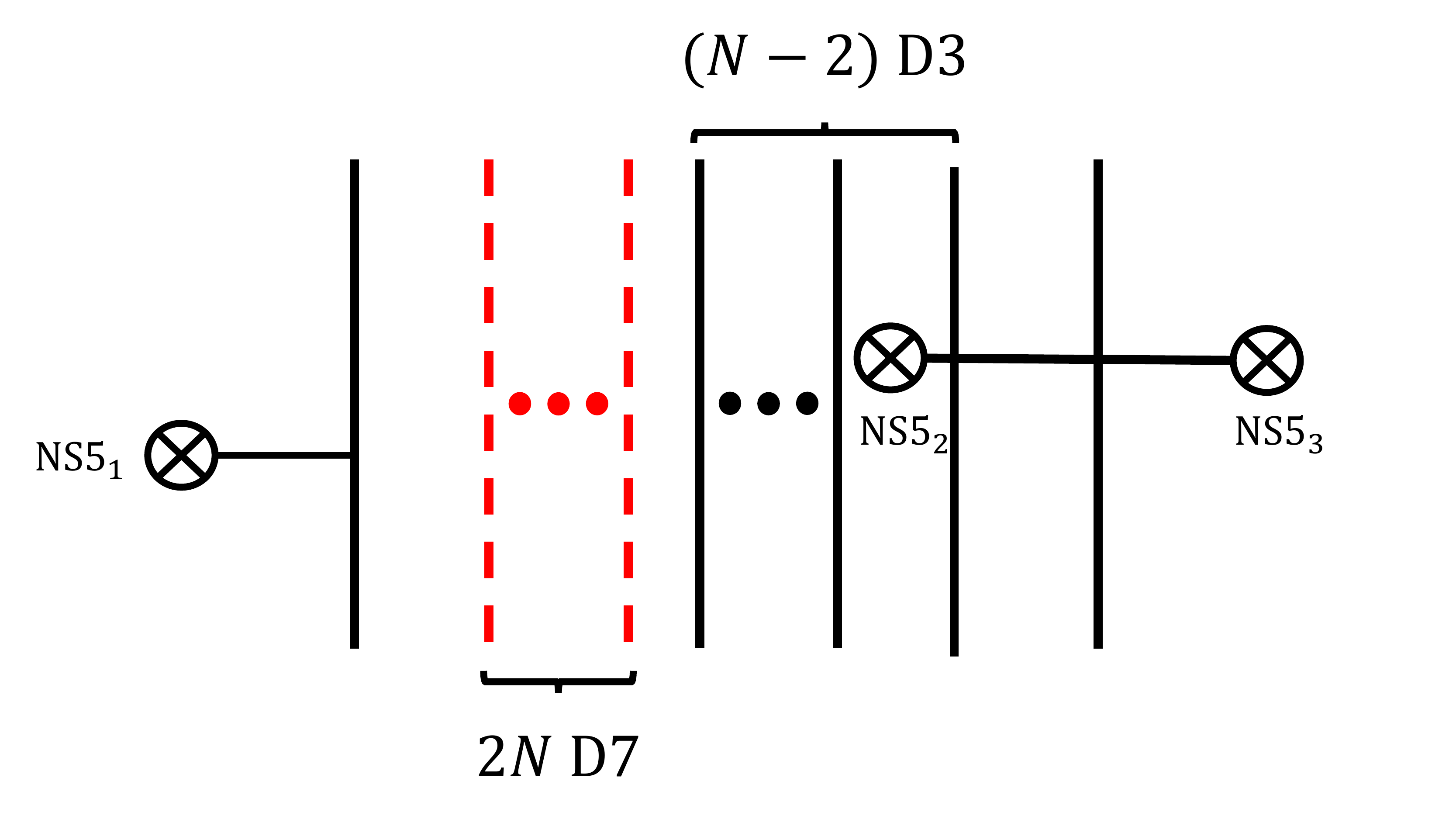}}
\subfigure[]{\label{fig:quiver102v111}
\includegraphics[width=8cm]{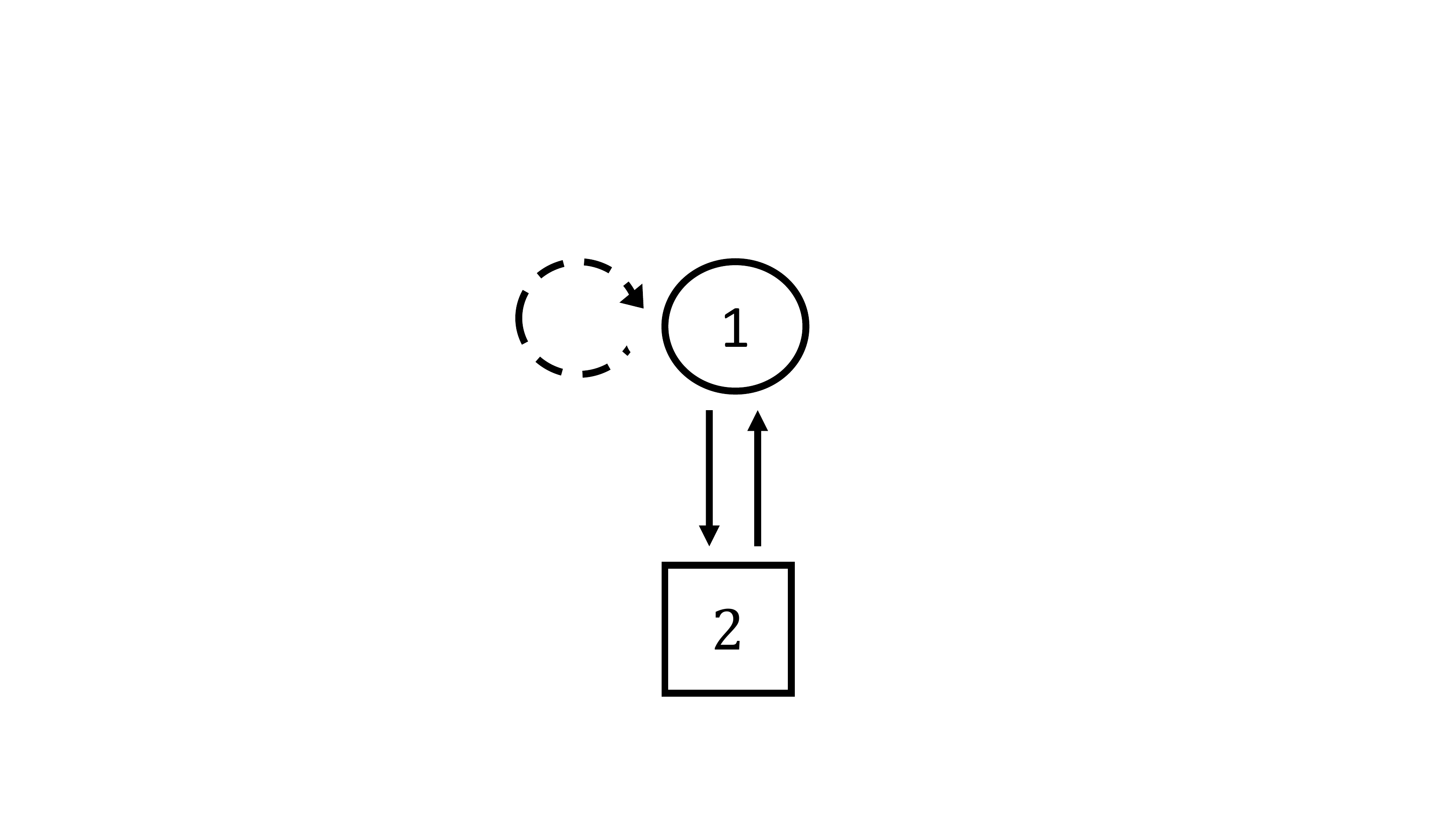}}
\caption{(a):  An introduction of a smooth monopole to Figure \ref{fig:B102no1}. (b):  We adjust the position of the D1-brane to realize the screening sector ${\bm v} = -{\bm e}_{1} + {\bm e}_{N-1} + {\bm e}_N$. (c): The brane configuration  that allows us to read off the SQM for the sector.
(d): The  corresponding quiver diagram.
 }
\label{fig:B102v111}
\end{figure}

\paragraph{Sector ${\bm v} = -{\bm e}_1 + {\bm e}_{N-1} + {\bm e}_N$.}
This sector can be realized by introducing a smooth monopole with magnetic charge $
{\bm e}_{N-1} - {\bm e}_N$ as in Figure \ref{fig:B102v111no2}. 
Then tuning the position of the introduced D1-brane between D3-branes screens the magnetic charge, leading to the sector ${\bm v} = -{\bm e}_1 + {\bm e}_{N-1} + {\bm e}_{N}$ as in Figure \ref{fig:B102v111no2to3}. Arranging the position of an NS5-brane gives a diagram in Figure \ref{fig:B102v111no3} from which we can read off the SQM for the monopole screening contribution in the sector ${\bm v} =  -{\bm e}_1 + {\bm e}_{N-1} + {\bm e}_{N}$. The quiver diagram of the SQM is depicted in Figure \ref{fig:quiver102v111}. The quiver theory has an FI parameter $\zeta$ from the single $U(1)$ gauge node. The FI parameter is related to the position of two of the three NS5-branes as $\zeta = s_3 - s_2$. Namely the position of the left NS5-brane is not involved in the FI parameter.  

Note here that the quiver diagram is in fact identical to the one in Figure \ref{fig:quiver02}. Therefore the Witten index of the SQM is simply given by \eqref{B011result1} or \eqref{B011result2}, depending on the sign of the FI parameter $\zeta$. The results \eqref{B011result1} and \eqref{B011result2} are actually equal to each other and they do not depend on the sign of $\zeta$. The results perfectly agree with the monopole screening contribution in \eqref{TFFAFvm111}, which were computed by the Moyal product.

In this case there is no wall-crossing although the Moyal product involves two types of operators. This is actually natural  as seen from the brane configuration in~Figure \ref{fig:B102v111}. 
The FI parameter is given by $\zeta = s_3 - s_2$, where $s_2$ and $s_3$ are the locations of the two NS5-branes on the right side in Figure~\ref{fig:B102v111}.  These NS5-branes yield the same operator~$T_{\Box}$. Hence changing the sign of the FI parameter in fact corresponds to  exchanging the operators of the same type. This happens because the magnetic charge of the operator $T_{\overline{\Box}}$ is not screened in the sector ${\bm v } = -{\bm e}_1 + {\bm e}_{N-1} + {\bm e}_N$, which implies that the monopole screening contribution $Z_{\text{mono}}({\bm v} = -{\bm e}_1 + {\bm e}_{N-1} + {\bm e}_N)$ is  independent of the position of the operator $T_{\overline{\Box}}$ in the Moyal product.

\begin{figure}[t]
\centering
\subfigure[]{\label{fig:B102no2}
\includegraphics[width=8cm]{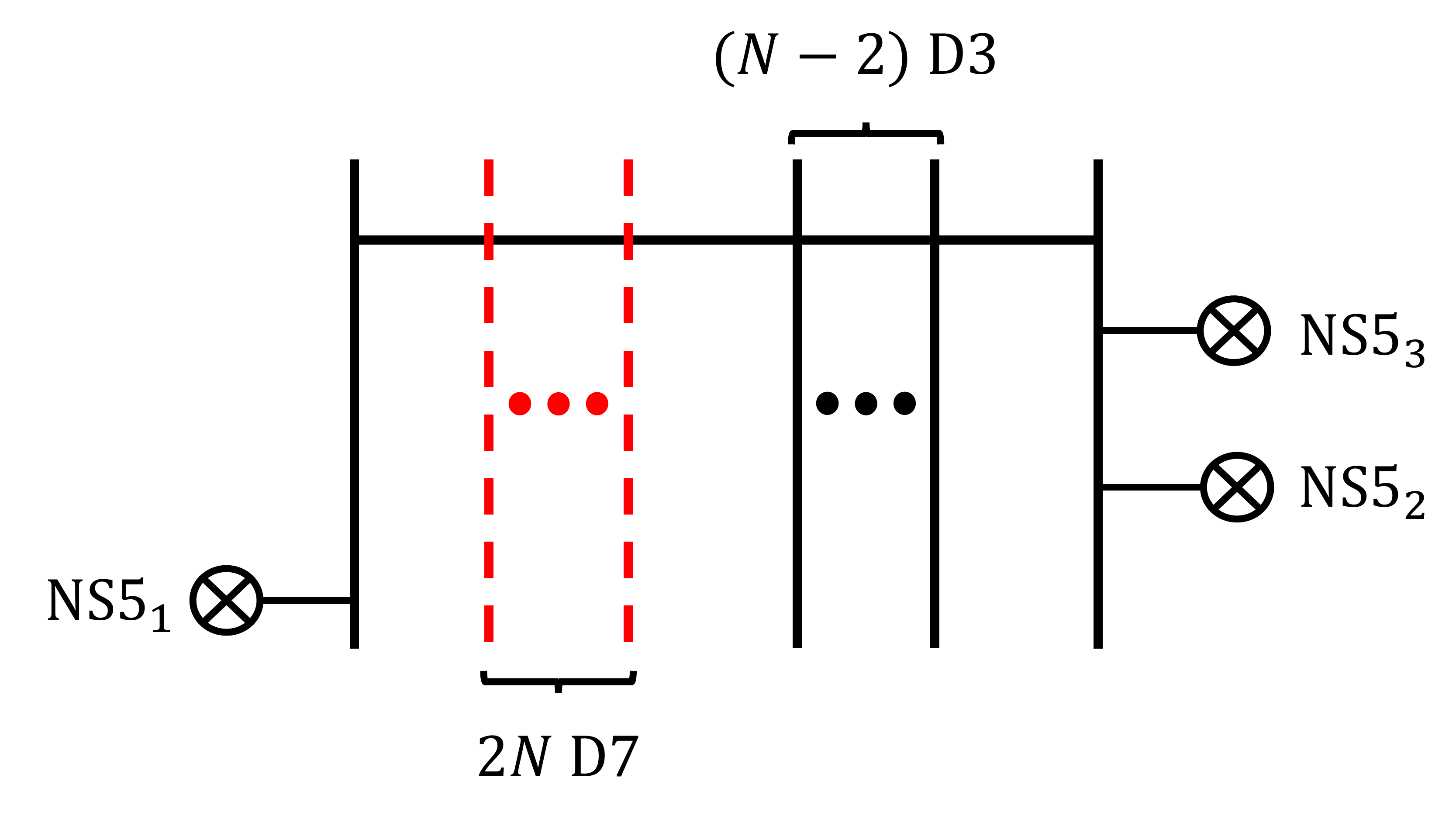}}
\subfigure[]{\label{fig:B102no2to3}
\includegraphics[width=8cm]{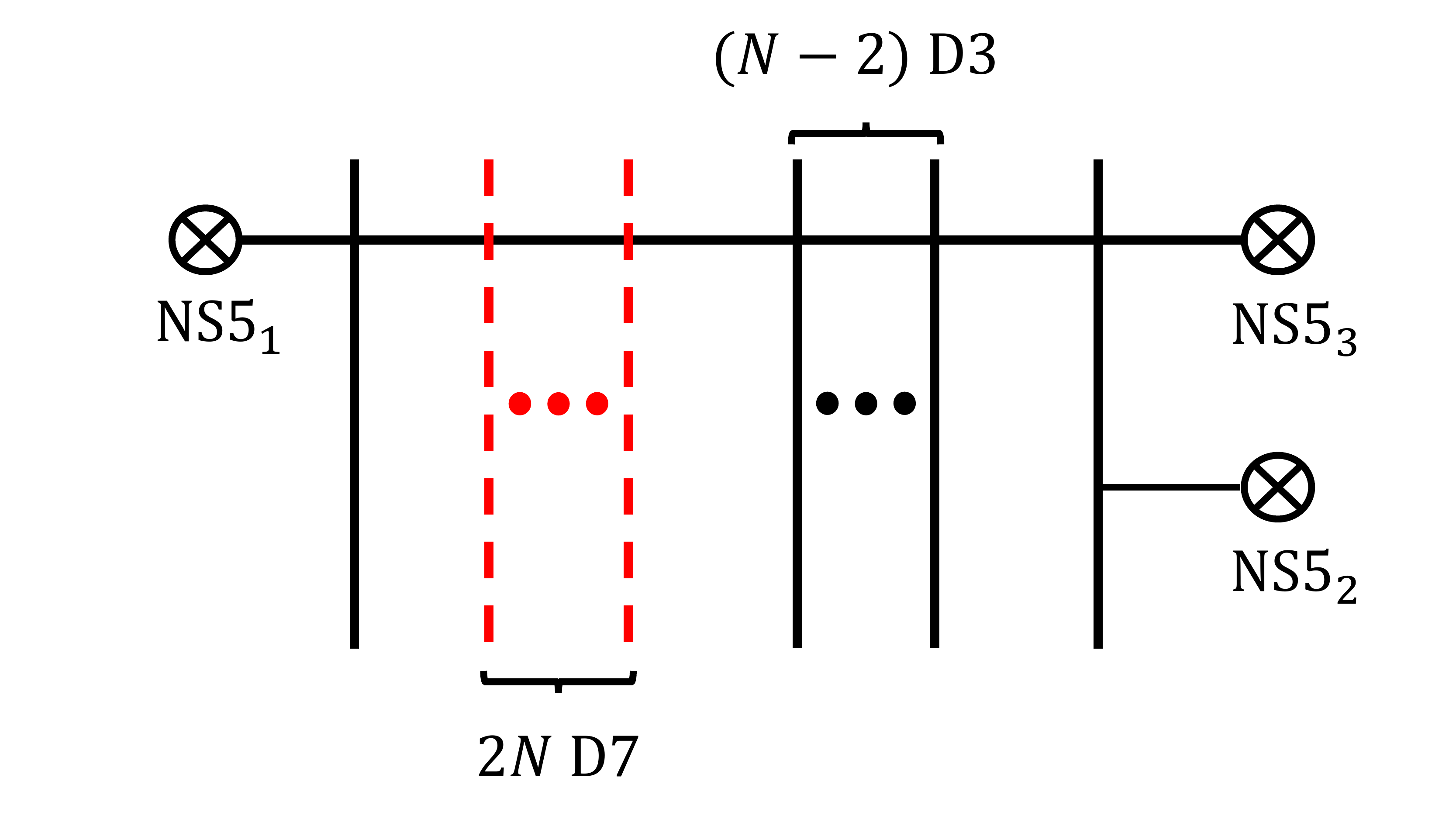}}
\subfigure[]{\label{fig:B102no3}
\includegraphics[width=8cm]{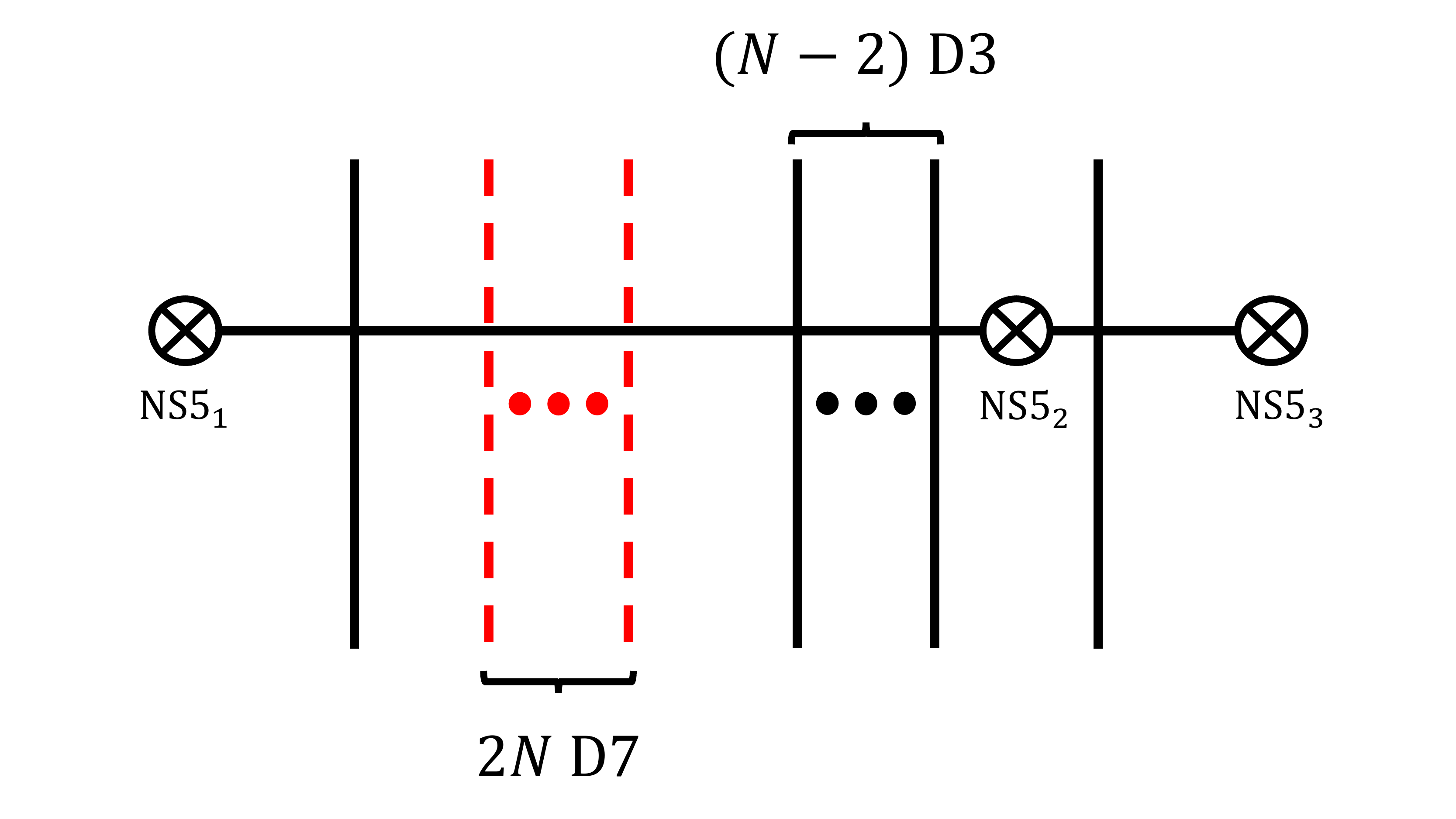}}
\subfigure[]{\label{fig:quiver102}
\includegraphics[width=8cm]{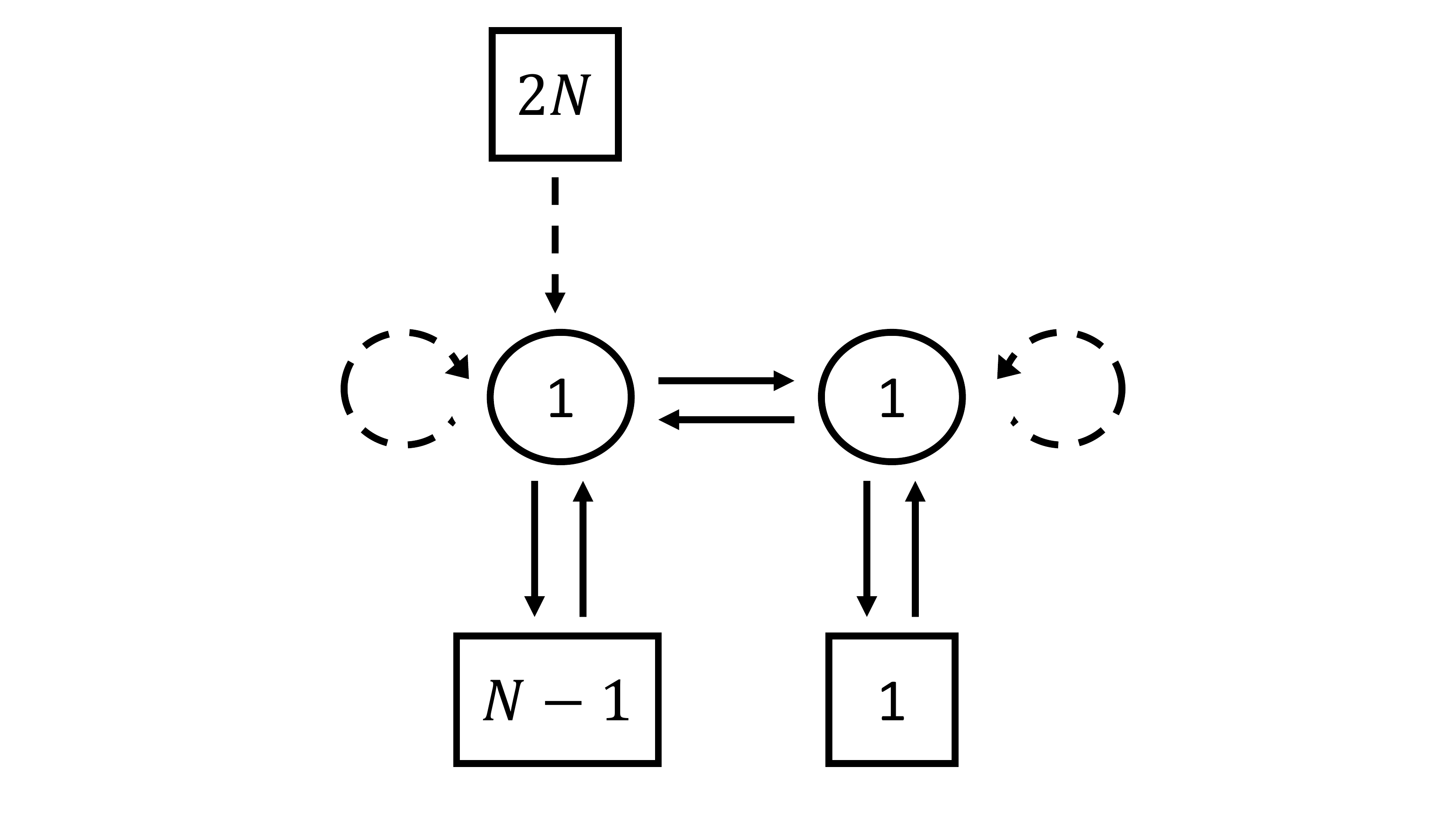}}
\caption{(a):  An introduction of a smooth monopole with magnetic charge $
{\bm e}_1 - {\bm e}_N$ to Figure~\ref{fig:B102no1}. 
(b):  We adjust the position of the D1-brane to realize
 the screening sector ${\bm v} = {\bm e}_N$. (c): The brane configuration from which we can read off  the SQM for  the sector.
  (d): The  corresponding quiver diagram.
   }
\label{fig:B102}
\end{figure}
\paragraph{Sector ${\bm v} = {\bm e}_N$.}
Finally we consider the sector ${\bm v} = {\bm e}_N$ in the products $\Braket{T_{\Box}}\ast\Braket{T_{\Box}}\ast\Braket{T_{\overline{\Box}}}$, $\Braket{T_{\Box}}\ast\Braket{T_{\overline{\Box}}}\ast\Braket{T_{\Box}}$ and $\Braket{T_{\overline{\Box}}}\ast\Braket{T_{\Box}}\ast\Braket{T_{\Box}}$. This case is interesting as we will see wall-crossing corresponding to the ordering of the Moyal product. In order to realize the sector ${\bm v} = {\bm e}_N$, we introduce a smooth monopole with magnetic charge $
{\bm e}_1 - {\bm e}_N$ as in Figure \ref{fig:B102no2}. Tuning the position of the introduced D1-brane between D3-branes gives a configuration in Figure \ref{fig:B102no2to3}, leading to the sector ${\bm v} = {\bm e}_N$. Moving the right lower NS5-brane yields a diagram in Figure \ref{fig:B102no3} and the worldvolume theory on the D1-brane gives an SQM for the monopole screening contribution in the sector ${\bm v} = {\bm e}_N$. The quiver diagram of the SQM is depicted in Figure \ref{fig:quiver102}. We denote the $U(1)$ gauge node with the $2N$ short Fermi multiplets by $U(1)_1$ and the other $U(1)$ gauge node by $U(1)_2$. The FI parameters of the $U(1)_1, U(1)_2$ are denoted by $\zeta_1, \zeta_2$ respectively. In terms of the position of the NS5-branes they are given by $\zeta_1 = s_2 - s_1, \zeta_2 = s_3 - s_2$. 

We can use the quiver diagram in Figure \ref{fig:quiver102} to compute the Witten index of the SQM. The Witten index is given by
\begin{align}
&Z({\bm B}=-{\bm e}_1 + 2{\bm e}_N, {\bm v}={\bm e}_N; {\bm \zeta})\nn\\
= &\oint_{JK({\bm \zeta})}\frac{d\phi_1}{2\pi i}\frac{d\phi_2}{2\pi i}\frac{\left(2\sinh\epsilon_+\right)^2}{2\sinh\frac{\phi_1 - \phi_2 + \epsilon_+}{2}2\sinh\frac{\phi_2 - \phi_1 + \epsilon_+}{2}}\nn\\
&\times\frac{\prod_{f=1}^{2N}2\sinh\frac{\phi_1 - m_f}{2}}{\prod_{i=1}^{N-1}2\sinh\frac{\phi_1 - a_i + \epsilon_+}{2}2\sinh\frac{-\phi_1 + a_i + \epsilon_+}{2}}\frac{1}{2\sinh\frac{\phi_2 - a_N + \epsilon_+}{2}2\sinh\frac{-\phi_2 + a_N + \epsilon_+}{2}}. \label{Z102v01}
\end{align}
The integral can be evaluated by the JK residue prescription with choices of the FI parameter ${\bm \zeta} = \zeta_1{\bm e}_1 + \zeta_2{\bm e}_2 \in \mathfrak{h}^{\ast}_{U(1)_1} \oplus \mathfrak{h}_{U(1)_2}^{\ast}$. ${\bm e}_1$ is the normalized basis in $\mathfrak{h}^{\ast}_{U(1)_1}$ and ${\bm e}_2$ is the normalized basis in $\mathfrak{h}^{\ast}_{U(1)_2}$. 
The choices of the FI parameter ${\bm \zeta}$ is the same as the ones depicted in Figure~\ref{fig:JK1}. For the cones $C_{1a}, C_{1b}$ and $C_{1c}$ in \eqref{cone1}, the contributions of the poles are
\begin{align}
&\widetilde Z({\bm B}=-{\bm e}_1 + 2{\bm e}_N, {\bm v}={\bm e}_N;C_{1a})\nn\\
&\hspace{5mm}= \sum_{i=1}^{N-1}\frac{1}{2\sinh\frac{a_i - a_N + \epsilon_+}{2}2\sinh\frac{-a_i + a_N + \epsilon_+}{2}}\frac{\prod_{f=1}^{2N}2\sinh\frac{a_i - m_f - \epsilon_+}{2}}{\prod_{1 \leq j \neq i \leq N-1}2\sinh\frac{a_i - a_j}{2}2\sinh\frac{-a_i + a_j + 2\epsilon_+}{2}},\\
&\widetilde Z({\bm B}=-{\bm e}_1 + 2{\bm e}_N, {\bm v}={\bm e}_N;C_{1b}) \nn\\
&\hspace{5mm}= \sum_{i=1}^{N-1}\frac{\prod_{f=1}^{2N}2\sinh\frac{a_i - m_f - \epsilon_+}{2}}{\prod_{1\leq j\neq i \leq N-1}2\sinh\frac{a_i - a_j}{2}2\sinh\frac{-a_i + a_j + 2\epsilon_+}{2}}\frac{1}{2\sinh\frac{a_i - a_N - \epsilon_+}{2}2\sinh\frac{-a_i + a_N + 3\epsilon_+}{2}},\\
&\widetilde Z({\bm B}=-{\bm e}_1 + 2{\bm e}_N, {\bm v}={\bm e}_N;C_{1c}) = \frac{\prod_{f=1}^{2N}2\sinh\frac{a_N - m_f - 2\epsilon_+}{2}}{\prod_{i=1}^{N-1}2\sinh\frac{a_N - a_i - \epsilon_+}{2}2\sinh\frac{-a_N + a_i + 3\epsilon_+}{2}}.
\end{align}
Therefore the Witten index for the  choice ${\bm \zeta}={\bm \zeta}_1$ becomes
\begin{align}
Z({\bm B}=-{\bm e}_1 + 2{\bm e}_N, {\bm v}={\bm e}_N; {\bm \zeta}_1;\epsilon_+) &= \sum_{\alpha=a,b,c}\widetilde Z({\bm B}=-{\bm e}_1 + 2{\bm e}_N, {\bm v}={\bm e}_N;C_{1\alpha}).
\label{Z102v01result1}
\end{align}
For the  choice ${\bm \zeta}={\bm \zeta}_2$, the Witten index is in fact equal to \eqref{Z102v01result1}, namely
\begin{align}
Z({\bm B}=-{\bm e}_1 + 2{\bm e}_N, {\bm v}={\bm e}_N ;{\bm \zeta}_1;\epsilon_+) = Z({\bm B}=-{\bm e}_1 + 2{\bm e}_N, {\bm v}={\bm e}_N ;{\bm \zeta}_2;\epsilon_+). \label{Z102v01result2}
\end{align}
However the choice ${\bm \zeta}={\bm \zeta}_3$ gives a different result from \eqref{Z102v01result1}. The cones which contain ${\bm \zeta}_3$ are given by
\begin{align}
&C_{3a} = \text{Cone}\left[(1, 0), (0, -1)\right],\quad C_{3b} = \text{Cone}\left[(1, -1), (-1, 0)\right], 
\nn\\
&\hspace{3cm}
C_{3c} = \text{Cone}\left[(0, -1), (1, -1)\right]. \label{cone3}
\end{align}
Then the evaluation of the poles corresponding to the cones \eqref{cone3} gives
\begin{align}
&\widetilde Z({\bm B}=-{\bm e}_1 + 2{\bm e}_N, {\bm v} = {\bm e}_N;C_{3a})\nn\\
&\hspace{5mm}= \sum_{i=1}^{N-1}\frac{1}{2\sinh\frac{a_i - a_N - \epsilon_+}{2}2\sinh\frac{-a_i + a_N + 3\epsilon_+}{2}}\frac{\prod_{f=1}^{2N}2\sinh\frac{a_i - m_f - \epsilon_+}{2}}{\prod_{1 \leq j \neq i \leq N-1}2\sinh\frac{a_i - a_j}{2}2\sinh\frac{-a_i + a_j + 2\epsilon_+}{2}},\\
&\widetilde Z({\bm B}=-{\bm e}_1 + 2{\bm e}_N, {\bm v} = {\bm e}_N;C_{3b})\nn\\
&\hspace{5mm}= \sum_{i=1}^{N-1}\frac{\prod_{f=1}^{2N}2\sinh\frac{a_i - m_f + \epsilon_+}{2}}{\prod_{1\leq j\neq i \leq N-1}2\sinh\frac{a_i - a_j + 2\epsilon_+}{2}2\sinh\frac{-a_i + a_j}{2}}\frac{1}{2\sinh\frac{a_i - a_N + 3\epsilon_+}{2}2\sinh\frac{-a_i + a_N - \epsilon_+}{2}},\\
&\widetilde Z({\bm B}=-{\bm e}_1+2{\bm e}_N, {\bm v} = {\bm e}_N;C_{3c}) = \frac{\prod_{f=1}^{2N}2\sinh\frac{a_N - m_f}{2}}{\prod_{i=1}^{N-1}2\sinh\frac{a_N - a_i + \epsilon_+}{2}2\sinh\frac{-a_N + a_i + \epsilon_+}{2}}.
\end{align}
Therefore the Witten index for the  choice ${\bm \zeta}={\bm \zeta}_3$ becomes
\begin{align}
Z({\bm B}=-{\bm e}_1 + 2{\bm e}_N, {\bm v} = {\bm e}_N;{\bm \zeta}_3;\epsilon_+) &= \sum_{\alpha=a,b,c}\widetilde Z({\bm B}=-{\bm e}_1 + 2{\bm e}_N, {\bm v} = {\bm e}_N;C_{3\alpha}). 
\label{Z102v01result3}
\end{align}
The other choices of the FI parameters are in fact related to \eqref{Z102v01result1}, \eqref{Z102v01result2} and \eqref{Z102v01result3} by replacing $\epsilon_+$ with $-\epsilon_+$. 
Namely the 
Witten index
 evaluated with ${\bm \zeta}_{4, 5, 6}$ taken as JK parameters satisfy the relations 
\begin{align}
&Z({\bm B}=-{\bm e}_1 + 2{\bm e}_N, {\bm v} = {\bm e}_N; {\bm \zeta}_4;\epsilon_+) = Z({\bm B}=-{\bm e}_1 + 2{\bm e}_N,  {\bm v} = {\bm e}_N;{\bm \zeta}_1;-\epsilon_+),\\
&Z({\bm B}=-{\bm e}_1+2{\bm e}_N, {\bm v} = {\bm e}_N;{\bm \zeta}_5 ;\epsilon_+) = Z({\bm B}=-{\bm e}_1 + 2{\bm e}_N,  {\bm v} = {\bm e}_N;{\bm \zeta}_2;-\epsilon_+),\\
&Z({\bm B}=-{\bm e}_1 + 2{\bm e}_N, {\bm v} = {\bm e}_N; {\bm \zeta}_6; \epsilon_+) = Z({\bm B}=-{\bm e}_1 + 2{\bm e}_N, {\bm v} = {\bm e}_N;{\bm \zeta}_3;-\epsilon_+)\nn\\
 &
 \hspace{6.2cm}
 = Z({\bm B}=-{\bm e}_1 + 2{\bm e}_N,  {\bm v} = {\bm e}_N;{\bm \zeta}_3;\epsilon_+).
\end{align}

Then comparing $Z({\bm B}=-{\bm e}_1 + 2{\bm e}_N, {\bm v} = {\bm e}_N;{\bm \zeta}_i)\, (i=1, \cdots, 6)$ with \eqref{TFFAFv01result1}-\eqref{TAFFFv01result1} yields the relations 
\begin{align}
\Braket{T_{\Box}}\ast\Braket{T_{\Box}}\ast\Braket{T_{\overline{\Box}}}\Big|_{Z_{\text{mono}}({\bm v}={\bm e}_N)} =&  Z({\bm B}=-{\bm e}_1 + 2{\bm e}_N,  {\bm v} = {\bm e}_N;{\bm \zeta}_1) \nn\\
=& Z({\bm B}=-{\bm e}_1 + 2{\bm e}_N,  {\bm v} = {\bm e}_N;{\bm \zeta}_2),\label{B102opSQM1}\\
\Braket{T_{\Box}}\ast\Braket{T_{\overline{\Box}}}\ast\Braket{T_{\Box}}\Big|_{Z_{\text{mono}}({\bm v}={\bm e}_N)} =&   Z({\bm B}=-{\bm e}_1 + 2{\bm e}_N,  {\bm v} = {\bm e}_N;{\bm \zeta}_3) \nn\\
=& Z({\bm B}=-{\bm e}_1 + 2{\bm e}_N,  {\bm v} = {\bm e}_N;{\bm \zeta}_6),\\
\Braket{T_{\overline{\Box}}}\ast\Braket{T_{\Box}}\ast\Braket{T_{\Box}}\Big|_{Z_{\text{mono}}({\bm v}={\bm e}_N)} =&   Z({\bm B}=-{\bm e}_1 + 2{\bm e}_N,  {\bm v} = {\bm e}_N;{\bm \zeta}_4) \nn\\
=& Z({\bm B}=-{\bm e}_1 + 2{\bm e}_N,  {\bm v} = {\bm e}_N;{\bm \zeta}_5).\label{B102opSQM3}
\end{align}
To summarize, for $ {\bm v} = {\bm e}_N$ the three different orderings for the Moyal product give three different monopole screening contributions, and correspond to different values of $\bm\zeta$ (or, more precisely, different FI-chambers) for the SQM.

\begin{figure}[t]
\centering
\includegraphics[width=10cm]{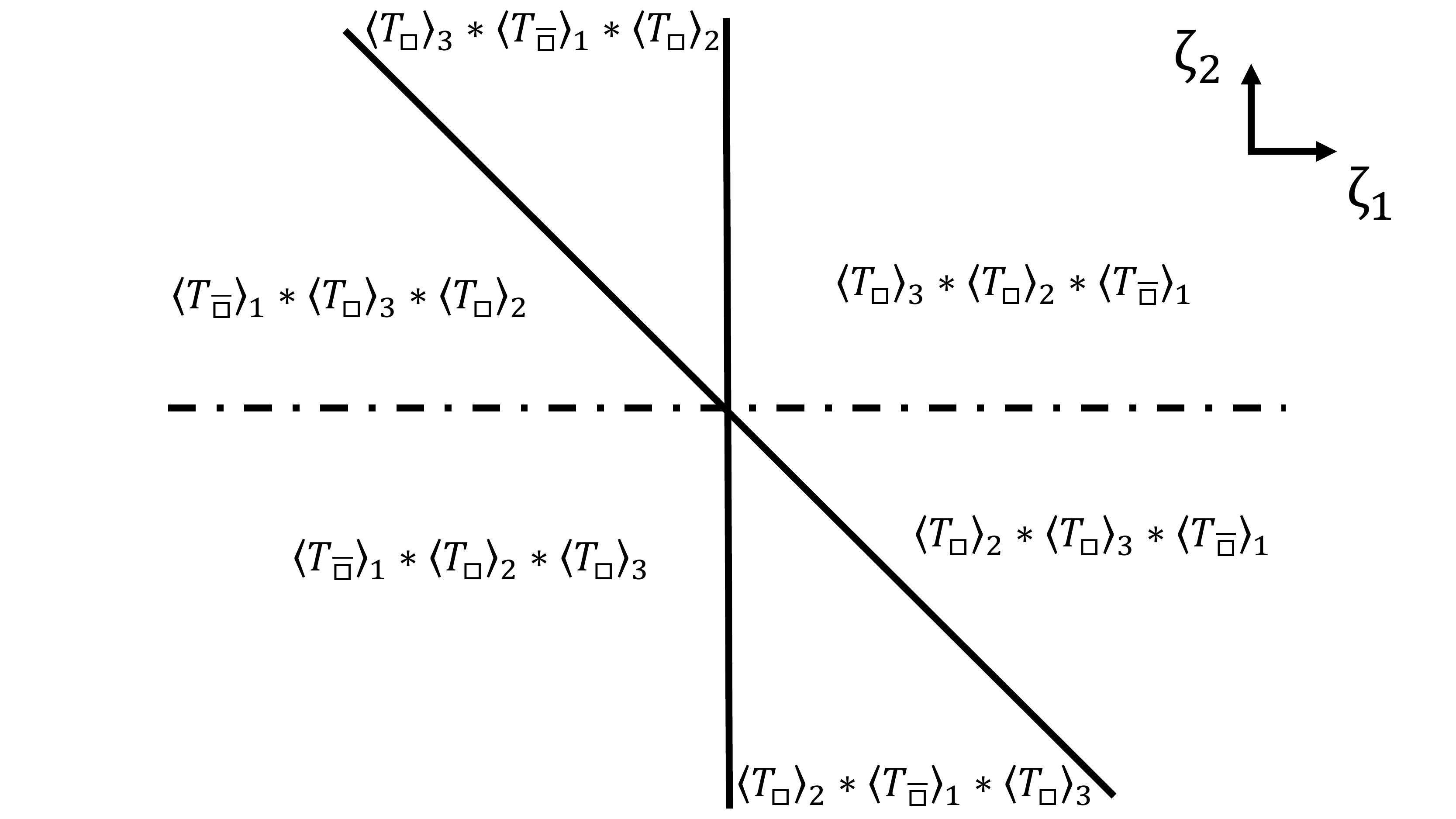}
\caption{The relation between the order of the Moyal product and the space of the FI parameters in the case of the ${\bm v} = {\bm e}_N$ sector.  
Crossing the solid lines causes a discrete change while no such change occurs across the broken lines.
The solid and broken lines separate the plane into six FI-chambers, while the solid ones alone separate the plane into four chambers.
}
\label{fig:phase2}
\end{figure}

The equalities \eqref{B102opSQM1}-\eqref{B102opSQM3} are indeed what we expect from the brane construction of 't~Hooft operators.
To see this
let $\Braket{T_{\overline{\Box}}}_1$, $\Braket{T_{\Box}}_2$, and $\Braket{T_{\Box}}_3$ denote the expectation values of the minimal 't~Hooft operators corresponding to NS5$_i$ ($i=1,2,3$) in Figure~\ref{fig:B102}, respectively.
Then the brane configuration in Figure \ref{fig:B102no2} corresponds to the Moyal product $\Braket{T_{\Box}}_3\ast\Braket{T_{\Box}}_2\ast \Braket{T_{\overline{\Box}}}_1$ because $s_3 > s_2$ and $s_2 > s_1$. In terms of the FI parameters the product corresponds to the region $\zeta_1 > 0$ and $\zeta_2 > 0$. We then increase the value of the parameter~$s_1$ with $s_3 > s_2$ kept.  The ordering of the operators changes when $s_1 = s_2$ and when $s_1 = s_3$. The ordering $\Braket{T_{\Box}}_3\ast \Braket{T_{\overline{\Box}}}_1\ast\Braket{T_{\Box}}_2$ corresponds to the region $\zeta_1 + \zeta_2 > 0$ and $\zeta _1 < 0$, and the ordering $\Braket{T_{\overline{\Box}}}_1\ast\Braket{T_{\Box}}_3\ast \Braket{T_{\Box}}_2$ corresponds to the region $\zeta_1 + \zeta_2 < 0$ and $\zeta _2 > 0$. 
The operator ordering does not change when we exchange $\Braket{T_{\Box}}_2$ with $\Braket{T_{\Box}}_3$. 
Hence we are in the same FI-chamber when we flip the sign of $\zeta_2$ with $s_1$ kept. 
Then, the relations between the ordering and the chamber structure is given by
\begin{align}
\Braket{T_{\Box}}_3\ast\Braket{T_{\Box}}_2\ast\Braket{T_{\overline{\Box}}}_1 \qquad &\longleftrightarrow \qquad \zeta_1 > 0, \quad \zeta_2 > 0,\label{B102relation1}\\
\Braket{T_{\Box}}_2\ast\Braket{T_{\Box}}_3\ast\Braket{T_{\overline{\Box}}}_1 \qquad &\longleftrightarrow \qquad \zeta_1 + \zeta_2 > 0, \quad \zeta_2 < 0,\\
\Braket{T_{\Box}}_2\ast\Braket{T_{\overline{\Box}}}_1\ast\Braket{T_{\Box}}_3 \qquad &\longleftrightarrow \qquad \zeta_1 >  0, \quad \zeta_1 + \zeta_2 < 0,\\
\Braket{T_{\overline{\Box}}}_1\ast\Braket{T_{\Box}}_2\ast\Braket{T_{\Box}}_3 \qquad &\longleftrightarrow \qquad \zeta_1 <  0, \quad  \zeta_2 < 0,\\
\Braket{T_{\overline{\Box}}}_1\ast\Braket{T_{\Box}}_3 \ast\Braket{T_{\Box}}_2\qquad &\longleftrightarrow \qquad \zeta_1+ \zeta_2 <  0, \quad  \zeta_2 > 0,\\
\Braket{T_{\Box}}_3 \ast\Braket{T_{\overline{\Box}}}_1\ast\Braket{T_{\Box}}_2\qquad &\longleftrightarrow \qquad \zeta_1 < 0, \quad  \zeta_1 + \zeta_2 > 0.\label{B102relation6}
\end{align}
The right-hand sides of the relations \eqref{B102relation1}-\eqref{B102relation6} are precisely the FI-chambers that ${\bm \zeta}_i\, (i=1, \ldots, 6)$ in \eqref{B102FI1}-\eqref{B102FI6} belong to. Therefore the relations \eqref{B102relation1}-\eqref{B102relation6} are consistent with the equalities \eqref{B102opSQM1}-\eqref{B102opSQM3}. 
A schematic picture of the relation between the ordering and the chambers in the FI parameter space is given in Figure \ref{fig:phase2}.

\section{$U(N)$ SQCD with $N_F<2N$ flavors}\label{sec:NFsmaller}


The discussions so far have focused on the cases with $N_F = 2N$. In this section we consider the 4d $\mathcal{N}=2$ $U(N)$ gauge theory with $N_F < 2N$ hypermultiplets in the fundamental representation. The expectation values of 't Hooft operators in the cases with $N_F < 2N$ can be obtained by decoupling flavors from the results of $N_F = 2N$. We will see that discrete changes of the expectation values depending on the orderings in products also occur in these cases.  

As a guide for understanding the dependence on $N_F$, we start from $\langle T_\Box \rangle$ and $\langle T_{\overline\Box} \rangle$ in the theory with $2N$ flavors and then integrate out $2N-N_F$ hypermultiplets in the fundamental representation by giving them masses with large absolute values.
Let $k_{\pm}$ be the number of hypermultiplets whose real parts of the masses are sent to $\pm\infty$.
Explicitly,
\begin{equation} \label{eq:decoupling}
\begin{array}{lll}
{\rm Re}[m_f] \rightarrow +\infty & \text{ for } & f\in\{N_F+1,\ldots,N_F+k_+\}, \\
{\rm Re}[m_f] \rightarrow -\infty & \text{ for } & f\in\{N_F+k_+ +1,\ldots,N_F+k_++k_-=2N\}.
\end{array}
\end{equation}
Removing a multiplicative divergent constant%
\footnote{%
Integrating  out a hypermultiplet with $\pm m_f\gg 0$ contributes $(\mp e^{\pm \frac12 m_f} e^{\mp\frac{1}{2}\sum_i a_i})^{1/2}$.
}, 
we obtain from $\langle T_\Box \rangle$ in~(\ref{TF})
\begin{align} \label{T-F-NFsmaller}
\langle T_\Box \rangle \to
\braket{L_1} = 
\sum_{i=1}^N e^{b_i -\frac{k}{4} a_i}
 \left(\frac{\prod_{f=1}^{N_F}2\sinh\frac{a_i -m_f}{2}}{\prod_{1\leq j (\neq \text{ fixed }i) \leq N}2\sinh\frac{a_i - a_j + \epsilon_+}{2}2\sinh\frac{-a_i + a_j + \epsilon_+}{2}}\right)^{\frac{1}{2}},
\end{align}
where we defined $k=k_+-k_-$. Note that $k$ cannot be zero when $N_F$ is an odd number. 
We denoted by $L_1$ the line operator that descends from $T_\Box$ in the theory with $2N$ flavors.
We can apply the same procedure to $\Braket {T_{\overline \Box}}$ in \eqref{TAF}, which gives the expectation value of another line operator $L_2$:
\begin{align} \label{T-AF-NFsmaller}
\Braket {T_{\overline \Box}}\to
\braket{L_2} = 
\sum_{i=1}^N 
e^{-b_i -\frac{k}{4} a_i}
 \left(\frac{\prod_{f=1}^{N_F}2\sinh\frac{a_i -m_f}{2}}{\prod_{1\leq j (\neq \text{ fixed }i) \leq N}2\sinh\frac{a_i - a_j + \epsilon_+}{2}2\sinh\frac{-a_i + a_j + \epsilon_+}{2}}\right)^{\frac{1}{2}}.
\end{align}
When $k=0$, $L_1$ and $L_2$ are simply $T_\Box$ and $T_{\overline{\Box}}$ in the theory with $N_F$ flavors, respectively.

Let us consider the possible discrete changes in the correlator of $L_1$ and $L_2$
 when their ordering changes.
The correlator
is again given by the Moyal product \eqref{moyal} of \eqref{T-F-NFsmaller} and \eqref{T-AF-NFsmaller}.
The difference of the correlators with two orderings is obtained by applying the decoupling procedure above
to \eqref{TFTAFDiff}. 
The end result depends on $k$ and is given by
\begin{align}\label{dyonicDiff}
&\quad
\Braket{L_1} \ast \Braket{L_2} 
-
\Braket{L_2} \ast \Braket{L_1} 
\nonumber
\\
&=
\left\{
\begin{array}{ccc}
2\sinh\epsilon_+(-1)^N e^{\sum_{i=1}^N a_i - \frac{1}{2} \sum_{f=1}^{N_F} m_f } & \text{ for }& k = N_F-2N, \\
0 & \text{ for }&  N_F - 2N  < k < 2N-N_F ,\\
2\sinh\epsilon_+(-1)^{N+N_F+1}e^{- \sum_{i=1}^N a_i + \frac{1}{2} \sum_{f=1}^{N_F} m_f } & \text{ for }& k = 2N -N_F.
\end{array}
\right.
\end{align}
Hence, the discrete changes occurs when $|k| = 2N - N_F$. 

As in the case of $N_F = 2N$, the discrete change~(\ref{dyonicDiff}) can be understood as a wall-crossing phenomenon for the Witten index of an SQM in the case with $N_F < 2N$ flavors. 
The monopole screening contributions in the products
 $\Braket{L_1 \cdot L_2}$ and  $\Braket{L_2 \cdot L_1}$  can be obtained by applying the limit~(\ref{eq:decoupling}) to the SQM for the theory with $2N$ flavors and by
integrating out $2N-N_F$ short Fermi multiplets.
Integrating out a short Fermi multiplet of mass $m$ induces a shift of the Chern-Simons level $\frac{1}{2}\text{sign}\left(\text{Re}[m]\right)$%
\footnote{%
Integrating out an $\mathcal{N}=(0, 4)$ hypermultiplet of mass $a$ shifts the Chern-Simons level by $-\text{sign}\left(\text{Re}[a]\right)$. 
}.
The presence of a Chern-Simons term affects the form of the Witten index of an SQM by the factor~\eqref{CSterm}. 
When we decouple the matter in the same way as we obtained \eqref{T-F-NFsmaller} and \eqref{T-AF-NFsmaller}, the induced Chern-Simons level for the corresponding SQM is $\kappa = \frac{k}{2}$. 
Therefore, the Witten index for the monopole screening contribution in the product
$\Braket{L_1} * \Braket{L_2}$ or $\Braket{L_2} * \Braket{L_1}$
 is given by\footnote{We can obtain the same result by applying the decoupling procedure directly to \eqref{Z101}.}
\begin{align}
Z_{N_F, k}(\zeta) = 2\sinh\epsilon_+\oint_{JK(\zeta)}\frac{d\phi}{2\pi i}\frac{ e^{-\frac{k}{2}\phi} \prod_{f=1}^{N_F}2\sinh\frac{\phi - m_f}{2}}{\prod_{i=1}^{N}2\sinh\frac{\phi - a_i + \epsilon_+}{2}2\sinh\frac{-\phi + a_i + \epsilon_+}{2}}
\end{align}
with $\zeta>0$ or $\zeta<0$ respectively.
The factor $e^{-\frac{k}{2}\phi}$ is due to the fact that the SQM has the Chern-Simons level $\kappa = \frac{k}{2}$. 
The potential discrete change $Z_{N_F, k}(\zeta > 0) - Z_{N_F, k}(\zeta < 0)$ can be attributed to the contributions from the poles at $u=0$ and $u=\infty$ when the integral is written in terms of $u=e^{\phi}$;
the sum of the residues gives the result \eqref{dyonicDiff}. 

Indeed when $N_F$ is odd a half-integer Chern-Simons level is required to cancel the global anomaly associated with the $N_F$ short Fermi multiplets in the fundamental representation~\cite{Hori:2014tda}.
Integrating out the Fermi multiplets induces an extra half-integer Chern-Simons level.  For the final theory to have an integer Chern-Simons level as necessary for gauge invariance, the original theory needs to have a half-integer level.

\section{Conclusion and discussion}\label{sec:discussion}

In this paper we investigated the correspondence between the ordering of minimal 't~Hooft operators and the choice of a chamber in the space of FI parameters of the SQMs that describe monopole screening, in the case of the 4d $\mathcal{N}=2$ $U(N)$ gauge theory with $2N$ flavors.
The correspondence is inferred from the brane realization of the product of minimal 't~Hooft operators. 
There are two kinds of minimal operators, $T_\Box$ and $T_{\overline\Box}$.
The correspondence naturally leads to the conjecture that the monopole screening contributions read off from the Moyal product of the vevs of the minimal operators, which is easy to compute, should match the Witten indices of the SQMs.
We confirmed the conjecture by explicitly evaluating the Witten indices for all the possible combinations and the orderings of two and three minimal 't~Hooft operators.

For the product of two operators, the case of one $T_{\Box}$ and one $T_{\overline{\Box}}$ was studied earlier in~\cite{Assel:2019iae}.
In addition we analyzed the product of $T_{\Box}$ and $T_{\Box}$. 
Since the operator ordering is unique, we do not expect wall-crossing in the Witten index. Indeed the explicit evaluation of the Witten index, which involves the JK residues of degenerate poles, yielded the same result regardless of the sign of the FI parameter. 

For the product of three operators we have two cases. One case involves three~$T_{\Box}$'s, which has two sectors for the monopole screening. We found that the Witten index for the monopole screening contributions gives the same result as expected for the both sectors. The other case involves two $T_{\Box}$'s and one $T_{\overline{\Box}}$, which also have two sectors for the monopole screening. One sector is characterized by ${\bm v} = -{\bm e}_1 + {\bm e}_{N-1} + {\bm e}_N$. In this case, the different orderings give the same result. This is because the monopole screening relates only~$T_{\Box}$ with~$T_{\Box}$, which is also reflected into the fact that the corresponding SQM has only one FI parameter related to the difference between the position of $T_{\Box}$ and that of the other~$T_{\Box}$. The evaluation of the Witten index indeed had nothing to do with the sign of the FI parameter. The other sector is given by ${\bm v} = {\bm e}_N$. In this case all three different orderings yield different results. The corresponding SQM has two FI parameters. We evaluated the Witten index for all the possible (in total six) FI-chambers. We found that the FI parameter space is divided into four chambers and each chamber corresponds to one of the three orderings, as summarized in Figure~\ref{fig:phase2}.

It is natural to ask if, in the set-up of Section~\ref{sec:proposal} (where $N_F=2N$) and for a given charge of the form $\bm{B}= -n_- \bm{e}_1 + n_+ \bm{e}_N$ ($n_\pm\geq 1$), one can write down a formula that determines the $\bm{v}$'s for which the corresponding SQMs exhibit wall-crossing.
The examples we studied explicitly suggest the following.
For wall-crossing to occur in the sector specified by $\bm{v}$, the corresponding brane configuration must include a set of finite D1-branes that can reconnect to form a single D1-brane stretched between an NS5-brane on the left, and another on the right, of the whole stack of D3- and D7-branes.
In terms of the SQM quiver that results from this configuration via Hanany-Witten transitions, this means that the quiver includes a total of $2N$ $\mathcal{N}=(0,4)$ short Fermi multiplets coupled to various gauge nodes, and that all of the $\ell-1$ potential gauge nodes are actually present, {\it i.e.}, the gauge group is $\prod_{s=1}^{\ell-1} U(n_a)$ with $n_a\geq 1$ for all $a=1,\ldots,\ell-1$
(see the discussion at the end of Section~\ref{sec:SQMbrane}).
In terms of $\bm{v}$ the condition seems to be that $\sum_i |v_i| < \sum_i |B_i|$.
It would be interesting to decide for $N_F=2N$ if this is the necessary and sufficient condition for wall-crossing to actually occur.

It seems possible to make statements similar to conjectures (i) and (ii) in the introduction for dyonic line operators.
Our analysis in Section~\ref{sec:NFsmaller} and the study of dyonic operators in~\cite{Assel:2019iae} suggest that there exist SQMs that capture magnetic screening contributions for dyonic operators, and that they involve a 1d Chern-Simons coupling (Wilson loop).
In the special case where dyonic operators arise from integrating out hypermultiplets in the presence of 't~Hooft operators,  (\ref{dyonicDiff}) can be interpreted wall-crossing for dyonic operators.
It would be interesting to study wall-crossing for more general dyonic operators.

The assignment of electric charges to the line operators $L_1$ and $L_2$ in Section~\ref{sec:NFsmaller} is subtle, especially for $N_F$ odd.
When $k$ is a non-zero even integer, the vevs given in~(\ref{T-F-NFsmaller}) and~(\ref{T-AF-NFsmaller}) indicate that the line operators $L_1$ and $L_2$ obtained from $T_\Box$ and $T_{\overline\Box}$ by integrating out hypermultiplets are dyonic.
We note that integrating out hypermultiplets can modify the theta angle $\theta$; moreover the resulting value $\theta$ mod $2\pi$ depends on $k=k_+-k_-$, where $k_+$ ($k_-$) is the number of hypermultiplets integrated out with a positive (negative) large mass.
Indeed we have
\begin{equation}
\begin{aligned}
\frac{
\det( i D \hspace{-2.7mm} /+ M)
}{
\det( i D \hspace{-2.7mm} /- M)
}
= \left(\frac{M}{-M}\right)^{n_0^+ + n_0^-}
\hspace{-1mm}
{\prod_i}' \frac{(\lambda_i +M)(-\lambda_i +M)}{(\lambda_i-M)(-\lambda_i-M)}
=\exp \left(
\pi i
\frac{1}{8\pi^2}
\int {\rm Tr} \, F\wedge F
\right) ,
\end{aligned}
\end{equation}
where $n_0^\pm$ are the numbers of zero-modes with $\pm$ chiralities and $\prod'_i$ indicates the product over the pairs $(\lambda_i,-\lambda_i)$ of eigenvalues of the Dirac operator $ i D \hspace{-2.7mm}/$.
The last expression is obtained by writing $(-1)^{n_0^+ + n_0^-}=(e^{\pi i})^{n_0^+ - n_0^-=\text{index}}$ in terms of the integral using the Atiyah-Singer index theorem.
Thus integrating out a Dirac fermion with a large (time-reversal preserving) mass $M>0$ yields the value of $\theta$ that differs by $+\pi$ mod $2\pi$ from the value of $\theta$ that arises from the mass $-M<0$.
The relative shift is $\pi = (+\pi/2) - (-\pi/2)$.
It is then tempting to attribute the $+\pi/2$ ($-\pi/2$) to integrating out a fermion of positive (negative) mass, and to regard the electric charges of $L_1$ and $L_2$ as due to the Witten effect caused by a shift in $\theta$ ~\cite{Witten:1979ey}.
The interpretation superficially works if we consider only $L_2$.
A shift $\theta^\text{new} = \theta + k\pi/2$ in~(\ref{T-F-NFsmaller}) modifies $b_i$ to $b_i^\text{new}= b_i + k a_i/4$, so the $-k a_i/4$ in~$\Braket{L_2}$ of~(\ref{T-AF-NFsmaller}) is absorbed into $-b_i^\text{new}$.
But in $\Braket{L_1}$ of~(\ref{T-F-NFsmaller}) there remains $-k a_i/2$, so the apparent electric charge of $L_1$ cannot be explained by the Witten effect.
Moreover the shift of $\theta$ is only defined mod $2\pi$, {\it i.e.}, we have $+\pi = -\pi$ mod $2\pi$%
\footnote{%
This is in contrast with the shift by $\frac{1}{2} \text{sign}(\text{mass})$ of a Chern-Simons level induced by integrating out a massive Dirac fermion in odd dimensions.
There is no periodic identification for the Chern-Simons level.
}.
Another shift $\theta^\text{new} = \theta - k\pi/2$ would give $b_i^\text{new}= b_i - k a_i/4$ and absorb the $-k a_i/4$ in~$\Braket{L_1}$ of~(\ref{T-F-NFsmaller}) into $b_i^\text{new}$, but then $\Braket{L_2}$ of~(\ref{T-AF-NFsmaller}) is left with $-k a_i/2$.
In neither choice the electric charges are completely absorbed into $b_i^\text{new}$ by the Witten effect, and neither choice is more natural.
On top of this, $k$ is odd for $N_F$ odd, so the term $-k a_i/2$ that remains in one of $\Braket{L_1}$ and $\Braket{L_2}$ seemingly corresponds to an electric charge not quantized in a gauge invariant way.
Note that this is true even for $N=1$, {\it i.e.}, the $U(1)$ gauge group.
We leave the investigation of these interesting issues to the future.

\section*{Acknowledgements}
We thank B.~Assel and A.~Sciarappa for useful comments on a draft.
The work of H.H. is supported in part by JSPS KAKENHI Grant Number JP18K13543, and that of T.O. by Grant Number JP16K05312.
The work of Y.Y. is supported in part by JSPS KAKENHI Grant Number JP16H06335 and also by World Premier International Research Center Initiative (WPI), MEXT Japan.

\appendix

\section{Useful facts and formulae}\label{sec:formulas}


The expectation value of an 't~Hooft line operator with a magnetic charge ${\bm B}$  in a 4d $\mathcal{N}=2$ gauge theory is in general written as \cite{Alday:2009fs, Drukker:2009id, Ito:2011ea}
\begin{align}
\Braket{T_{\bm B} } = \mathop{\sum_{\bm{v} \in \bm{B}+ \Lambda_{\text{cort}}}}_{|\bm v|\leq |\bm B|} e^{\bm{v}\cdot \bm{b}} Z_\text{1-loop}(\bm{a}, {\bm m}, \epsilon_+;\bm{v}) Z_\text{mono}({\bm a}, {\bm m}, \epsilon_+;\bm B,\bm v), \label{TB.general} 
\end{align}
where ${\bm v}$ labels monopole screening sectors. The definitions of the parameters are explained in Section~\ref{sec:Moyal}. Namely the expectation value consists of two parts $Z_\text{1-loop}(\bm{a}, {\bm m}, \epsilon_+;\bm{v})$ and $Z_\text{mono}({\bm a}, {\bm m}, \epsilon_+ ;\bm B,\bm v)$. The first part, $Z_\text{1-loop}(\bm{a}, {\bm m}, \epsilon_+ ;\bm{v})$, can be computed by the one-loop determinant of the 4d path integral. The second part, $Z_\text{mono}({\bm a}, {\bm m}, \epsilon_+;\bm B,\bm v)$, may be calculated by applying a localization method to the 4d path integral \cite{Ito:2011ea} but it can be also expressed as a Witten index
 of a supersymmetric quantum mechanics \cite{Brennan:2018yuj, Brennan:2018moe, Brennan:2018rcn}. In this appendix, we summarize formulae for computing the one-loop determinant of the 4d path integral and the Witten index of the SQM.

\subsection{One-loop determinants of the 4d theory}
\label{sec:locfor4d}
 The one-loop determinant that appears in the expectation value of an 't~Hooft operator \eqref{TB.general} is the product of     
 the contribution $Z_{\text{1-loop}}^{\text{vm}} ({\bm a}, \epsilon_+ ;\bm{v} )$ of a vector multiplet and the contributions $Z_{\text{1-loop}}^{\text{hm}}( \bm{a}, m, \epsilon_+ ;\bm{v})$ of hypermultiplets, namely, 
\begin{align}
Z_\text{1-loop}(\bm{a}, {\bm m}, \epsilon_+ ;\bm{v})=Z_{\text{1-loop}}^{\text{vm}} ( {\bm a}, \epsilon_+; \bm{v} ) \prod_f Z_{\text{1-loop}}^{\text{hm}}( \bm{a}, m_f, \epsilon_+ ;\bm{v}), 
\end{align}
where the product $\prod_f$ runs over  the hypermultiplets and $m_f$'s are the mass parameters. 
 Closed formulas are available for the one-loop determinants 
 \cite{Ito:2011ea}. The one-loop determinant of  the vector multiplet  for gauge group $G$ is given by  
\begin{align}
Z_{\text{1-loop}}^{\text{vm}}  ({\bm a}, \epsilon_+ ;\bm{v} )
= \prod_{ {\bm \alpha} \in \text{root}, |{\bm \alpha}\cdot \bm{v}| \neq 0}\prod_{k=0}^{|{\bm \alpha}\cdot \bm{v}| - 1}\left[2\sinh\left(\frac{{\bm \alpha}(\bm{a}) + \left(|{\bm \alpha}(\bm{v})|- 2k\right)\epsilon_+}{2}\right)\right]^{-\frac{1}{2}} , \label{1loopvm}
\end{align}
where 
 the symbol ``$\text{root}$'' stands for the set of the roots of $G$.
On the other hand, the one-loop determinant of the hypermultiplet in a representation $R_G$ of $G$ with a mass parameter $m$ is given by 
\begin{align}
Z_{\text{1-loop}}^{\text{hm}}  ({\bm a}, m, \epsilon_+ ;\bm{v} )= 
\mathop{\prod_{{\bm \rho} \in \Delta(R_G)}}_{ |{\bm \rho} \cdot \bm{v}|\neq 0}
\prod_{k=0}^{|{\bm \rho}\cdot \bm{v}| - 1}\left[2\sinh\left(\frac{{\bm \rho}(\bm{a}) - m + \left(|{\bm \rho}(\bm{v})|-1 - 2k\right)\epsilon_+}{2}\right)\right]^{\frac{1}{2}}, \label{1loophm}
\end{align}
where  $\Delta (R_G)$ is the set of the weights  in $R_G$.

\subsection{Review of the Witten index of the SQM}
\label{sec:locforSQM}

The Witten index
of an SQM obtained by a dimensional reduction of a 2d $\mathcal{N}=(0, 4)$ gauge theory may be computed by the localization method in \cite{Hwang:2014uwa, Cordova:2014oxa, Hori:2014tda}.
The formal definition of the Witten index is 
\begin{align}
Z_{\text{SQM}} = \text{Tr}_{\mathcal{H}}(-1)^Fe^{-\beta H + \epsilon_+ J_+ + {\bm a}\cdot {\bm J}_a+ {\bm m}\cdot{\bm J}_F}, \label{windex}
\end{align}
where $\mathcal{H}$ is the Hilbert space of the SQM, $H$ is the Hamiltonian, $J_+$ is   the Cartan generator of the $SU(2)$ R-symmetry.  Collectively~${\bm J}_a$ denotes the Cartan part of the generators of the flavor symmetries that act on $\mathcal{N}=(0,4)$ hypermultiplets,
 and~${\bm J}_F$ denotes the Cartan part of the generators of the flavor symmetries that act on $\mathcal{N}=(0,4)$ short Fermi multiplets.
When SQM is a quiver quantum mechanics associated with monopole screening, ${\bm a}$ is identified with the Cartan part of  complexified gauge holonomies $a_i$  \eqref{aandb} and 
${\bm m}$ is identified with the masses~$m_f$ of the hypermultiplets in four dimensions.

\subsubsection{One-loop determinants}
We summarize the result of the localization computation. For our purpose, it is enough to focus on an SQM which consists of $U(n)$ vector multiplets and other multiplets in the fundamental or the bi-fundamental representation of unitary gauge groups. The Witten index is given by an integral of a product of contributions from each multiplet. Each $\mathcal{N} = (0, 2)$ multiplet yields one factor in the integrand. The decomposition of an $\mathcal{N}=(0, 4)$ multiplet relevant in our computation into $\mathcal{N}= (0, 2)$ multiplets is given by
\begin{equation}
\begin{aligned}
\text{$\mathcal{N}=(0,4)$ vector} & \quad \rightarrow \quad
\text{$\mathcal{N}=(0,2)$ vector}\quad 
\oplus\quad
\text{$\mathcal{N}=(0,2)$ Fermi} \,,
\\
\text{$\mathcal{N}=(0,4)$ hyper} & \quad \rightarrow \quad
\text{$\mathcal{N}=(0,2)$ chiral}\quad
\oplus\quad
\text{$\mathcal{N}=(0,2)$ chiral} \,,
\\
\text{$\mathcal{N}=(0,4)$ short Fermi} & \quad \rightarrow \quad
\text{$\mathcal{N}=(0,2)$ Fermi} \,.
\end{aligned}
\end{equation}
We consider complex fields $\phi_k = \varphi_k + iA_k, k=1, \cdots, n$ where $\varphi_k$ is the scalar in the 1d vector multiplet of the Cartan subalgebra $\mathfrak{h}$ and $A_k$ is the holonomy of the gauge field of the Cartan subalgebra $\mathfrak{h}$ on the circle. $n$ is the rank of a unitary gague group. Then, the contribution from an $\mathcal{N}=(0, 4)$ $U(n)$ vector multiplet is 
\begin{align}
Z_{\text{vec}}({\bm \phi}, \epsilon_+) = \left[\prod_{i=1}^n\frac{d\phi_i}{2\pi i}\right]\left[\prod_{1 \leq i \neq j \leq n}2\sinh\frac{\phi_i - \phi_j}{2}\right]\left[\prod_{1 \leq i, j \leq n}2\sinh\frac{\phi_i - \phi_j + 2\epsilon_+}{2}\right]. \label{Zvec}
\end{align}
When the Cherns-Simons level is $\kappa$, there is an additional factor
\begin{align}
Z_{\text{cs}}({\bm \phi}, \kappa) =  e^{-\kappa\sum_{i=1}^n\phi_i}. \label{CSterm}
\end{align} 
The contribution from an $\mathcal{N}=(0, 4)$ hypermutliplet in the bi-fundamental representation of $U(n) \times U(n')$ is given by
\begin{align}
Z_{\text{hyp}}({\bm \phi}, {\bm \phi}', \epsilon_+) = \frac{1}{\prod_{i=1}^{n}\prod_{j=1}^{n'}2\sinh\frac{\phi_i - \phi'_j + \epsilon_+}{2}2\sinh\frac{-\phi_i + \phi'_j + \epsilon_+}{2}} .\label{Zhyp}
\end{align}
Finally, the contribution from an $\mathcal{N}=(0, 4)$ short Fermi multiplet in the fundamental representation of $U(n)$ and in the anti-fundamental representation of $U(n')$ is 
\begin{align}
Z_{\text{Fermi}}({\bm \phi}, {\bm \phi}', \epsilon_+) = \prod_{i=1}^{n}\prod_{j=1}^{n'}2\sinh\frac{\phi_i - \phi'_j + \epsilon_+}{2}. \label{ZFermi}
\end{align}
For \eqref{Zhyp} and \eqref{ZFermi} a factor in $U(n) \times U(n')$ can be a flavor symmetry group. 
For example, when $U(n')$ is a flavor symmetry in \eqref{Zhyp}, $\phi'_i$ are interpreted as the chemical potentials $a_i$. 
 When  $U(n')$ is a flavor symmetry in \eqref{ZFermi}, $\phi'_i$ are interpreted as the chemical potentials~$m_f$.
The Witten index of the SQM is given by an integral of the product of the contributions~\eqref{Zvec}-\eqref{ZFermi}
\begin{align}
Z_{\text{SQM}} = \frac{1}{|W|}\oint \prod Z_{\text{cs}}  \prod Z_{\text{vec}} \prod Z_{\text{hyp}} \prod Z_{\text{Fermi}}, \label{partfn}
\end{align}
where $|W|$ is the order of the Weyl group of the gauge group and the products are taken over all the multiplets.

Although the FI parameters of the 1d gauge theory do not appear in the integrand of~(\ref{partfn}), the Witten index in fact depends on them in a subtle way \cite{Hwang:2014uwa, Cordova:2014oxa, Hori:2014tda}.
To illustrate the subtlety let us consider a
$U(1)$ gauge theory such as the one specified by the quiver in Figure~\ref{fig:quiver02}. 
The $U(1)$ gauge theory has a single FI parameter $\zeta$. 
In the convention of~\cite{Hori:2014tda} the poles associated to positively (negatively) charged fields contribute when $\zeta > 0$ ($\zeta<0$).
In general 
the Witten indices for $\zeta>0$ and $\zeta<0$ take different values.

\subsubsection{JK residues for non-degenerate and degenerate poles}\label{sec:JK}

The situation becomes more complicated for a higher rank gauge theory or a quiver gauge theory with multiple gauge nodes. 
To evaluate the integral~\eqref{partfn} for such a theory we use the Jeffrey-Kirwan (JK) residue prescription, which we review here.

The {\it charge vectors} ${\bm Q}_i \in \mathfrak{h}^{\ast}$ are the weights of the gauge group in the representations carried by the $\mathcal{N}=(0,2)$ chiral multiplets in the theory;
they appear as 
  $\prod Z_{\text{hyp}} = \prod_i ( 2 \sinh \frac{1}{2} \left( {\bm Q}_i ( {\bm \phi})-{c}_i \right) )^{-1} $ in \eqref{partfn}. 
Here $c_i$ is a linear function of $\epsilon_+$ and $a_j$, ${\bm \phi } \in \mathfrak{h}$ collectively denotes all $\phi_j$'s, and ${\bm Q}_i ( {\bm \phi}) $ is the inner product of ${\bm Q}_i$ and  ${\bm \phi}$. 
For each value of $i$ the locus ${\bm Q}_i ({\bm \phi} )-{c}_i=0$ in the ${\bm \phi}$-space is called a {\it singular hyperplane}. 
Suppose that exactly $l$ hyperplanes, given (after a $\bm{\phi}_*$-dependent relabeling of the $i$'s) by ${\bm Q}_1({\bm \phi}-{\bm \phi}_{\ast})=0, \ldots, {\bm Q}_l({\bm \phi}-{\bm \phi}_{\ast})=0$, intersect at a point ${\bm \phi}={\bm \phi}_{\ast}$; we necessarily have $l\geq \dim \mathfrak{h}=:n$.
We set ${\bm Q}_{\ast}(\bm{\phi}_*)= \{{\bm Q}_1, \ldots, {\bm Q}_l \}$. 
We will often write~$ {\bm Q}_{\ast}$ for~${\bm Q}_{\ast}(\bm{\phi}_*)$, keeping the $\bm{\phi}_*$-dependence implicit.
We assume that ${\bm Q}_{\ast}$ satisfies the projectivity condition, which states that all the elements of ${\bm Q}_{\ast}$ are contained in a half space of $\mathfrak{h}^*$. 
All the relevant poles of the SQMs treated in the paper satisfy the projectivity condition.
When
$l=n$ the intersection point (pole) $\bm{\phi}_*$  is called non-degenerate. 
When $l>n$ it is called degenerate.

At a non-degenerate pole $\bm{\phi}_*$
the JK residue
is defined as \cite{Benini:2013xpa}
\begin{align}
&\mathop{\text{JK-Res}}_{{\bm \phi}
= {\bm \phi}_{\ast}}({\bm Q}_{\ast}, {\bm \eta}) \frac{d\phi_1 \wedge \cdots \wedge d\phi_n}{{\bm Q}_{1}({\bm \phi} - {\bm \phi}_{\ast}) \cdots  {\bm Q}_{n}({\bm \phi} - {\bm \phi}_{\ast})}
\nn\\
&\qquad
 = 
\begin{cases}
\left|\det({\bm Q}_1, \ldots, {\bm Q}_n) \right|^{-1} \;\; &\text{if}\;\;{\bm \eta } \in \text{Cone}[{\bm Q}_1, \ldots, {\bm Q}_n],\\
0& \text{otherwise},
\end{cases} \label{JK-def-non-degenerate}
\end{align}
where $\text{Cone}[{\bm Q}_1, \ldots, {\bm Q}_n]= \{
\sum_{i=1}^nz_i{\bm Q}_i | z_i > 0 \text{ for } i=1, \ldots, n\}$.

 At a degenerate pole ${\bm \phi}_*$ we use the constructive ``definition'' of the JK residue, which we review following~\cite{Benini:2013xpa}. 
First, let $\Sigma {\bm Q}_{\ast} $ be the set of partial sums of elements in ${\bm Q}_{\ast}$ defined as
\begin{align}
{\Sigma {\bm Q}_{\ast}  } = \Bigl\{ \sum_{i \in \pi } {\bm Q}_{i} \Big| \pi \subset \{1, \ldots, l \}  \Bigr\}. 
\end{align}
We assume that
${\bm \eta} \in \mathfrak{h}^* $ satisfies the {\it strong regularity condition} 
\begin{align}
{\bm \eta} \notin {\rm Cone}_{\rm sing} [\Sigma {\bm Q}_{\ast}].  \label{eq:strongref}
\end{align} 
Here ${\rm Cone}_{\rm sing} [\Sigma {\bm Q}_{\ast}]$ is the union of all the cones  spanned by $(n-1)$ elements of $\Sigma {\bm Q}_{\ast}$. 
Second, let  ${\cal F L} ({\bm Q}_{\ast})$ be the set of flags  
\begin{align}
F=[\{0 \}=F_{0} \subset F_1 \subset \cdots \subset F_{n} = \mathfrak{h}^{\ast}], \quad \dim F_i = i 
\end{align}
such that ${\bm Q}_{\ast}$ contains a basis of $F_i$ for $i=1,2, \ldots, n$; we let $\mathfrak{B}(F)=\{ {\bm Q}_{j_1}, \cdots, {\bm Q}_{j_n} \}$
be the ordered set whose first $i$ elements  form a basis of $F_i$ for $i=1, \ldots, n$.
For each flag $F \in {\cal F L} ({\bm Q}_{\ast})$ the iterated residue ${\rm Res}_F$ of an $n$-form $ \omega$ is defined by 
\begin{align}
{\rm Res}_{F} \, \omega= \oint_{\tilde{\phi}_{j_n}=0} \frac{d \tilde{\phi}_{j_n}} {2 \pi i}  \cdots  \oint_{\tilde{\phi}_{j_1}=0}  \frac{d \tilde{\phi}_{j_1}}{2 \pi i}\tilde{\omega}_{j_1 \cdots j_n } , 
\end{align}
where $\tilde{\phi}_i= {\bm Q}_i ({\bm \phi}-{\bm \phi}_{\ast})$ and $\omega = \tilde{\omega}_{j_1 \cdots j_n } d \tilde{\phi}_{j_1} \wedge \cdots \wedge d \tilde{\phi}_{j_n}$.
Third, for any flag $F \in {\cal F L} ({\bm Q}_{\ast})$ let us introduce the vectors
\begin{align}
 {\bm \kappa}^{F}_i=\sum_{k=1}^i  {\bm Q_{j_k}}
\end{align}
and the closed cone
\begin{align}
\mathfrak{s}^+ (F, {\bm Q}_{\ast}) := \sum_{i=1}^n \mathbb{R}_{\ge 0} {\bm \kappa}^{F}_{i}.
\end{align}
Using $\mathfrak{s}^+ (F, {\bm Q}_{\ast}) $ we also define 
\begin{align}
{\cal F L}^+ ({\bm Q}_{\ast}, {\bm \eta}) :=  \{ F\in {\cal F L} ({\bm Q}_{\ast})| {\bm \eta} \in \mathfrak{s}^+ (F, {\bm Q}_{\ast})\}.
\end{align}
Then the JK residue at the pole ${\bm \phi}={\bm \phi}_*$ is defined by
\begin{align}
\mathop{{\rm JK} \mathchar `- {\rm Res} }_{{\bm \phi}={\bm \phi}_*} ({\bm Q}_{\ast}, {\bm \eta} )
 =\sum_{F \in {\cal F L}^+ ({\bm Q}_{\ast},{\bm \eta}) } \nu (F) {\rm Res}_F  ,
 \label{eq:consJKres}
\end{align}
where $\nu (F) $ is defined as $\nu(F):= {\rm sign} \det ({\bm \kappa}^{F}_{1},\cdots {\bm \kappa}^{F}_{n})$ with the understanding that ${\rm sign}\, x$ is $+1$ for $x>0$, 0 for $x=0$, and $-1$ for $x<0$.

For gauge group $U(n)$ the FI parameter $\zeta$ determines an element  $\widetilde{\bm{\zeta}}=\zeta \sum_{i=1}^N \bm{e}_i 
\in \mathbb{R}^N\simeq \mathfrak{h}^{\ast}$~\cite{Hori:2014tda}.
For a product gauge group $U(n_1)\times U(n_2)\times\ldots\times U(n_L)$ the FI parameters $\bm{\zeta}=(\zeta_1,\zeta_2,\ldots,\zeta_L)$ determine an element $\widetilde{\bm \zeta}:=\sum_{a=1}^L \zeta_a 
( \bm{e}_{\tilde{n}_{a-1}+1} + \ldots +\bm{e}_{\tilde{n}_a} )
\in\mathfrak{h}^*$, 
where 
$\{\bm{e}_{\tilde{n}_{a-1}+1} ,\ldots,\bm{e}_{\tilde{n}_a}\}$
is an orthonormal basis of $\mathfrak{h}^*_{U(n_a)}\simeq \mathbb{R}^{n_a}$, $\tilde{n}_a = n_1+\ldots+n_{a}$, and $\tilde{n}_0=0$.
An example appears in~(\ref{zeta-tilde-U2U1}).

When all the zero-dimensional intersections of the singular hyperplanes are non-degenerate, the Witten index can be written as~\cite{Hwang:2014uwa, Cordova:2014oxa, Hori:2014tda}
\begin{align}
&
Z_{\text{SQM}} 
\nn\\
=& \frac{1}{|W|}\oint_{JK({\bm \zeta})}   \prod Z_{\text{cs}} \prod Z_{\text{vec}} \prod Z_{\text{hyp}} \prod Z_{\text{Fermi}} \label{partitionfunction} \\ 
:=&\frac{1}{|W|} \sum_{{\bm \phi}_* } \mathop{{\rm JK} \mathchar `- {\rm Res} }_{{\bm \phi}={\bm \phi}_*} ({\bm Q}_{\ast}, {\bm \eta} = { \widetilde{\bm \zeta}} ) 
\prod Z_{\text{cs}} \prod Z_{\text{vec}}^{\prime} \prod Z_{\text{hyp}} \prod Z_{\text{Fermi}} \,
 d \phi_1 \wedge \cdots \wedge  d \phi_{\mathrm{dim} \;\mathfrak{h} },   \nn
\end{align}
where $\prod Z_{\text{vec}}^{\prime}$ is defined by removing $\prod_{j=1}^{\mathrm{dim} \;\mathfrak{h}} (d\phi_j/2\pi i)$ from $\prod Z_{\text{vec}}$ and the JK-residues are computed according to~(\ref{JK-def-non-degenerate}).
We emphasize that the simple relation~(\ref{partitionfunction}) only holds when we set the JK parameter~$\bm{\eta}$ to the FI-parameter~$\widetilde{\bm{\zeta}}$.

When some of the zero-dimensional intersections are degenerate and when they satisfy the strong regularity condition~(\ref{eq:strongref}), we follow a proposal of~\cite{Hori:2014tda} and compute $Z_{\text{SQM}}$ using~(\ref{partitionfunction}) by applying the constructive definition~(\ref{eq:consJKres}) to the degenerate poles.
We take the summation in  \eqref{partitionfunction} over all the degenerate poles ${\bm \phi}_*$ with ${\bm \eta} \in \mathfrak{s}^+ (F, {\bm Q}_{\ast} (\bm{\phi}_*))$ {for some $F\in {\cal F L}^+ ({\bm Q}_{\ast} (\bm{\phi}_*),{\bm \eta})$
and also over all the non-degenerate poles ${\bm \phi}_*$ with 
 ${\bm \eta} \in \text{Cone}[{\bm Q}_1, \cdots, {\bm Q}_n]$.

When some of the zero-dimensional intersections are degenerate and when some of them violate the strong regularity condition~(\ref{eq:strongref}), we compute $Z_{\text{SQM}}$ for~${\bm \zeta}$ in the interior of an FI-chamber~(\ref{def-FI-chamber}) as follows.
We use almost the same formula~(\ref{partitionfunction}) and apply the constructive definition~(\ref{eq:consJKres}) to the degenerate poles and sum the JK-residues as in the previous paragraph, but at a degenerate pole that violates the strong regularity condition~(\ref{eq:strongref}), we use as $\bm{\eta}$ not $\widetilde{\bm \zeta}$ itself but a vector $\widetilde{\bm\zeta}'$ that is obtained by infinitesimally shifting~$\widetilde{\bm \zeta}$ and that satisfies the strong regularity condition.
An example is given in~(\ref{eq:shiftFI}).

In this paper we use the terminology
\begin{align}
\text{{\it JK-chamber} } := \text{ a connected component of }\mathfrak{h}^{\ast} \backslash  \mathrm{Cone}_{\text{sing}} [\cup_{{\bm \phi}_*} {\bm Q}_{*}] \label{def-JK-chamber}
\end{align}
to distinguish it from an FI-chamber defined in~(\ref{def-FI-chamber}).
Here
$\cup_{{\bm \phi}_*} {\bm Q}_*$ is the union of the~${\bm Q}_*{ (\bm{\phi}_*)}$'s for all  the poles ${\bm \phi}_*$  
and 
$\mathrm{Cone}_{\text{sing}} [\cup_{{\bm \phi}_*} {\bm Q}_{*}] $  is the union of all the cones generated by subsets of $\cup_{{\bm \phi}_*} {\bm Q}_*$ with $n-1$ elements.
$\mathrm{Cone}_{\text{sing}} [\Sigma {\bm Q}_*]$ divides a JK-chamber into  subchambers.
The prescription above to shift $\widetilde{\bm \zeta}$ to $\widetilde{\bm \zeta}'$ is motivated by the fact that the expression
$$
\frac{1}{|W|} \sum_{{\bm \phi}_* } \mathop{{\rm JK} \mathchar `- {\rm Res} }_{{\bm \phi}={\bm \phi}_*} ({\bm Q}_{\ast}, {\bm \eta} ) 
\prod Z_{\text{cs}} \prod Z_{\text{vec}}^{\prime} \prod Z_{\text{hyp}} \prod Z_{\text{Fermi}} \,
 d \phi_1 \wedge \cdots \wedge  d \phi_{\mathrm{dim} \;\mathfrak{h} }
 $$ 
 as a function of $\bm{\eta}$ is constant as long as~$\bm{\eta}$ stays within the same JK-chamber~\cite{Szenes-Vergne,Benini:2013xpa}.

\section{SQMs from branes for $U(2)$ SQCD with four flavors}\label{sec:branes}


\begin{figure}[t]
\begin{center}
\subfigure[]{\label{fig1:negative-p1-and-p2_1}
\includegraphics[width=5cm]{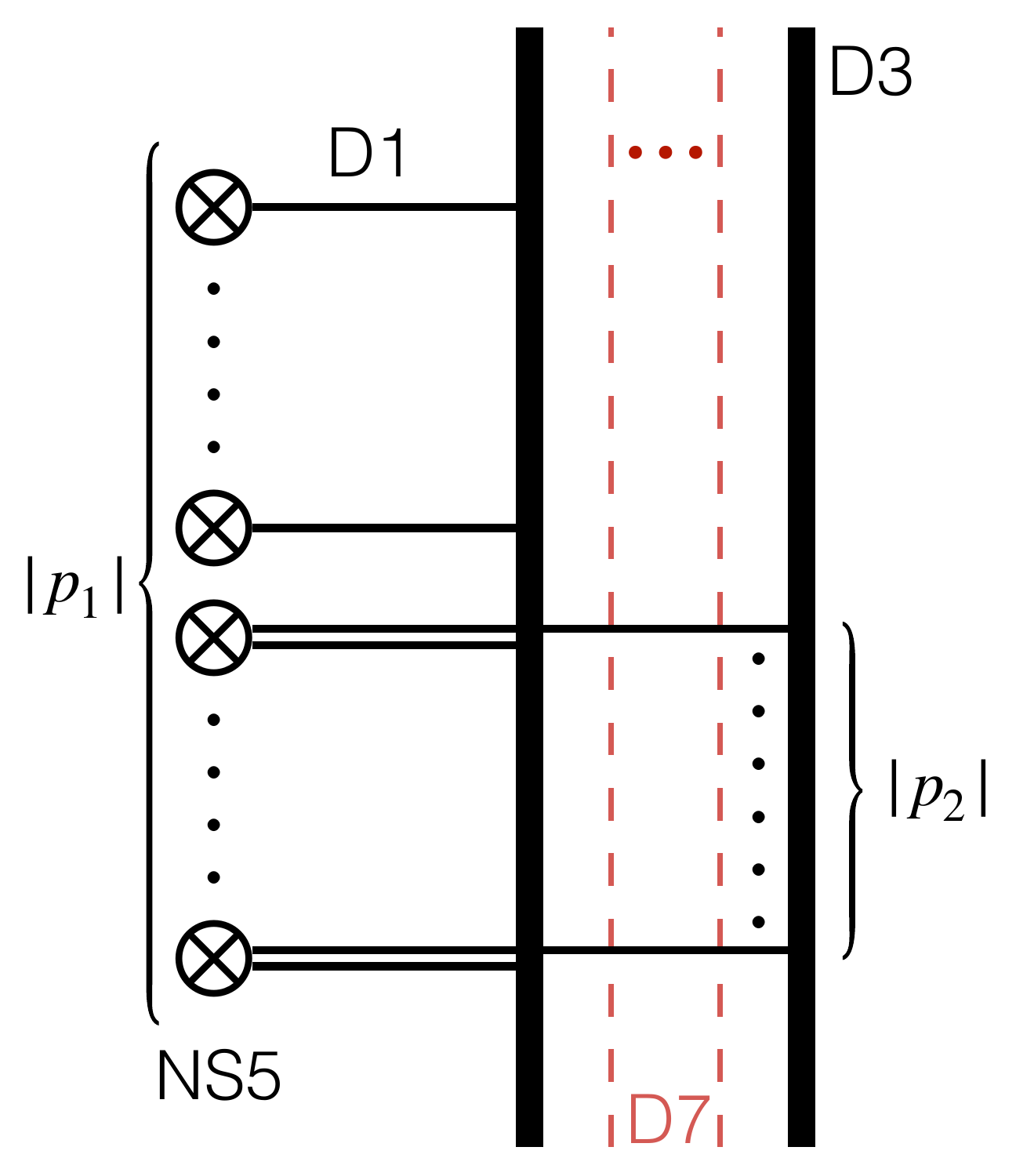}}
\hspace{0mm}
\subfigure[]{\label{fig2:negative-p1-and-p2_1}
\includegraphics[width=5cm]{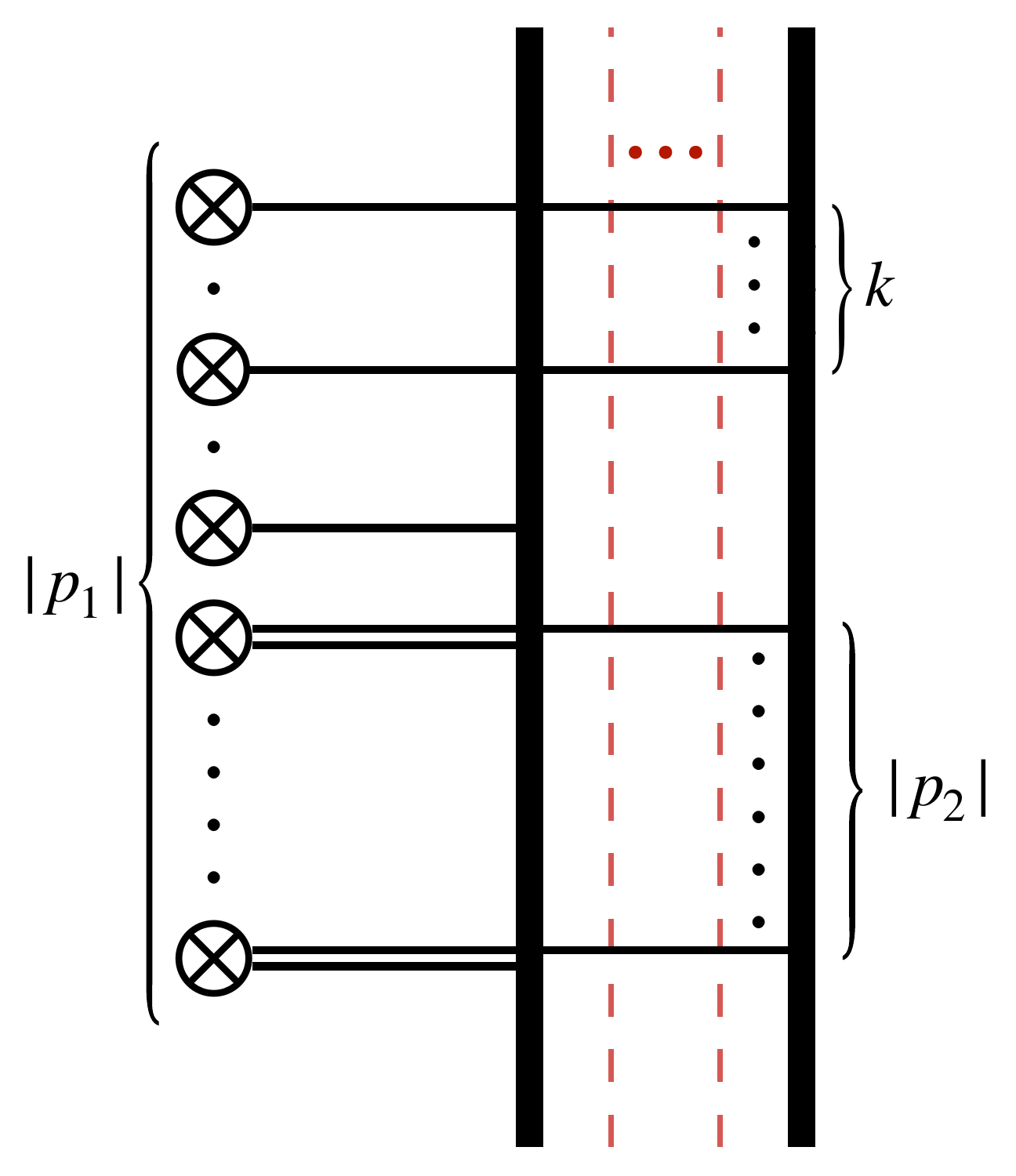}}
\hspace{0mm}
\subfigure[]{\label{figure:negative-p1-and-positive-p2}
\raisebox{1.5cm}{\includegraphics[width=4.5cm]{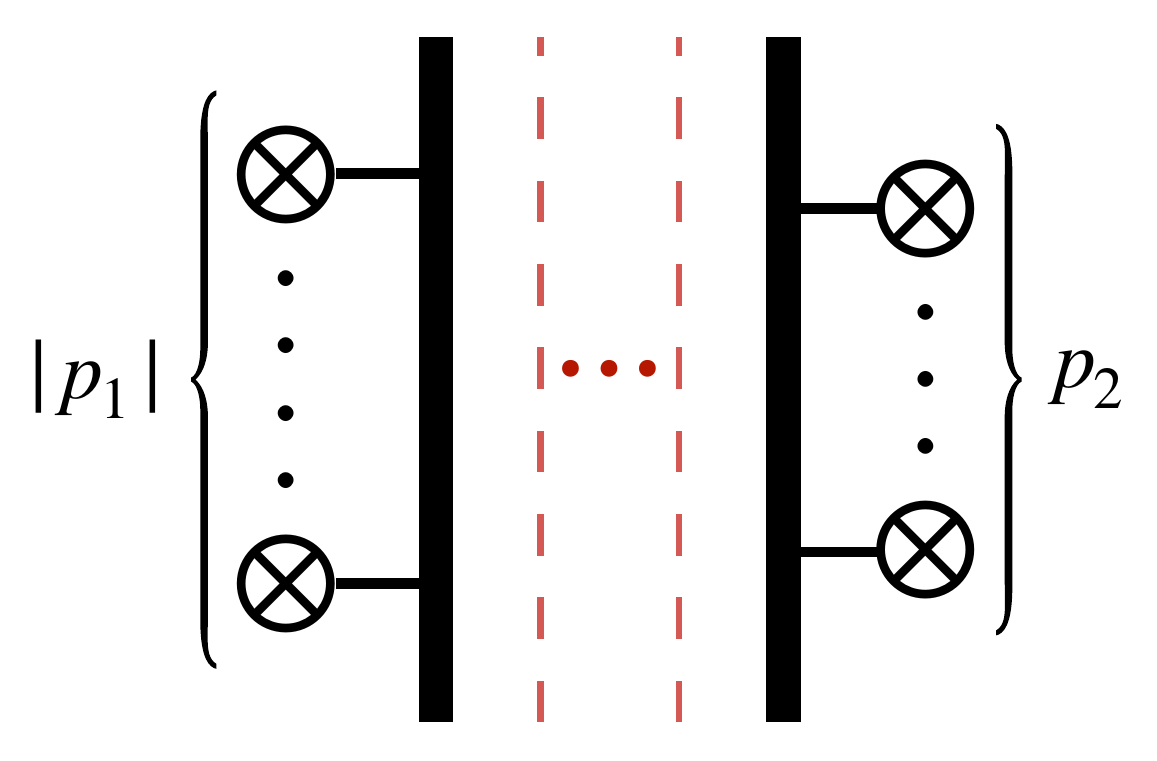}}}
\end{center}
\caption{(a): A brane configuration that realizes the 't Hooft operator with $\bm B=(p_1,p_2)$ such that $p_1\leq p_2\leq 0$.  
(b): A brane configuration corresponding to the screening of $\bm B=(p_1,p_2)$ to $\bm v= (p_1+k,p_2-k)$. 
(c): A brane configuration that realizes the 't Hooft operator with $\bm B=(p_1,p_2)$ such that $p_1\leq 0 \leq p_2$.
}\label{figure:negative-p1-and-p2_1}
\end{figure}

In this appendix we consdier SQMs which describe monopole screening contributions for a product of 't Hooft operators of the 4d $\mathcal{N}=2$ $U(2)$ gauge theory with $N_F = 4$ flavors. We label the magnetic charge of the 't Hooft operator by ${\bm B} = (p_1, p_2)$ where $p_1$, $p_2$ are integers and satisfy $p_1 \leq p_2$. The monopole screening sectors are specified by ${\bm v} = (p_1 + k, p_2 - k)$. We here assume $1 \leq k \leq \frac{n}{2}$ where $n = |p_1| - |p_2|$. 
The SQM describing the monopole screening sectors can be read off from the brane configuration considered in Section \ref{sec:SQMbrane}. We have two D3-branes, four D7-branes for realizing the 4d $\mathcal{N}=2$ $U(2)$ gauge theory with four flavors. Depending on $p_1$, $p_2$, we introduce NS5-branes on the leftmost side and/or rightmost side. Our convention is that a D1-brane ending on a D3-brane from the right (or left) injects a Dirac monopole of charge $+1$ (or respectively $-1$) and we label D3-brane from left to right in the diagram. 
There are $k$ finite D1-branes between the two D3-branes; they realize smooth monopoles.

Our aim here is to derive the quivers of the SQMs that capture all the monopole screening sectors which satisfy the conditions above, paying special attention to which gauge node short Fermi multiplets couple to.
In particular we find that Fermi multiplets can couple to a gauge node that is not necessarily at the center of the quiver if we choose an appropriate charge~${\bm B}$.
In the following discussion we do not distinguish brane configurations that differ by the $x^3$-values of the NS5-branes, although they give rise to different expectation values of  't Hooft operator correlation functions as we demonstrate in the main text. 
We also assume that the Chern-Simons level in each gauge node is zero.

\begin{figure}[t]

\begin{center}
\subfigure[]{\label{figure:negative-p1-and-p2-SQM-frame}
\raisebox{-5mm}{\includegraphics[width=11cm]{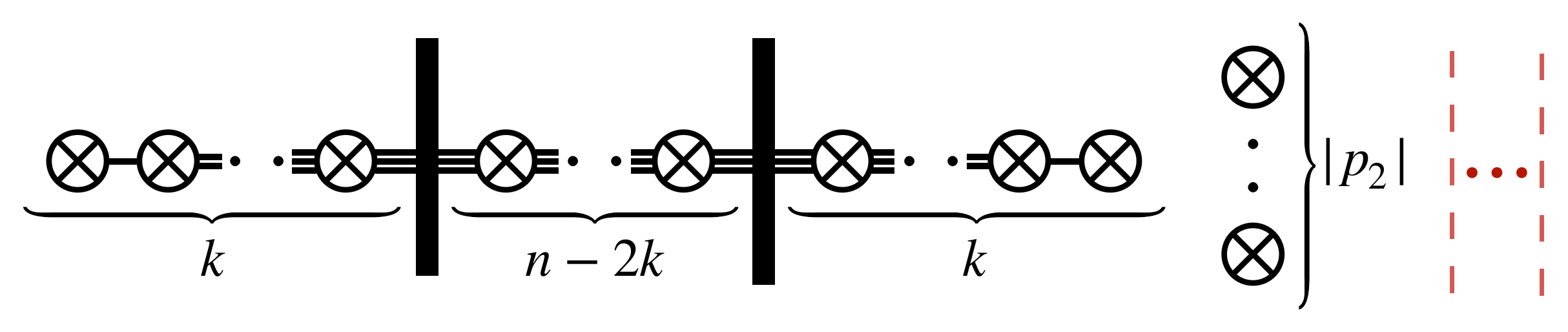}}}
\subfigure[]{\label{figure:negative-p1-and-p2-quiver}
\raisebox{-5mm}{\includegraphics[width=11cm]{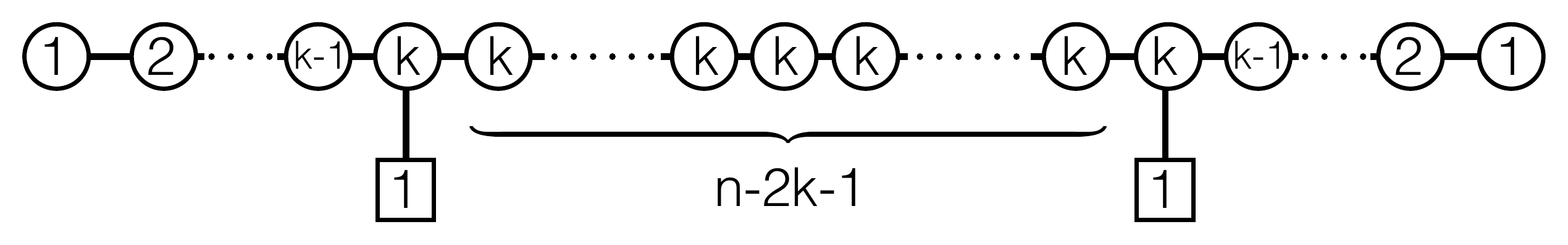}}}
\end{center}
\caption{(a):The $|p_2|$ NS5-branes and the D7-branes on the right are decoupled from the rest of the system.
(b): The SQM quiver realized on the D1-branes in the brane configuration in Figure~\ref{figure:negative-p1-and-p2-SQM-frame}. There are $(n-2k+1)$ $U(k)$ gauge nodes. There is no Fermi multiplet. }
\end{figure}

\paragraph{Case $p_1\leq p_2 \leq 0$.}

The 't Hooft operator $T_{\bm B}$ in this case is realized by the brane configuration shown in Figure~\ref{fig1:negative-p1-and-p2_1}.
The vertical dashed lines represent four D7-branes.
It is important that in the initial configuration, the NS5-branes are placed at the leftmost side
of the system. The monopole screening sector ${\bm v} = (p_1 + k, p_2 - k)$ is described by the diagram in Figure~\ref{fig2:negative-p1-and-p2_1} where we introduced $k$ D1-branes to the diagram in Figure~\ref{fig1:negative-p1-and-p2_1}.

To read off the SQM, we perform a sequence of Hanany-Witten transitions~\cite{Hanany:1996ie} to the configurations in Figures~\ref{fig2:negative-p1-and-p2_1}. Note that D7-branes commute with D3-branes since they are both point-like in a two-dimensional place. 
After a series of HW transitions
the brane configuration becomes the one in Figure~\ref{figure:negative-p1-and-p2-SQM-frame} from which we can read off an SQM living on the D1-branes. Each D3-brane yields an $\mathcal{N}=(0, 4)$ hypermultiplet and it is coupled to the degrees of freedom on the D1-branes which intersect with the D3-brane. The resulting SQM quiver theory is depicted in Figure \ref{figure:negative-p1-and-p2-quiver}\footnote{%
The case $k>n/2$ can also be analyzed by exchanging the two D3-branes and leads to the same SQM.
}.
Since D7-branes are all decoupled from the system, we see that there is no Fermi multiplet in the quiver in Figure \ref{figure:negative-p1-and-p2-quiver}.

\begin{figure}[t]
\begin{center}
\subfigure[]{\label{figure:negative-p1-and-positive-p2-SQM-frame}\includegraphics[width=12cm]{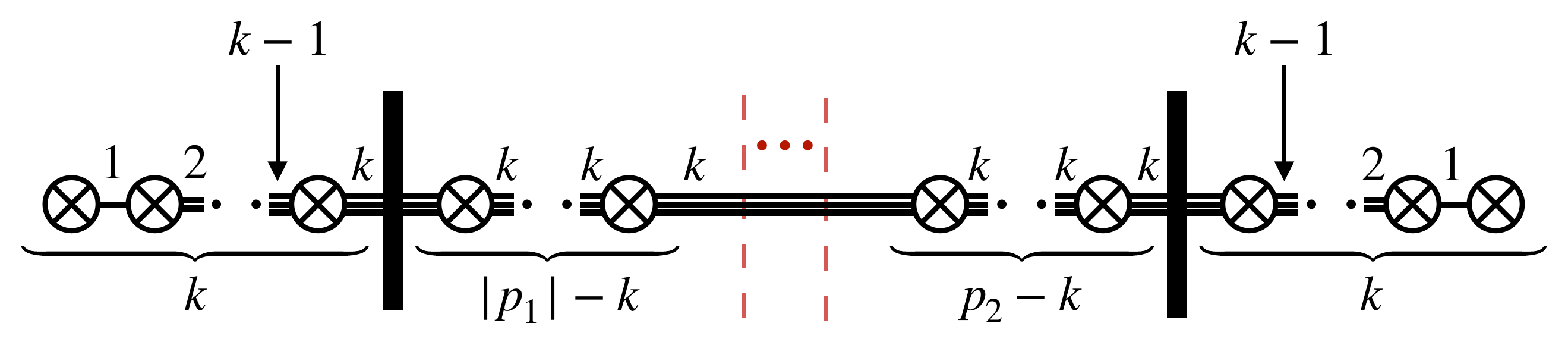}}
\subfigure[]{\label{figure:U2-quiver-with-Fermi}\includegraphics[width=12cm]{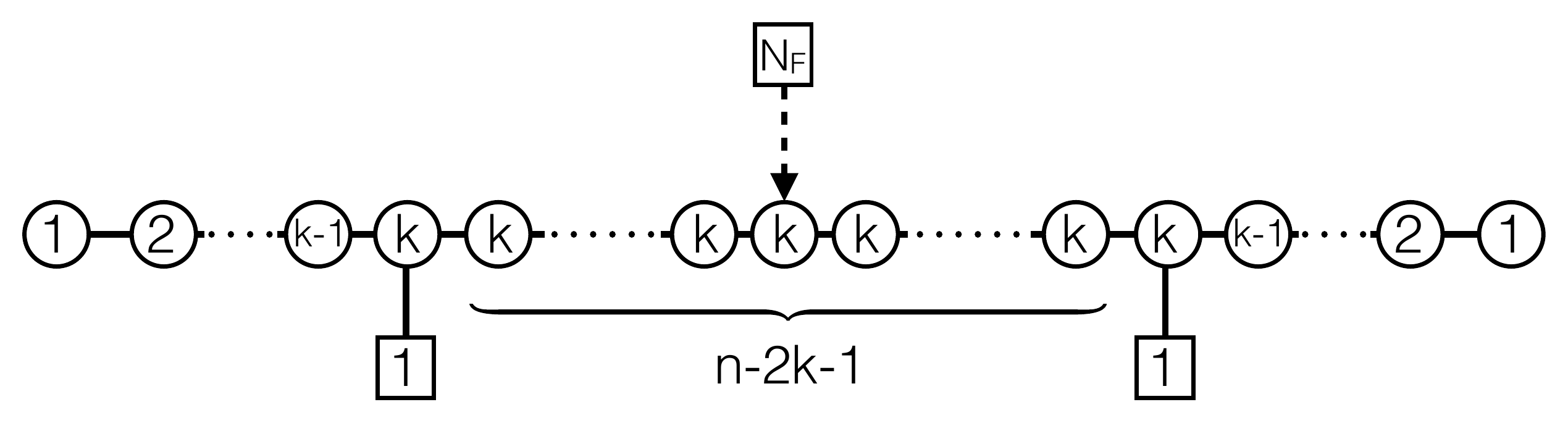}}
\end{center}
\caption{(a): A HW frame from which we can read off the SQM quiver for  $ k \leq \text{min}\{ |p_1|,p_2\}$. (b): The SQM quiver theory realized on the D1-branes in Figure \ref{figure:negative-p1-and-positive-p2-SQM-frame}.}
\end{figure}


\paragraph{Case $p_1\leq 0\leq p_2 $.}

We start with brane system, shown in Figure~\ref{figure:negative-p1-and-positive-p2},  which realizes the 't Hooft operator $T_{\bm B}$.
It is important that in this initial configuration the NS5-branes are either to the left or the right of the rest of the system. Then the monopole screening sector ${\bm v} = (p_1 + k, p_2 -k)$ is realized by introducing $k$ D1-branes between the two D3-branes. In this case, we can obtain three different types of quiver theories.

\begin{enumerate}
\item Case $ k \leq \text{min}\{ |p_1|,p_2\}$.

Via a series of transitions and moving D1-branes and NS5-braner, paying attention to the fact that NS5-branes cannot pass over D7-branes, we can obtain the brane configuration shown in Figure~\ref{figure:negative-p1-and-positive-p2-SQM-frame}. Then the SQM living on the D1-branes in Figure~\ref{figure:negative-p1-and-positive-p2-SQM-frame} yields the quiver theory in Figure~\ref{figure:U2-quiver-with-Fermi}. In this case, the Fermi multiplet is coupled to the $|p_1|$-th gauge node from the left, which is one of the $U(k)$ nodes in the middle.

\item Case $k\geq  \text{min}\{ |p_1|,p_2\}$.

\begin{figure}[t]
\begin{center}
\subfigure[]{\label{figure:negative-p1-and-positive-p2-SQM-frame2}\includegraphics[width=12cm]{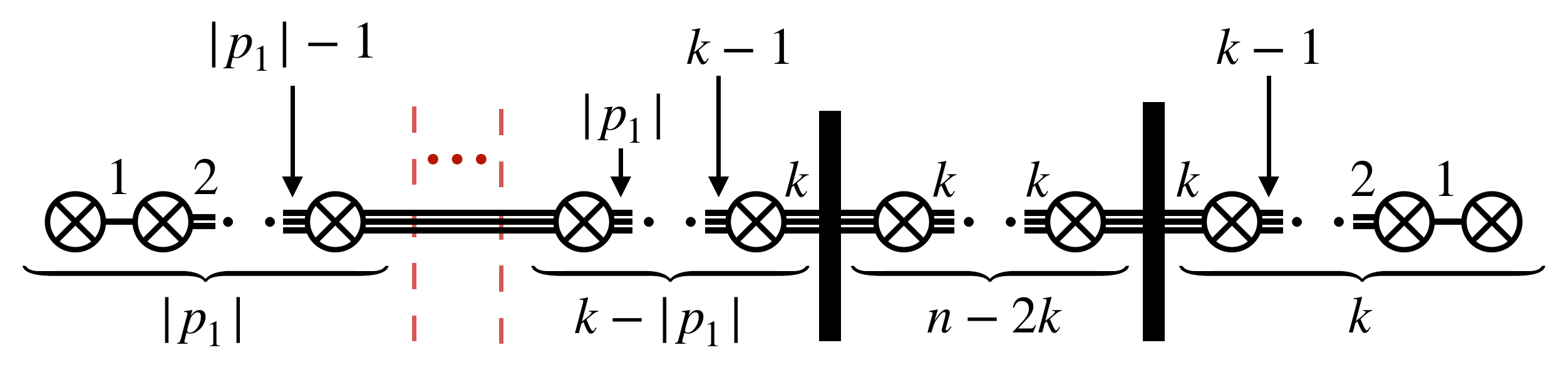}}
\subfigure[]{\label{figure:U2-quiver-with-Fermi2}\includegraphics[width=12cm]{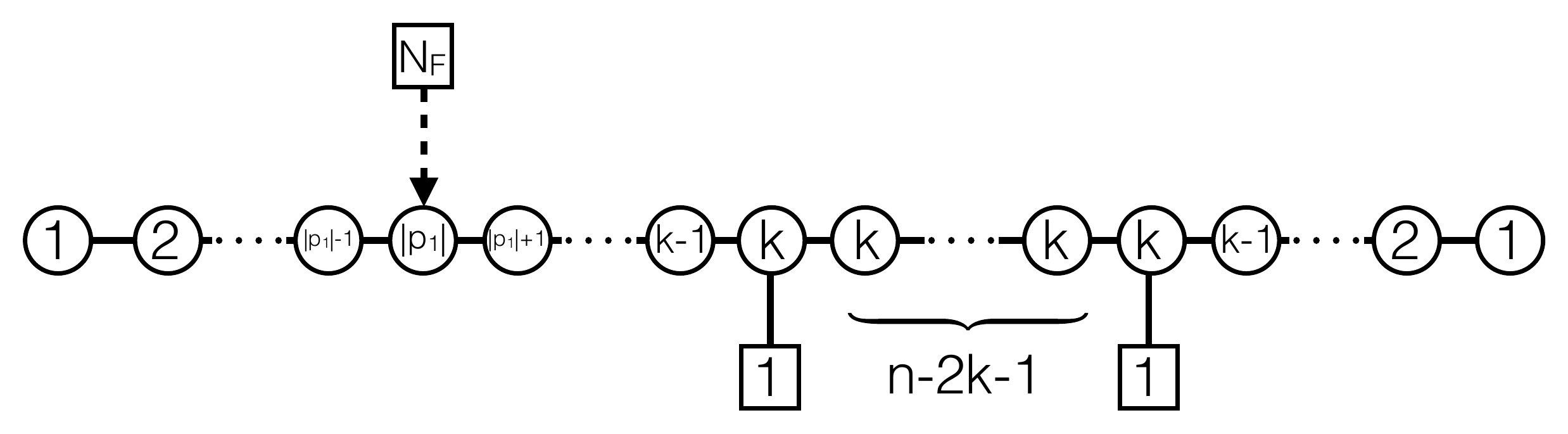}}
\end{center}
\caption{(a): A HW frame from which we can read off the SQM quiver for the case with $k\geq  \text{min}\{ |p_1|,p_2\}$ and $|p_1|\leq  p_2$. (b): The SQM quiver corresponding to the brane configuration in Figure~\ref{figure:negative-p1-and-positive-p2-SQM-frame2}.  Four short Fermi multiplets are coupled to the $U(|p_1|)$ node in the left half of the quiver.}
\end{figure}

\begin{figure}[t]
\begin{center}
\includegraphics[scale=.5]{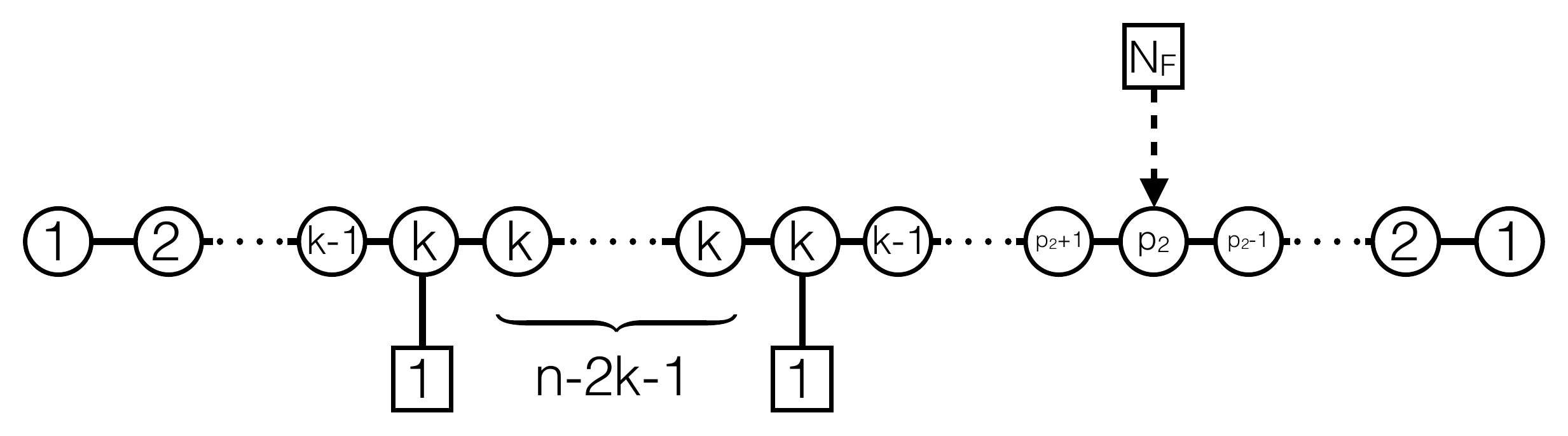}
\end{center}
\caption{The SQM quiver corresponding for $k\geq  \text{min}\{ |p_1|,p_2\}$ and $|p_1|\geq  p_2$.  The Fermi multiplets are coupled to the $U(p_2)$ node in the right half of the quiver. }\label{figure:U2-quiver-with-Fermi3}
\end{figure}

\begin{enumerate}
\item Case $|p_1|\leq  p_2$.

Through a sequence of transitions and moving D1-branes and NS5-branes, we can arrive at the configuration in Figure~\ref{figure:negative-p1-and-positive-p2-SQM-frame2}. The SQM quiver is the one shown in Figure~\ref{figure:U2-quiver-with-Fermi2}. The Fermi multiplet is now coupled to the $U(|p_1|)$ node in the left half of the quiver.

\item Case $|p_1|\geq  p_2$.

This is related to the case $|p_1|\leq  p_2$ by $(p_1,p_2) \rightarrow (-p_2,-p_1)$. The SQM quiver is shown in Figure~\ref{figure:U2-quiver-with-Fermi3}. The Fermi multiplet is coupled to the $U(p_2)$ node in the right half of the quiver.

\end{enumerate}

\end{enumerate}

\paragraph{Case $0\leq p_1\leq p_2 $.}

This is related to the case with $p_1\leq p_2 \leq 0$ by a replacement $(p_1,p_2) \rightarrow (-p_2,-p_1)$.
The SQM quiver in the case with $p_1\leq p_2 \leq 0$ is given by the quiver in Figure~\ref{figure:negative-p1-and-p2-quiver}. Since the quiver diagram is invariant under the replacement and hence the case with $0\leq p_1\leq p_2 $ also gives rise to the quiver in Figure~\ref{figure:negative-p1-and-p2-quiver}. As in the case with $p_1\leq p_2 \leq 0$, there is no Fermi multiplet either in this case.

\bibliography{refs}

\end{document}